\documentclass[iop,apj]{emulateapj}%
\usepackage{natbib}
\usepackage{epsf}
\usepackage{graphicx}
\usepackage{amssymb}
\usepackage{footnote}
\usepackage{aasmacros}
\usepackage[flushleft]{threeparttable}
\usepackage{color}

\usepackage{mathtools}
\usepackage[squaren, Gray, cdot]{SIunits}
\definecolor{Pink}{rgb}{1.0,0.05,0.5}
\definecolor{Orange}{rgb}{1.0,0.05,0.15}
\definecolor{Green}{rgb}{0.15,0.45,0.25}
\definecolor{Blue}{rgb}{0.0,0.08,0.65}
\definecolor{Brown}{rgb}{0.7,0.25,0.0}

\shortauthors{Laigle et al.}
\shorttitle{COSMOS2015}
\makeatother

 \usepackage{hyperref}
  \hypersetup{breaklinks,colorlinks,urlcolor=Blue,citecolor=black}
  \usepackage[hyphenbreaks]{breakurl} % Has to be loaded after hyperref

\begin{document}
\title{The COSMOS2015 catalog: exploring the $1<z<6$ Universe with half a million galaxies}

\author{C.~Laigle\altaffilmark{1},
  H. J. McCracken\altaffilmark{1},
  O. Ilbert\altaffilmark{2}, B.~C.~Hsieh\altaffilmark{3}, I. Davidzon\altaffilmark{2}, P. Capak\altaffilmark{4}, G. Hasinger\altaffilmark{5},
  J.~D.~Silverman\altaffilmark{6}, C.~Pichon\altaffilmark{1}$^{,}$\altaffilmark{7},
  J. Coupon\altaffilmark{8}, H. Aussel\altaffilmark{9}, D. Le
  Borgne\altaffilmark{1}, K.~Caputi\altaffilmark{15}, P.~Cassata\altaffilmark{20}, Y.-Y. Chang\altaffilmark{9},  F. Civano\altaffilmark{17}$^{,}$\altaffilmark{18}, J. Dunlop\altaffilmark{14},  J.~Fynbo\altaffilmark{13},  J.~S.~Kartaltepe\altaffilmark{22},
A. Koekemoer\altaffilmark{16}, O. Le
  F\`evre\altaffilmark{2},  E. Le
  Floc'h\altaffilmark{9}, A. Leauthaud\altaffilmark{6}, S. Lilly\altaffilmark{22},
  L. Lin\altaffilmark{3}, S. Marchesi\altaffilmark{17}$^{,}$\altaffilmark{19}, B.~Milvang-Jensen\altaffilmark{13},
  M. Salvato\altaffilmark{10}, D.~B.~Sanders\altaffilmark{5}, N. Scoville\altaffilmark{4}, V. Smolcic\altaffilmark{21},
  M. Stockmann\altaffilmark{13},
  Y. Taniguchi\altaffilmark{12}, L.~Tasca\altaffilmark{2},
  S.~Toft\altaffilmark{13}, M. Vaccari\altaffilmark{23},
  J.~Zabl\altaffilmark{13} }

\altaffiltext{1}{Sorbonne Universit\'e, UPMC Univ Paris 06, and CNRS, UMR 7095, IAP, 98b bd Arago, F-75014, Paris, France}
\altaffiltext{2}{LAM, Universit\'e d'Aix-Marseille \& CNRS, UMR7326, 38
rue F. Joliot-Curie, 13388 Marseille Cedex 13, France}
\altaffiltext{3}{Institute of Astronomy \& Astrophysics, Academia
  Sinica, P.O. Box 23-141, Taipei 106, Taiwan, R.O.C.}
\altaffiltext{4}{California Institute of Technology,
 Pasadena, USA}
\altaffiltext{5}{Institute for Astronomy, 2680 Woodlawn Drive
 Honolulu, HI 96822-1839 USA}
\altaffiltext{6}{Kavli Institute for the Physics and Mathematics of the Universe (WPI), Todai Institutes for Advanced Study, The University of Tokyo, Kashiwa, Chiba 277-8583, Japan}
\altaffiltext{7}{Institute of Astronomy, University of Cambridge,
 Madingley Road, Cambridge, CB3 0HA, United Kingdom}
\altaffiltext{8}{Astronomical Observatory of the University of Geneva,
ch. d'Ecogia 16, 1290 Versoix, Switzerland}
\altaffiltext{9}{Laboratoire AIM Paris-Saclay, UMR 7158, CEA, CNRS,
 Universit\'e Paris VII, CE-SACLAY, Bat 709, F91191 Gif-sur-Yvette, France}
\altaffiltext{10}{Max-Planck-Institut f{\"u}r extraterrestrische Physik,
 Giessenbachstrasse, D-85748 Garching, Germany}
\altaffiltext{11}{Institute of Astronomy, Department of Physics, ETH
 Zurich, Switzerland}
\altaffiltext{12}{ Faculty of Science and Engineering, Kindai
  University, Japan}
\altaffiltext{13}{Dark Cosmology Centre, Niels Bohr Institute, Copenhagen University,
Juliane Maries Vej 30, 2100 Copenhagen O, Denmark}
\altaffiltext{14}{Institute for Astronomy, University of Edinburgh,
  Royal Observatory, Edinburgh, EH9 3HJ, UK}
\altaffiltext{15}{Kapteyn Astronomical Institute, University of Groningen, P.O. Box 800,
9700AV Groningen, The Netherlands}
\altaffiltext{16}{Space Telescope Science Institute, Baltimore, USA}
\altaffiltext{17}{Yale Center for Astronomy and Astrophysics, 260
  Withney Avenue, New Haven, CT 06520, USA}
\altaffiltext{18}{Smithsonian Astrophysical Observatory, 60 Garden
  Street, Cambridge, MA 02138, USA}
\altaffiltext{19}{Dipartimento di Fisica e di Astronomia, Universita
  di Bologna, Via Ranzini 1, 40127, Bologna, Italy}
 \altaffiltext{20}{Instituto de fisica Y astronomia, Universidad de Valparaiso, Av. Gran Bretana, 1111, Valparaiso, Chile}
  \altaffiltext{21}{Department of Physics, University of Zagreb, Bijeniÿcka cesta 32, HR-10000 Zagreb, Croatia}
  \altaffiltext{22}{School of Physics and Astronomy,  Institute of
    Technology, 84 Lomb Memorial Drive, Rochester, NY 14623, USA}
 \altaffiltext{23}{Astrophysics Group, Physics
Department, University of the Western Cape, Private Bag X17, 7535,
Bellville, Cape Town, South Africa}
 %}

%
%\maketitle

%-----------------------
\begin{abstract}

  We present the COSMOS2015\footnote{Based on data obtained
      with the European Southern Observatory Very Large Telescope,
      Paranal, Chile, under Large Programs 175.A-0839 (zCOSMOS),
      179.A-2005 (UltraVista) and 185.A-0791 (VUDS).} catalog which
  contains precise photometric redshifts and stellar masses for more
  than half a million objects over the 2deg$^{2}$ COSMOS
  field. Including new $YJHK_{\rm s}$ images from the UltraVISTA-DR2
  survey, $Y$-band from Subaru/Hyper-Suprime-Cam and infrared data from
 the Spitzer Large Area Survey with the Hyper-Suprime-Cam \textit{Spitzer} legacy program, this near-infrared-selected catalog
  is highly optimized for the study of galaxy evolution and
  environments in the early Universe. To maximise catalog completeness
  for bluer objects and at higher redshifts, objects have been
  detected on a $\chi^{2}$ sum of the $YJHK_{\rm s}$ and $z^{++}$
  images. The catalog contains $\sim 6\times 10^5$ objects in the 1.5
  deg$^{2}$ UltraVISTA-DR2 region, and $\sim 1.5\times 10^5$ objects are
  detected in the ``ultra-deep stripes'' (0.62 deg$^{2}$) at
  $K_{\rm s}\leq 24.7$ (3$\sigma$, 3$\arcsec$, AB magnitude).
  Through a comparison with the zCOSMOS-bright spectroscopic redshifts, we
  measure a photometric redshift precision of
  $\sigma_{\Delta z/(1+z_s)}$ = 0.007 and a catastrophic failure
  fraction of $\eta=0.5$\%. At $3<z<6$, using the unique database of
  spectroscopic redshifts in COSMOS, we find
  $\sigma_{\Delta z/(1+z_s)}$ = 0.021 and $\eta=13.2\% $. The deepest
  regions reach a 90\% completeness limit of 10$^{10}M_\odot$ to
  $z=4$. Detailed comparisons of the color distributions, number
  counts, and clustering show excellent agreement with the literature
  in the same mass ranges. COSMOS2015 represents a unique, publicly
  available, valuable resource with which to investigate the evolution of galaxies
  within their environment back to the earliest stages of the history
  of the Universe.  The COSMOS2015 catalog is distributed
    via anonymous
    ftp\footnote{\url{ftp://ftp.iap.fr/pub/from_users/hjmcc/COSMOS2015/}}
    and through the usual astronomical archive systems (CDS, ESO, IRSA).

\end{abstract}

%-----------------------

%%%%%%%%%%%%%%%
\section{Introduction}
\label{sec-preamble}
%%%%%%%%%%%%%%%%%%%%%%%%%%%%%%%%%%%%%%%%%%%%%%%%%%%%%%%
%%Context
%%%%%%%%%%%%%%%%%%%%%%%%%%%%%%%%%%%%%%%%%%%%%%%%%%%%%%%

Our understanding of the formation, evolution and large-scale
distribution of galaxies has been revolutionized in the past decade
by the availability of large, multi-wavelength data sets accurately
calibrated with densely sampled spectroscopic training sets. In
parallel, the availability of exponentially increasing computing power
has led to the development of {\it ab initio} cosmological simulations
which can now include most of the known baryonic physics processes
down to relatively small scales \citep[approximately kiloparsecs or less,
e.g.][]{2014MNRAS.444.1453D,2015Khandai,2014MNRAS.444.1518V,2015MNRAS.446..521S}
raising the possibility of detailed comparison with observational
surveys.  Such simulations can now reproduce the rich diversity of
observed colors, morphologies and star formation activity though a
complex combination of internal and external processes (such as
feedback, turbulence, smooth accretion, dry minor mergers, and mergers)
occurring at different scales and times. However, the exact balance
between all of these processes and how they affect galaxy evolution and
shape galaxy properties is still actively debated.

%%%%%%%%%%%%%%%%%%%%%%%%%%%%%%%%%%%%%%%%%%%%%%%%%%%%%%%
%%questions that we address about SF and quenching
%%%%%%%%%%%%%%%%%%%%%%%%%%%%%%%%%%%%%%%%%%%%%%%%%%%%%%%

Observationally, it is now clear that by $z\sim1$ most of the mass has
already assembled into galaxies. At high redshifts, star formation
occurs vigorously in blue, massive galaxies and with the passage of
cosmic time the peak of star formation activity shifts to
progressively lower-mass objects at lower redshifts
\citep[e.g][]{Cowie:1996p8471,Arnouts:2007p3665,2007A&A...474..443P,Noeske:2007p5802}.
%Lilly et al. 1996, Schiminocih et al. 2005, Le Floc'h et al. 2005
However, despite the success of phenomenological models in reproducing
at least some of these observational trends \citep{Peng:2010p11940},
the precise physical mechanisms of this ``quenching'' process remain
a topic of debate.  Since cold gas is the basic fuel for galaxies to form stars,
a better understanding of how gas accretion feeds galaxies and of the
effect of possible outflows -- which could stop the gas supply in
galaxies -- are crucial to explain both the peak of star formation at
high redshift and its quenching at lower redshifts.

%%%%%%%%%%%%%%%%%%%%%%%%%%%%%%%%%%%%%%%%%%%%%%%%%%%%%%%
%%what we know about gas fuelling in galaxies and star formation and how
%%to study it
%%%%%%%%%%%%%%%%%%%%%%%%%%%%%%%%%%%%%%%%%%%%%%%%%%%%%%%

%%%%1-The Main sequence suggests that accretion is smooth

The small dispersion in the galaxy ``main sequence'' (the observed
proportionality between star formation rate (SFR) and stellar mass)
found at $0<z<2$ \citep[e.g][]{Daddi:2007p2924} is reproduced in
hydrodynamical simulations and is now shown to exist up to $z\sim6.5$
\citep[e.g][]{2014ApJ...791L..25S,2015ApJ...799..183S} and
down to $\log M/M_{\odot}\sim 9.4$ \citep{2015Koch} , although the
different methods used to compute the stellar mass and SFR, in addition to
sample selection effects, are still producing partially inconsistent
results at high redshift \citep{2012ApJ...758L..31L}. This SFR-stellar
mass relation nonetheless clearly suggests that the mass assembly
should be smooth compared to a clumpy accretion driven by mergers.
However the privileged mode of smooth gas accretion remains unclear.

%%%%2-distinguish b/w cold and hot flows

The conventional model relied on the ``hot mode'' accretion scenario,
in which the infalling gas is shock-heated at the virial radius and
then radiatively cools starting from the central part and forming
centrifugally supported disk
\citep[e.g.][]{1977MNRAS.179..541R,1978MNRAS.183..341W}.  However, recent
hydrodynamical simulations now suggest however now that most of the gas is
accreted directly from cold dense filaments without being
shock-heated
\citep{2003ASSL..281..185K,2005MNRAS.363....2K,2008MNRAS.390.1326O,2009Natur.457..451D}
at least for lower-mass haloes at high redshift.  In this context, the
anisotropic large-scale environment of galaxies is therefore likely to
play an important role as it literally drives such cold flow
accretion.

Most observational analyses define ``environment'' as well-defined
structures \citep[clusters/groups/pairs and field galaxies,
e.g.][]{2014ApJ...782...33L} or using isotropic galaxy-density
estimators \citep[such as nearest neighbors
e.g.][]{1980ApJ...236..351D,Elbaz:2007p11978}. Galaxies are found to be more
massive and much less star-forming in high-density regions relative to low-density
regions \citep[e.g.][]{2004MNRAS.353..713K} which is consistent with
the clustering measurements of ultraviolet-selected galaxies
\citep{2007ApJS..173..503H,2007ApJS..173..494M}. Using local samples,
\cite{Peng:2010p11940} have demonstrated that quenching of star
formation activity can be separated into environmental (density
dependent) and internal (galaxy mass related) effects, suggesting that
nature and nurture both act in shaping galaxy properties.

Recent theoretical works have also predicted that there is a
significant connection between the dynamics within the intrinsically
anisotropic large-scale structures on the one hand, and the physical
properties of the galaxies embedded in them on the other hand. In
particular, the vorticity-rich filaments
\citep{2013ApJ...766L..15L,2015MNRAS.446.2744L} are the locus where
low-mass galaxies steadily grow in mass via quasi-polar cold gas
accretion, while their angular momentum (spin) is aligned with host
filaments \citep{2012MNRAS.427.3320C,2015MNRASCodis}. Mergers are
responsible for the spin flip along the filaments
\citep{2014MNRAS.445L..46W}, so that the flip should, in
principle, be traced in the distribution of the galaxy properties
(morphology, SFR) {\sl along} the ``cosmic web''
\citep{1996AAS...189.1303P}.  Correlations have already been found in
hydrodynamical simulations between the evolution of the physical
properties of galaxies (SFR, stellar mass, colors, metallicity) as a
function of the galaxy-spin alignment within the filaments
\citep{2014MNRAS.444.1453D}.

Notwithstanding some observational studies \citep[see also,
e.g.][]{2013ApJ...775L..42T,2013ApJS..206....3S,2014ApJ...796...51D},
accurately tracing the cosmic web remains challenging as long as we do
not observe a sufficiently large area (at least on the scale of a few
typical void sizes) with sufficiently precise galaxy redshifts to trace the
structures. Therefore, one of the outstanding challenges for the next
generation of deep multi-band surveys over wide fields is to enable
environmental studies while at the same time probing different epochs of
cosmic evolution to leverage their relative importance in building up
galaxies and also to detect the transition between different accretion
modes.
%%%s%3-

A method which could be more robust for constraining galaxy mass assembly
would be to
investigate the relationship between the integrated stellar
properties of galaxies (in particular, stellar mass, star formation
rate, and star formation history(SFH)) and their dark matter environment over a
range masses and redshifts. The gas accretion mode is expected to be
intimately connected to the halo mass and, depending on the dominant
scenario, the SFHs of galaxies will be
different due to the cooling delay implied by the ``hot mode''
accretion. In practice, the stellar-to-halo mass relation (SHMR) is
derived statistically by comparing the galaxy clustering measurement
with predictions from the phenomenological halo model
\citep[e.g.][]{Cooray:2002p846}. Already extensively studied up to
$z\sim2$
\citep[e.g.][]{Bethermin:2014dh,2015McCracken,2015MNRASCoupon},
this relationship is worth extending at higher redshift and for lower-mass galaxies, which requires sufficiently large and deep
data sets. Moreover, other halo-mass-dependent effects play a
non-negligible, if not crucial, role in regulating star formation,
especially feedback from active galactic nuclei (AGNs), either in a
negative sense \citep[e.g.][]{2006MNRAS.367..864C,2006ApJS..163....1H}
or a positive
\citep[e.g.][]{2012MNRAS.425..438G,2015ApJ...812L..36B}. This makes it
difficult to disentangle between all of the different mass-dependent
processes that affect star formation, unless robust
observations of the AGN population are available in the same field.

%%%%%%%%%%%%%%%%%%%%%%%%%%%%%%%%%%%%%%%%%%%%%%%%%%%%%%%
%%what we need and why COSMOS is the best field for this
%%%%%%%%%%%%%%%%%%%%%%%%%%%%%%%%%%%%%%%%%%%%%%%%%%%%%%%

%%%%1-what we need

Taking these considerations into account, it is
clear that new observational studies will require deep, near-infrared
(NIR)-and infrared (IR)-selected data. This will allow us to extend
stellar mass measurements and photometric redshift catalogs to higher
redshifts and lower stellar masses over the largest possible redshift range.
In particular, the challenge is to cover
\textit{simultaneously} in the same dataset the low-mass and
high-redshift ranges of the galaxy population. Especially the
redshift range $1<z<4$ where galaxies are most actively forming
stars. As most spectral features move into the rest-frame optical in
these redshift ranges, NIR data is essential for accurate photometric
redshift and stellar mass estimates.  Covering a large area is also
essential to derive robust statistical $N-$point functions or
count in cells, to probe a variety of galaxy environments, to trace
accurately the large-scale structure, and to minimize the effect of
cosmic variance. In addition, providing large numbers of bright, rare
objects is essential for ground-based follow-up spectroscopy. 

%%%%2-COSMOS field and this catalog

The COSMOS project has already pioneered the study of galactic
structures at intermediate to high redshifts as well as the evolution of the
galaxy and AGN populations, thanks to the unique combination of a
large area and precise photometric redshifts. However, early COSMOS
catalogs were primarily optically selected \citep{2007ApJS..172...99C}, although a subset of the COSMOS bands have been combined with WIRCAM
data \citep{2010ApJ...708..202M}. In \cite{Ilbert:2013dq} the first
UltraVISTA data release \citep{McCracken:2012gd} was used to derive an
NIR-selected photometric redshift catalog \citep[see
also][]{2013ApJS..206....8M}. In contrast to this earlier work, we now
add the optical $z^{++}$-band data to our object NIR-detection image, which
increases the catalog completeness for bluer objects. In addition,
this paper uses the deeper UltraVISTA-DR2 data release, a superior
method for homogenising the optical point-spread functions, much deeper
IR data from the \textit{Spitzer} Large Area Survey with Hyper-Suprime-Cam
(SPLASH) project, and new optical data from the Hyper-Suprime-Cam.

These improvements to the COSMOS catalog make it possible to create,
\textit{for the first time}, highly complete mass-selected samples to
high or very high redshifts subtending an area of
$54^{2}$Mpc$^{2}/h^{2}$ near $z\sim 1$. In particular, we are able
to extend the stellar-mass-halo-mass relationship to high
redshifts and to carefully study the connection between galaxies and
their large-scale environment throughout the transitional epoch of
mass accretion. This will be addressed in future works. Finally, this
catalog will also be invaluable in the preparation of simulated
catalogs for the \textit{Euclid} satellite mission and for defining what kind of
spectroscopic catalogs it will require.

%%%%%%%%%%%%%%%%%%%%%%%%%%%%%%%%%%%%%%%%%%%%%%%%%%%%%%%
%%outline
%%%%%%%%%%%%%%%%%%%%%%%%%%%%%%%%%%%%%%%%%%%%%%%%%%%%%%%

The paper is organized as
follows. Section~\ref{Sec:observ-data-reduct} describes the data set
and the preparation of the images. Section~\ref{section:prepIm}
details the galaxy detection and the photometric
measurements. Section~\ref{Sec:photoz} describes the computation of
the photometric redshift and the extraction of the physical
parameters.  Section~\ref{sec:char-glob-sample} summarizes the main
characteristics of the catalog. Section~\ref{Sec:conclusion}
presents our summary and outlines future data sets.
\\
We use a standard $\Lambda$CDM cosmology with Hubble constant
$H_{0}=70$ km s$^{-1}$Mpc$^{-1}$, total matter density $\Omega_{m}=0.3$ and dark
energy density $\Omega_{\Lambda}=0.7$. All magnitudes are expressed in
the AB \citep{Oke:1974p12716} system.
%%%%%%%%%%%%%%% 
%%%%%%%%%%%%%%%
\section{Observations and data reduction}
\label{Sec:observ-data-reduct}
\subsection{Overview of included data}
\label{Sec:summ}

The COSMOS field \citep{Scoville:2007p12720} offers a unique
combination of deep ($AB\sim 25-26$, multi-wavelength data
($0.25\mu m \rightarrow 24 \mu m$) covering a relatively large area of
$2 \deg^2$. The main improvement compared to previous
COSMOS catalog releases is the addition of new, deeper NIR and IR
data from the UltraVISTA and the SPLASH (\textit{Spitzer} Large Area
Survey with Hyper-Suprime-Cam) projects.

\begin{figure}
\begin{center}
\includegraphics[scale=0.44]{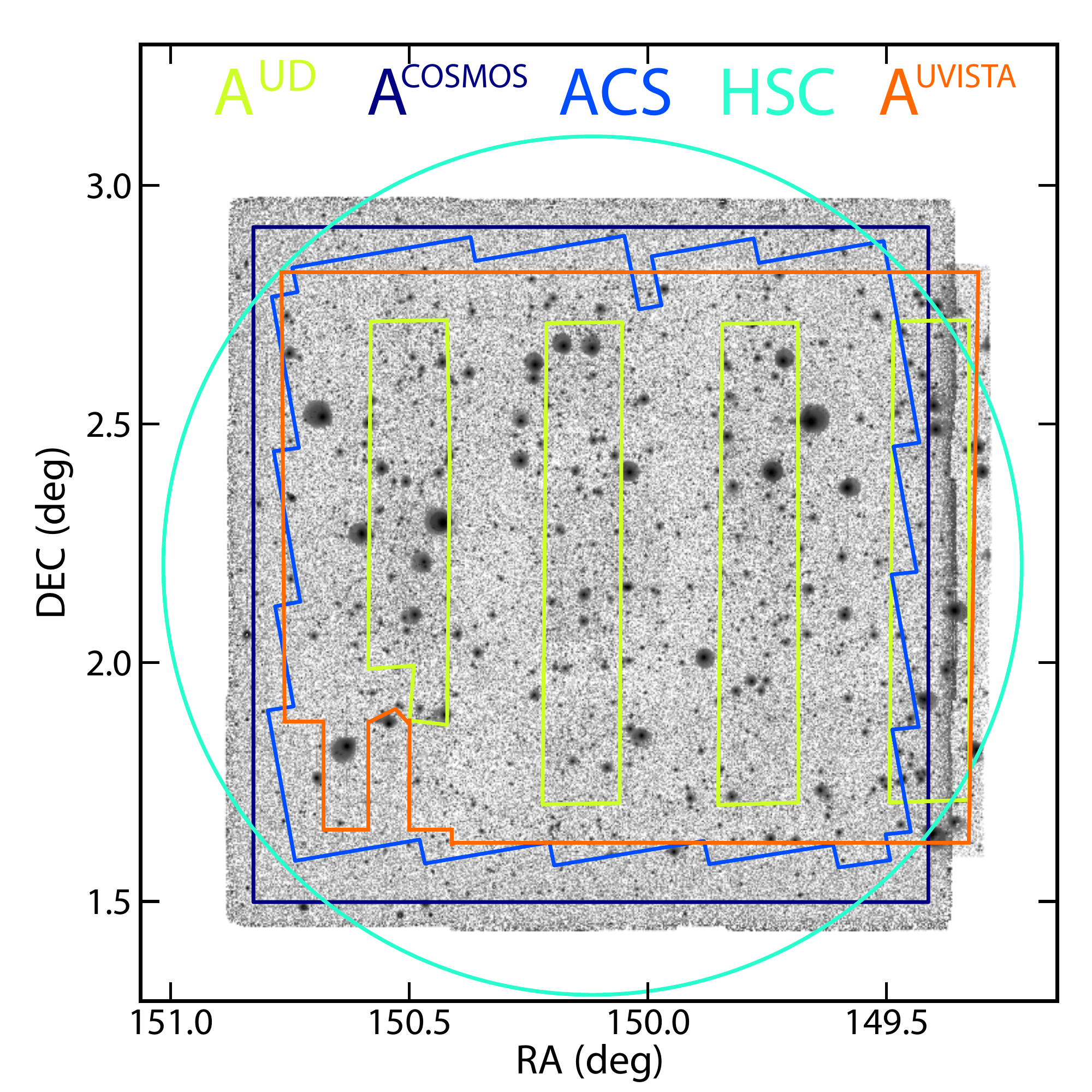}
\caption{Schematic of the COSMOS field showing all of the
  optical (dark blue and turquoise) and NIR (green and orange)
  observations used. The background image corresponds to the
  $\chi^{2}$ YKHK$_{\rm s}$-z$^{++}$ detection image (as described in
  section~\ref{section:prepIm}). For reference, the region covered by the
  COSMOS-Advanced Camera for Surveys (ACS) HST data \citep{2007ApJS..172..196K} is shown in cyan. ${\cal A}^{\rm
  COSMOS}$ defines the 2 deg$^{2}$ COSMOS square (dark blue). ${\cal A}^{\rm Uvista}$ (orange area) is the region covered by the UltraVISTA-DR2 observations. We define ${\cal A}^{\rm
  UD}$ as the light green area, corresponding to the ultra-deep stripes in the UltraVISTA-DR2 observations. ${\cal A}^{\rm
  Deep}$ is the difference between ${\cal A}^{\rm
  UVista }$ and ${\cal A}^{\rm
  UD}$.  In our analysis of the performance of the catalog, we limit ourselves to the intersection between ${\cal A}^{\rm UD}$ with ${\cal A}^{\rm COSMOS}$ and ${\cal A}^{\rm Deep}$ with ${\cal A}^{\rm COSMOS}$, after removing the masked objects in the optical bands (${\cal A}^{\rm !OPT}$, not shown on this figure). The effective areas are given in Table~\ref{Tab:coordinates}.}
\label{Fig:RegCos}
\end{center}
\end{figure}

\begin{figure*}
\begin{center}

\includegraphics[scale=0.6]{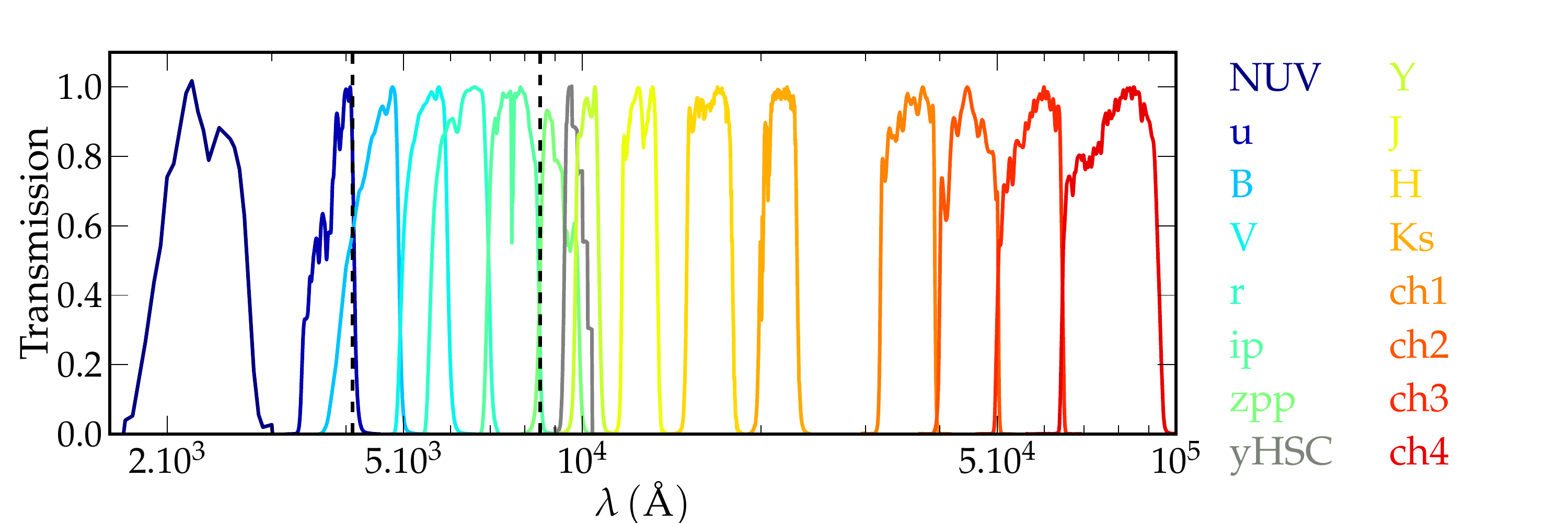}
\caption{Transmission curves for the photometric bands used. The
  effect of atmosphere, telescope, camera optics, filter, and the
  detector are included. Note that for clarity the profiles are
  normalized to a maximum throughput of one: therefore, the relative
  efficiencies of each telescope and detector system are \textit{not}
  shown. Intermediate and narrow bands are not
  represented, but the region of the spectrum covered by these bands
  is marked by dashed lines.}
\label{Fig:Curves}
\end{center}
\end{figure*}

\begin{figure}
\begin{center}

\includegraphics[scale=0.44]{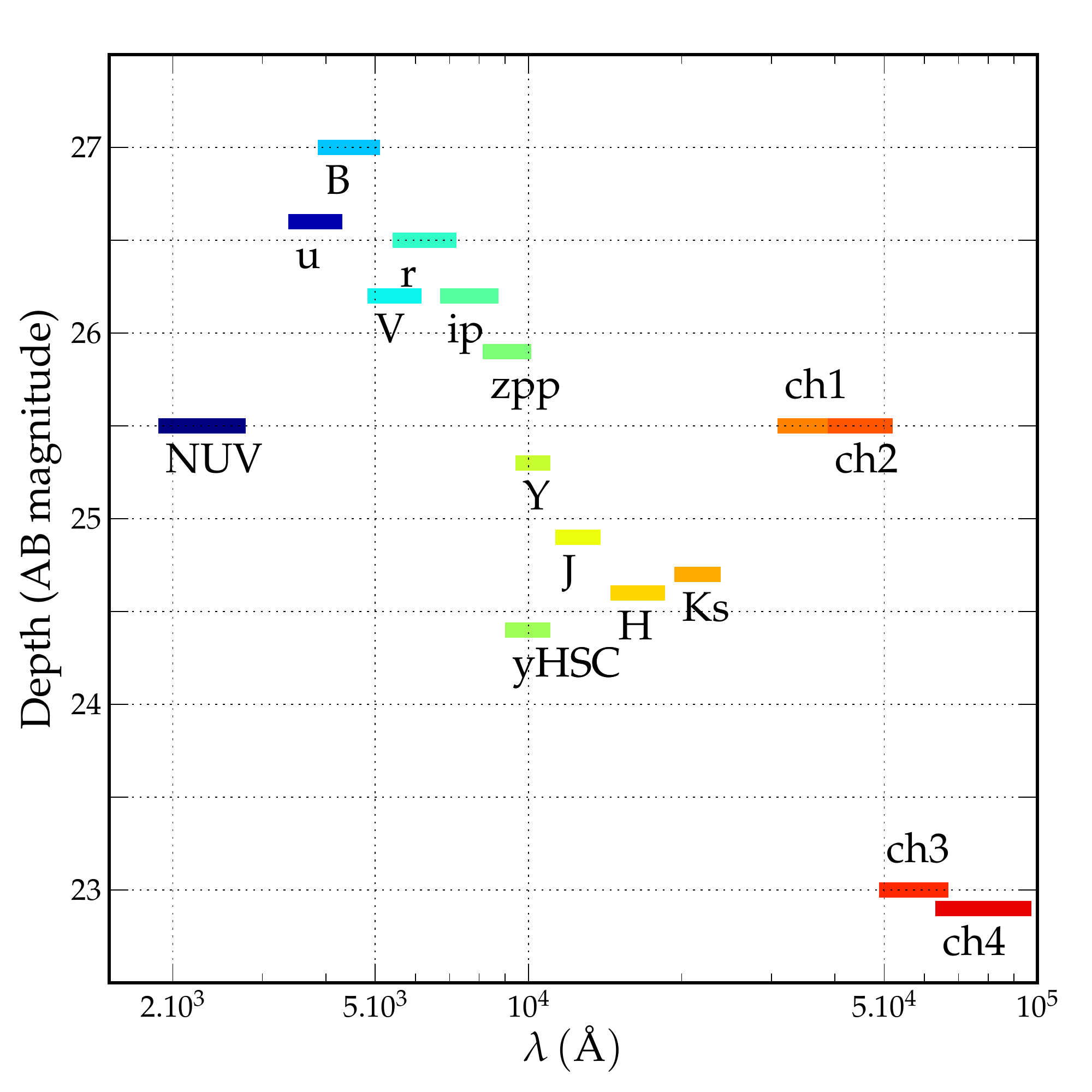}
\caption{Limiting magnitude at 3$\sigma$ in a 3$\arcsecond$ diameter
  aperture computed from empty aperture measurements for each band in
  COSMOS2015, with the exception for NUV filter \citep[value
  from][]{2007ApJS..172..468Z}. The length of each segment is the FWHM of the filter curves. For the $Y$, $J$, $H$,
  $K_{\rm s}$ bands; the limiting magnitudes which are specified
  correspond to ${\cal A}^{\rm UD}$.}
\label{Fig:limmag}
\end{center}
\end{figure}

As in previous COSMOS catalog papers, all of the images and noise maps have
been resampled to the same tangent point
RA,DEC$=(150.1163213,2.20973097)$. The entire catalog covers a square
of 2$\deg^2$ centered on this tangent point. When the images
were delivered as tiles, all of the data were assembled into a series of
$48096\times 48096$ images with an identical pixel scale of
0.15$\arcsecond$.  Figure~\ref{Fig:RegCos} shows the footprint of all of the
observations. Figure~\ref{Fig:Curves} shows the transmission curves of
all of the 
filters\footnote{\url{http://www.astro.caltech.edu/\~capak/filters/index.html}}
(filter, atmosphere and detector). COSMOS NIR data come from several
sources: WIRCam data \citep{2010ApJ...708..202M} covering the entire field,
and UltraVISTA \citep{McCracken:2012gd} data, covering the central
$1.5\deg^2$. The UltraVISTA data includes the DR2 ``deep'' and
``ultra-deep'' stripes. Note that this implies that the depth and
completeness in our final catalog are \textit{not} the same over the
whole COSMOS field because they are derived in part from these data.  The
COSMOS2015 catalog also offers a match with X-ray, near ultraviolet
(NUV), IR, and Far-IR data, coming, respectively, from \textit{Chandra}, GALEX,
MIPS/\textit{Spitzer}, PACS/\textit{Herschel} and SPIRE/\textit{Herschel}.  In this paper, we
limit ourselves to the inner, deep part covered by both UltraVISTA-DR2
and the $z^{++}$ band (which is flagged accordingly in our
%${\cal A}_{\rm UVISTA}^{\rm UD}$ 
catalog). We denote as ${\cal A}^{\rm UD}$ the part of the field covered
by the ``ultra-deep stripes'' ($K_{\rm s}=24.7$ at 3$\sigma$ in a
3$\arcsec$ diameter aperture) and as ${\cal A}^{\rm UVista}$ the
  full region covered by UltraVISTA-DR2 ($K_{\rm s}=24.0$ at
  3$\sigma$ in a 3$\arcsec$ diameter aperture). ${\cal A}^{\rm Deep}$
  is the difference between ${\cal A}^{\rm UVista }$ and
  ${\cal A}^{\rm UD}$.  In our analysis, we limit ourselves to the
  intersection of ${\cal A}^{\rm UD}$ and ${\cal A}^{\rm Deep}$ within
  the 2 deg$^{2}$ COSMOS area after removing the masked area in the
  optical. The effective areas corresponding to these intersections are 0.46
  deg$^{2}$ for ${\cal A}^{\rm UD}$ and 0.92 deg$^{2}$ for
  ${\cal A}^{\rm Deep}$.  Details of these flagged regions can be
  found in Table~\ref{Tab:coordinates} (section~\ref{Sec:Appendix1})
and on Figure~\ref{Fig:RegCos}. % and
All of the input data are summarized in Table~\ref{Tab:limmag}. The
limiting magnitudes can be observed in Figure~\ref{Fig:limmag}.

\subsubsection{Optical-ultraviolet data}
The optical-ultraviolet data set used here is similar to those used in previous
releases \citep{2007ApJS..172...99C, 2009ApJ...690.1236I}. It includes
near-UV ($0.23 \mu m$) observations from GALEX
\citep{2007ApJS..172..468Z}, $u^{*}$-band data from the Canada-France
Hawaii Telescope (CFHT/MegaCam), and the COSMOS-20 survey, which is composed of 6
broad bands ($B$, $V$, $g$, $r$, $i$, $z^{+}$), 12 medium bands
($IA427$, $IA464$, $IA484$, $IA505$, $IA527$, $IA574$, $IA624$,
$IA679$, $IA709$, $IA738$, $IA767$, and $IA827$), and two narrow bands
($NB711$, $NB816$), taken with Subaru Suprime-Cam
\citep{2007ApJS..172....9T,2015Taniguchi}. We have discarded poor seeing
($\sim1.3\arcsec$) $g$-band data. Finally, the initial COSMOS $z$-band
data were replaced by deeper $z^{++}$ band data taken with thinned
upgraded CCDs and a slightly different filter. At this stage, in each
band, image point-spread functions (PSFs) were homogenized to minimize
tile-to-tile variations \citep{2007ApJS..172...99C}. At the same time,
{\sc RMS\_MAP} and {\sc FLAG\_MAP} images were also generated, and
saturated pixels and bad areas were flagged. This release also contains
new $Y$-band data taken with Hyper-Suprime-Cam (HSC) Subaru
\citep{2012SPIE.8446E..0ZM}. The average exposure time per pixel is
2.1 hr. This data set is described fully in Hasinger G. et al. (in
preparation). The addition of the $Y-$band data is intended to
improve our stellar mass and redshift estimates in the important
$1<z<1.5$ range because it is slightly bluer than the $Y$ filter from
VIRCAM (see Figure~\ref{Fig:Curves}), but it is also intended to serve as a ``pilot
program'' to assess the utility of HSC data and to prepare for future
COSMOS data sets which will include much more HSC imaging.

\subsubsection{NIR data}
\label{Sec:NIR}

The $YJHK_{\rm s}$-band data used here were taken between 2009 December
2009 and 2012 May with the VIRCAM instrument on the VISTA telescope as
part of the UltraVISTA survey program and constitute the DR2
UltraVISTA
release\footnote{\url{http://www.eso.org/sci/observing/phase3/data\_releases/uvista\_dr2.pdf}}. The
UltraVISTA-DR2 processing steps are the same as those in the DR1 release
\citep{McCracken:2012gd}. Compared to DR1, the exposure time has been
increased significantly in the ultra-deep stripes, as shown in yellow in
Figure~\ref{Fig:RegCos}; these cover an area of 0.62 deg$^{2}$. An
important consequence of this is that the signal-to-noise ratio for an
object of a given magnitude is not constant across the image.  To
provide NIR photometry in zones not covered by UltraVISTA, we include
$H$ and $K$ WIRCAM data \citep{2010ApJ...708..202M} in our photometric
catalog. However, this paper does not discuss the performance of
photometric redshifts and physical parameters in these WIRCAM-only
areas.

\subsubsection{Mid-Infrared data}
\label{Sec:mid-infrared}
%%%%%Stop here   %%%%%%%%%%%%%%%%
The $3.6 \mu m$, $4.5 \mu m$, $5.8 \mu m$ and $8.0 \mu m$
(respectively, channel 1, 2, 3, and 4) IRAC data used in this paper
consist of the first two-thirds of the SPLASH COSMOS dataset together
with S-COSMOS \citep{2007ApJS..172...86S}, the \textit{Spitzer} Extended
Mission Deep Survey, the \textit{Spitzer}-Candels survey data, along
with a several smaller programs that observed the COSMOS field.  The
final processing is described in a companion paper (Capak et al. 2015
in prep).  The average exposure time per pixel is 3.8 hr, increasing
to 50hr in the central S-CANDELS coverage.  Before processing, a median
image was created for each AOR (observing block) and subtracted from
the frames to remove residual bias in the frames and persistence from
previous observations.  For the S-CANDELS data, a secondary median was
subtracted from the observations taken with repeats to remove the
``first frame effect'' residual bias. The resulting median-subtracted
images have a mean background near zero, so no overlap correction was
applied.  The median subtracted frames were then combined with the
MOPEX mosaic
pipeline\footnote{\url{http://irsa.ipac.caltech.edu/data/SPITZER/docs/dataanalysistools/tools/mopex/}}. 
The outlier and box-outlier modules were used to reject cosmic rays,
transients, and moving objects.  The data were then drizzled onto a
$0.6\arcsec$ pixel scale using a ``pixfrac'' of 0.65 and combined with an
exposure time weighted mean combination.  Mean, median, coverage,
uncertainty, standard-deviation, and color-term mosaics were also
created. Obviously, this variation as a function of position can be
expected to have an influence on the precision of the photometric
redshifts and stellar masses for the very highest redshift ($z>4$)
objects.

\begin{table}
\begin{center}
  \caption{Summary of available data in each band and the average
    limiting magnitudes computed from variance map in 2 and
    3$\arcsecond$ diameter apertures on the PSF-homogenised
    Images. The central wavelength is the median wavelength weighted
    by transmission and the widths are defined using the half-maximum
    transmission points.}
\begin{threeparttable}
\begin{tabular}{ ccccc}
\hline
 \textbf{Instrument} & \textbf{Filter} & \textbf{Central
                                                   } &
                                                          \textbf{Width}&\textbf{3$\sigma$
                                                                           depth}
                                                                           \tnote{a}\\
\textbf{/Telescope} & & \textbf{$\lambda$ ($\angstrom$)} &
                                                          \textbf{($\angstrom$)}&
                                                                                  \textbf{(3$\arcsecond$/2$\arcsecond$)}\\
\textbf{(Survey)} & & & &  \textbf{$\pm 0.1 $}\\\hline
GALEX & NUV & 2313.9 & 748 & 25.5 \tnote{b}\\\hline
MegaCam/CFHT & $u^{*}$ & 3823.3 & 670 & 26.6/ 27.2 \\
\hline
Suprime-Cam & $B$ & 4458.3 & 946 & 27.0/   27.6\\
 /Subaru & $V$ & 5477.8 & 955 & 26.2/  26.9\\ 
& $r$ & 6288.7 & 1382 & 26.5/  27.0 \\
 & $i^{+}$ & 7683.9 & 1497 & 26.2/ 26.9 \\
 & $z^{++}$ & 9105.7  & 1370  & 25.9/ 26.4 \\
 & $IA427$ & 4263.4 & 206.5 & 25.9/ 26.5  \\
 & $IA464$ & 4635.1 & 218.0 & 25.9 / 26.5  \\
 & $IA484$ & 4849.2 & 228.5 & 25.9/ 26.5 \\
 & $IA505$ & 5062.5 & 230.5 &  25.7/ 26.2 \\
& $IA527$ & 5261.1 & 242.0 & 26.1/ 26.6 \\
& $IA574$ & 5764.8 & 271.5 & 25.5/ 26.0 \\
& $IA624$ & 6233.1 & 300.5 & 25.9/ 26.4  \\
& $IA679$ & 6781.1  & 336.0 & 25.4/ 26.0 \\
& $IA709$ & 7073.6 & 315.5 & 25.7/ 26.2 \\
& $IA738$ & 7361.6 & 323.5 & 25.6/ 26.1 \\
 & $IA767$ & 7684.9 & 364.0 & 25.3/ 25.8 \\
 & $IA827$ & 8244.5 & 343.5 & 25.2/ 25.8 \\
 & $NB711$ & 7119.9 & 72.5 & 25.1/ 25.7  \\
 & $NB816$ & 8149.4 & 119.5 & 25.2/ 25.8  \\
\hline
HSC/Subaru & $Y$ & 9791.4  & 820 & 24.4/ 24.9 
 \\\hline
VIRCAM &$ Y^{\rm UD}$ & 10214.2  & 970 & 25.3/ 25.8   \\ 
/VISTA & $Y^{\rm Deep}$ &   & & 24.8/ 25.3 \\
(UltraVISTA-DR2)& \textbf{$J^{\rm UD}$} & 12534.6 & 1720  & 24.9/ 25.4 \\
& $J^{\rm Deep}$ & & &24.7/ 25.2 \\
& \textbf{$H^{\rm UD}$} & 16453.4  & 2900 & 24.6/ 25.0 \\
& $H^{\rm Deep}$ &  & & 24.3/ 24.9\\
& \textbf{$K_{\rm s}^{\rm  UD}$} & 21539.9 & 3090& 24.7/25.2 \\
& $K_{\rm s}^{\rm Deep}$ & & & 24.0/ 24.5
\\ \hline
 WIRCam & $K_{\rm s}$ & 21590.4 & 3120 & 23.4/ 23.9  \\
/CFHT & $H$ & 16311.4 & 3000  & 23.5/ 24.1 
  \\\hline
IRAC/\textit{Spitzer} & ch1 & 35634.3 & 7460 & 25.5/ o \tnote{c}\\
(SPLASH)& ch2 & 45110.1 & 10110 & 25.5/ o\tnote{c} \\
& ch3 & 57593.4 & 14140 & 23.0/ o\tnote{c}\\
 & ch4 & 79594.9 & 28760 & 22.9/ o \tnote{c} \\
\hline
\end{tabular}
\begin{tablenotes}
\item[$^{a}$] $3\sigma$ depth in $m_{\rm AB}$  computed on PSF-matched images from around
  800 apertures at 2 and 3$\arcsecond$.
\item[$^{b}$] Value given in \cite{2007ApJS..172..468Z} corresponding
  to a 3$\sigma$ depth.
\item[$^{c}$] $3\sigma$ depth in $m_{\rm AB}$ computed from the RMS
  maps, after masking the area containing an objects based on the
  segmentation map.
\end{tablenotes}
\label{Tab:limmag}

\end{threeparttable}
\end{center}
\end{table}

\subsection{Image Homogenisation}
\label{Sec:preparation-images}
In this paper, the variation of the PSF across individual
  images in a given band is neglected. This is reasonable because
  band-to-band variations are almost always greater than the variation
  within a single band. The residual impact of the PSF variation across the field is discussed in Appendix~\ref{Ap:seeing}.
 
  From $u$ to $K_{\rm s}$ the FWHM of the PSF has a range of values between $\sim0.5\arcsecond$
  and $1.02\arcsecond$ (corresponding to a Moffat fit). Therefore, the
  fraction of the total flux falling in a fixed aperture is
  band-dependent. One way to address this problem is to ``homogenize''
  the PSF so that it is the same in all of the bands (GALEX and IRAC bands
  are not homogenized, their photometry are extracted with a
  source-fitting technique, as detailed in
  section~\ref{section:prepIm}). In the first step in our
  homogenization process, {\sc SExtractor} \citep{Bertin:1996p13615}
  is used to extract a catalog of bright objects. Stars are identified
  by cross-matching with point sources in the COSMOS-Advanced Camera
  for Surveys (ACS) \textit{Hubble Space Telescope} (HST) catalog
  \citep{2007ApJS..172..196K,2007ApJS..172..219L}. Saturated or faint
  stars are removed by considering the position of each object in the
  FWHM versus m$_{\rm AB}$ diagram. For each star we extract a postage
  stamp using {\sc SExtractor}.  

The PSF is modelled in pixel space using {\sc PSFex}
\citep{2013ascl.soft01001B} as a linear combination of a limited number
of known basis functions:

\begin{equation}
\Psi_{c}=\sum_{b} c_{b}\psi_{ b}
\end{equation}

where the $c$ index reflects the dependance of $\Psi$ on the set of
coefficients $c_{b}$.   Given a basis, this PSF model can be entirely determined knowing
the coefficients $c_{b}$ of the linear combination.  The pixel basis
is the most ``natural'' basis but requires as many coefficients as the
number of pixels on the image postage stamp. We can then make some
assumptions to simplify the basis and to reduce the number of
coefficients. The adopted basis is the ``polar shapelet'' basis
\citep{2005MNRAS.363..197M}, for which the components have useful
explicit rotational symmetries.  We assume that the PSF is constant over
the field.  The global PSF of one band is then expressed as a function of the
coefficients $c_{b}$ at each pixel position $x_{i}$ on the postage
stamp image, which are derived by minimizing the $\chi^{2}$ sum over
all of the sources:

\begin{equation}
\chi^{2}(c)=\sum_{{\rm sources}\,s}\sum_{{\rm
    pixels}\,i}\frac{(p_{s}(x_{i})-f_{s}\Psi_{c}(x_{i}))^{2}}{\sigma_{i}^{2}},
\end{equation}
where $f_{s}$ is the total flux of the source $s$,
  $\sigma_{i}$ is the variance estimate of pixel $i$ of the source $s$,
  $p_{s}(x_{i})$ is the intensity of the pixel $i$, and $c$ refers to
  the set of PSF coefficients.   Once the global PSF has
  been determined in each band, we then decide on the ``target PSF'',
  corresponding to the desired PSF of all of the bands after
  homogenization. This is chosen so as to minimize the applied
  convolutions. We use a Moffat profile to represent the
  PSF \citep{Moffat:1969p12721}; this provides a better description of
  the inner and outer regions of the profile than a simple
  Gaussian. The stellar radial light profile is
\begin{equation}
I_{r}=I_{0}[1+(r/\alpha)^{2}]^{-\beta}
\end{equation}
with 
\begin{math}
\alpha={ \theta}/(2\sqrt{2^{1/\beta}-1}),
\end{math}
\begin{math}
I_{0}=(\beta -1 )(\pi\alpha^{2})^{-1}
\end{math}
and $\theta$ the FWHM.
Our target PSF is defined as a Moffat profile with \begin{math}
{\cal M}\left[\theta,\beta\right]={\cal M}\left[0.8\arcsecond,2.5\right].
\end{math}

The required convolution kernel is calculated in each band by finding
the kernel that minimizes the difference between the target PSF and
the convolution product of this kernel with the current PSF. The
images are then convolved with this kernel.

\begin{figure*}
\begin{center}
\includegraphics[scale=0.4]{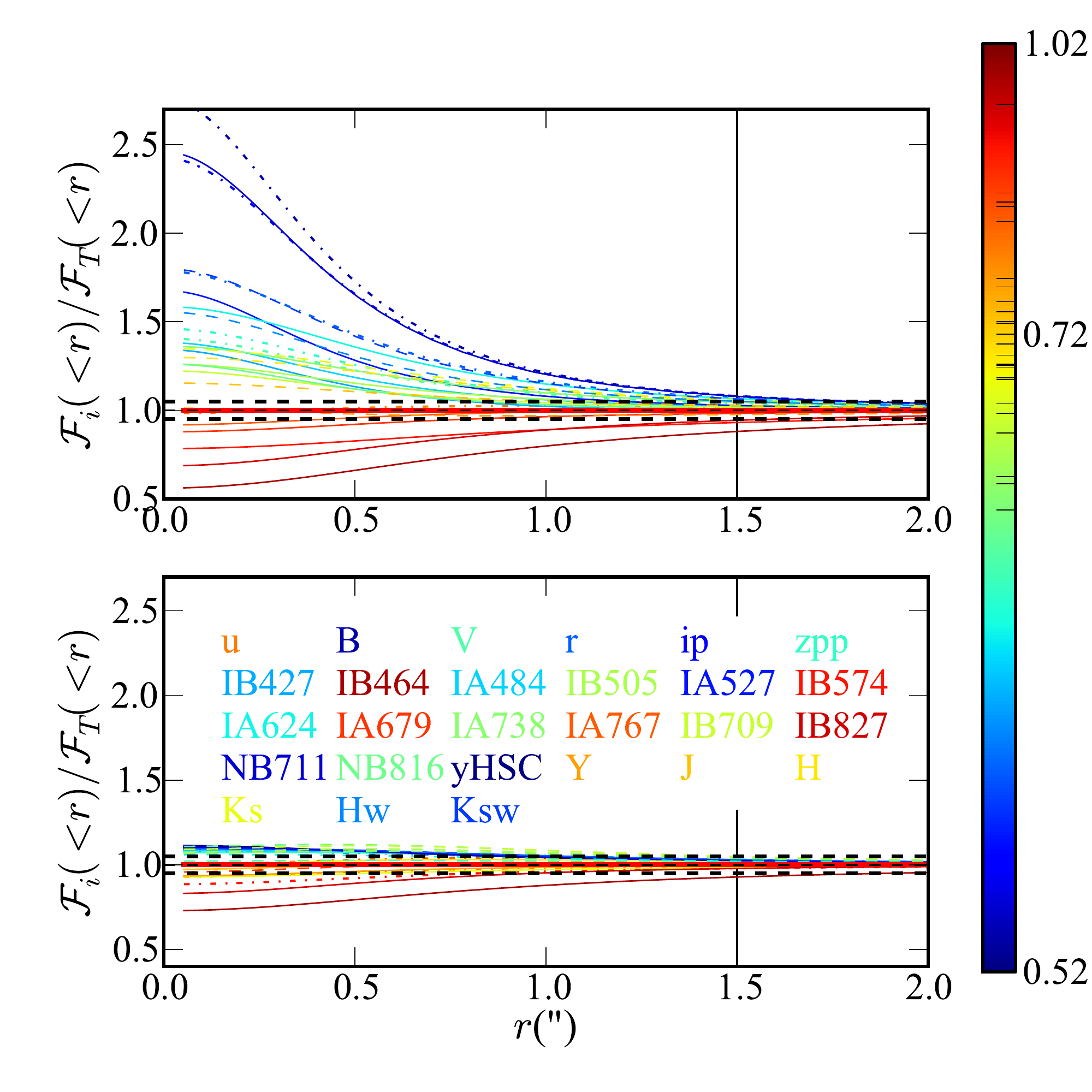}
\includegraphics[scale=0.4]{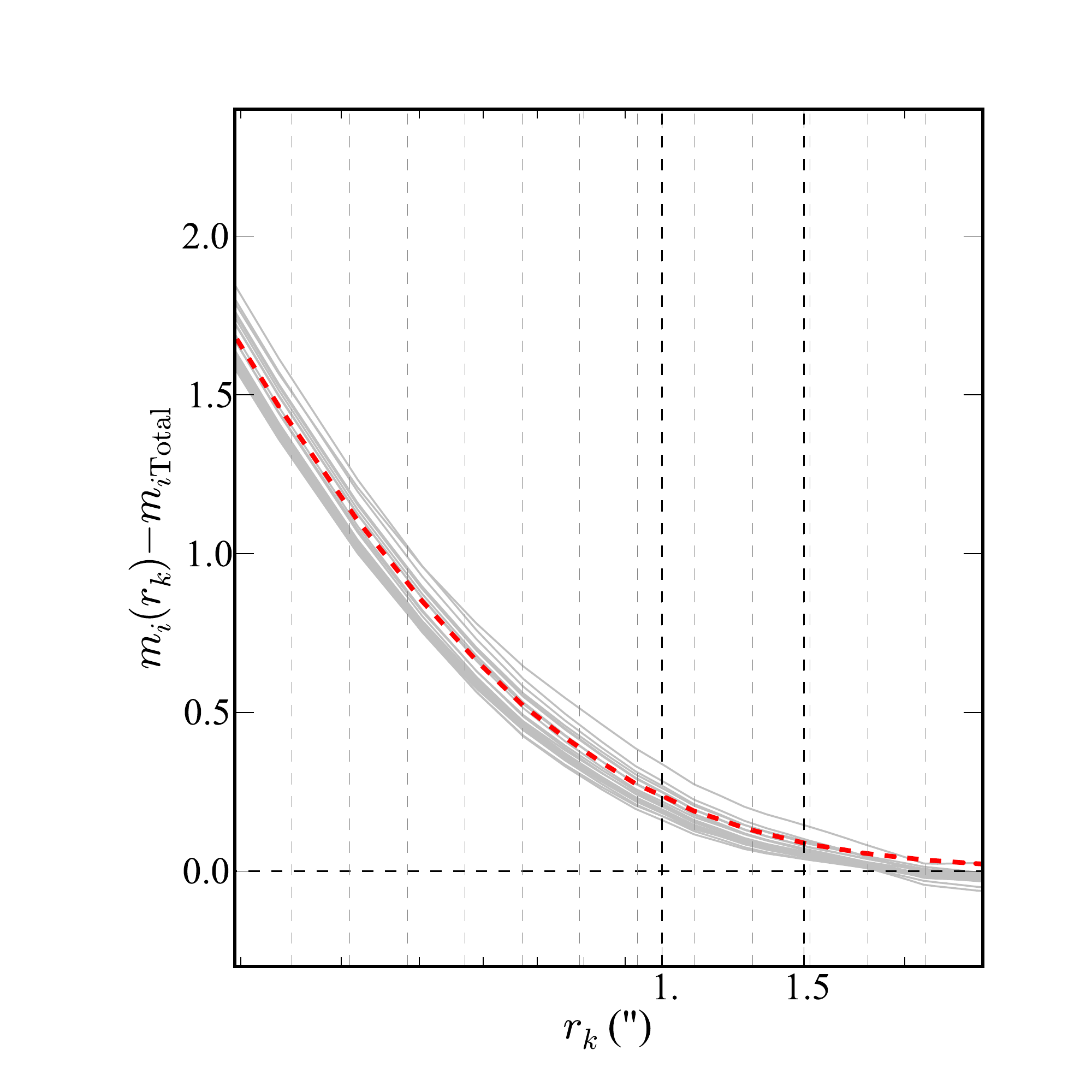}
\caption{\textit{Left}: best-fitting stellar PSFs for all bands before
  and after homogenization (left upper and lower panels respectively),
  normalized to the target PSF ${\cal F}_{T}$. The vertical black
  solid line corresponds to the 3$\arcsec$ diameter aperture used for
  photometric redshifts. The horizontal dashed lines show the 5\%
  relative error. The color map reflects the increase in seeing before
  homogenization. \textit{Right}: Median curves of growth (the difference
  between the magnitude in the $k$th aperture $r_{k}$ with the
  total magnitude for point-like sources, estimated from the
  4$\arcsec$ diameter aperture) as a function of aperture after
  homogenization for each band. The target Moffat profile with
  ($\theta$,$\beta$)=(0.8,2.5) is shown in red. The vertical dark
  dashed lines are the apertures provided in the final public catalog
  (2$\arcsecond$ and 3$\arcsecond$ diameters).}
\label{Fig:PSF}
\end{center}
\end{figure*}

To estimate the precision of our PSF matching procedure, the
photometry of the stars is extracted at 14 fixed apertures of radii
$r_{k}$, logarithmically spaced between 0.25$\arcsecond$ and
2.5$\arcsecond$. In each band, the difference between the magnitude of
the stars extracted in the aperture $r_{k}$ and the total magnitude
(computed from the 4$\arcsec$ diameter aperture) is plotted in
Figure~\ref{Fig:PSF} as a function of aperture.  For comparison, the
difference that would be obtained with the target profile
${\cal M}\left[0.8\arcsecond,2.5\right]$ is overplotted as a red
dashed line. The agreement is excellent up to a 2$\arcsecond$ radius
on the plot. 

\begin{figure*}
\begin{center}
\includegraphics[scale=0.4]{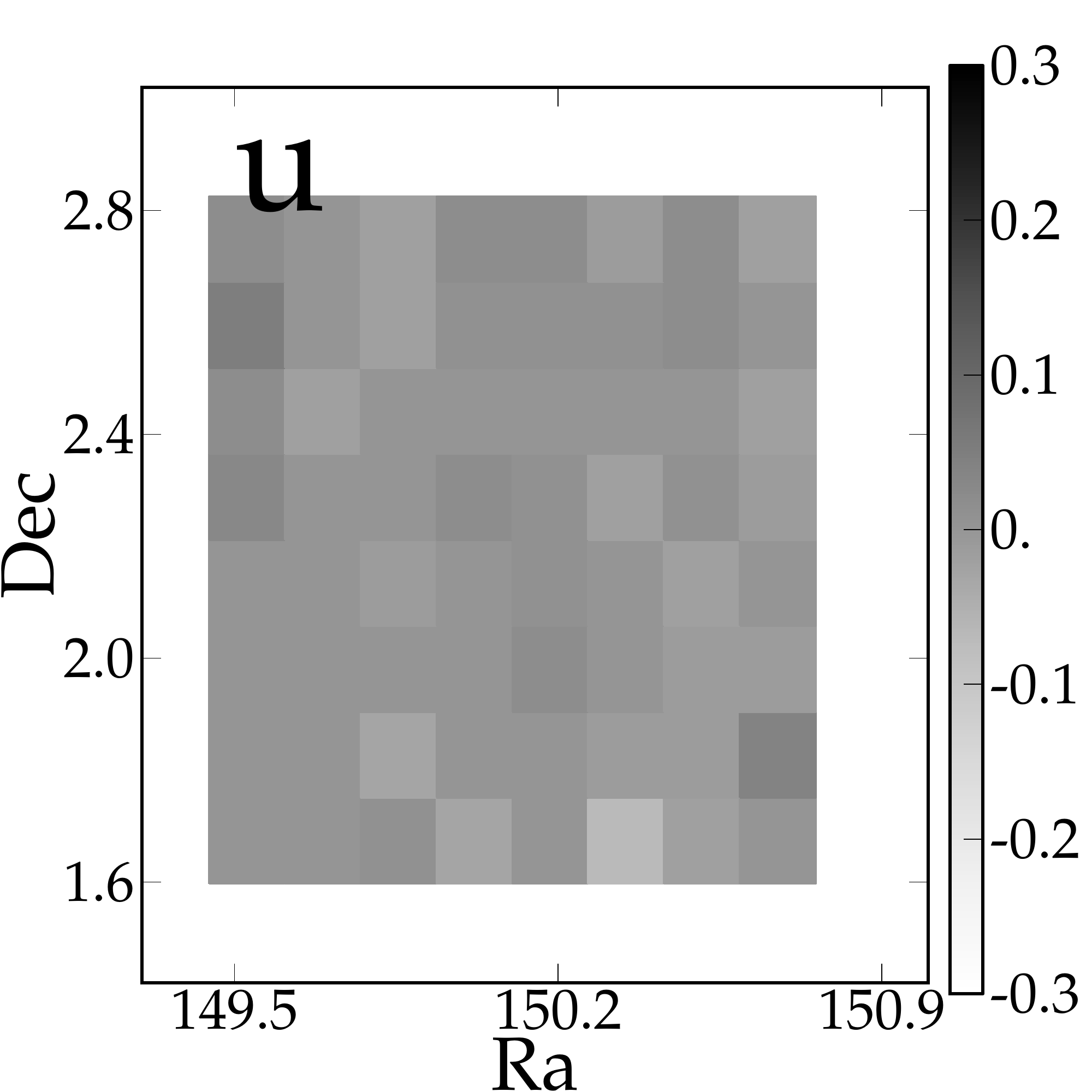}
\includegraphics[scale=0.4]{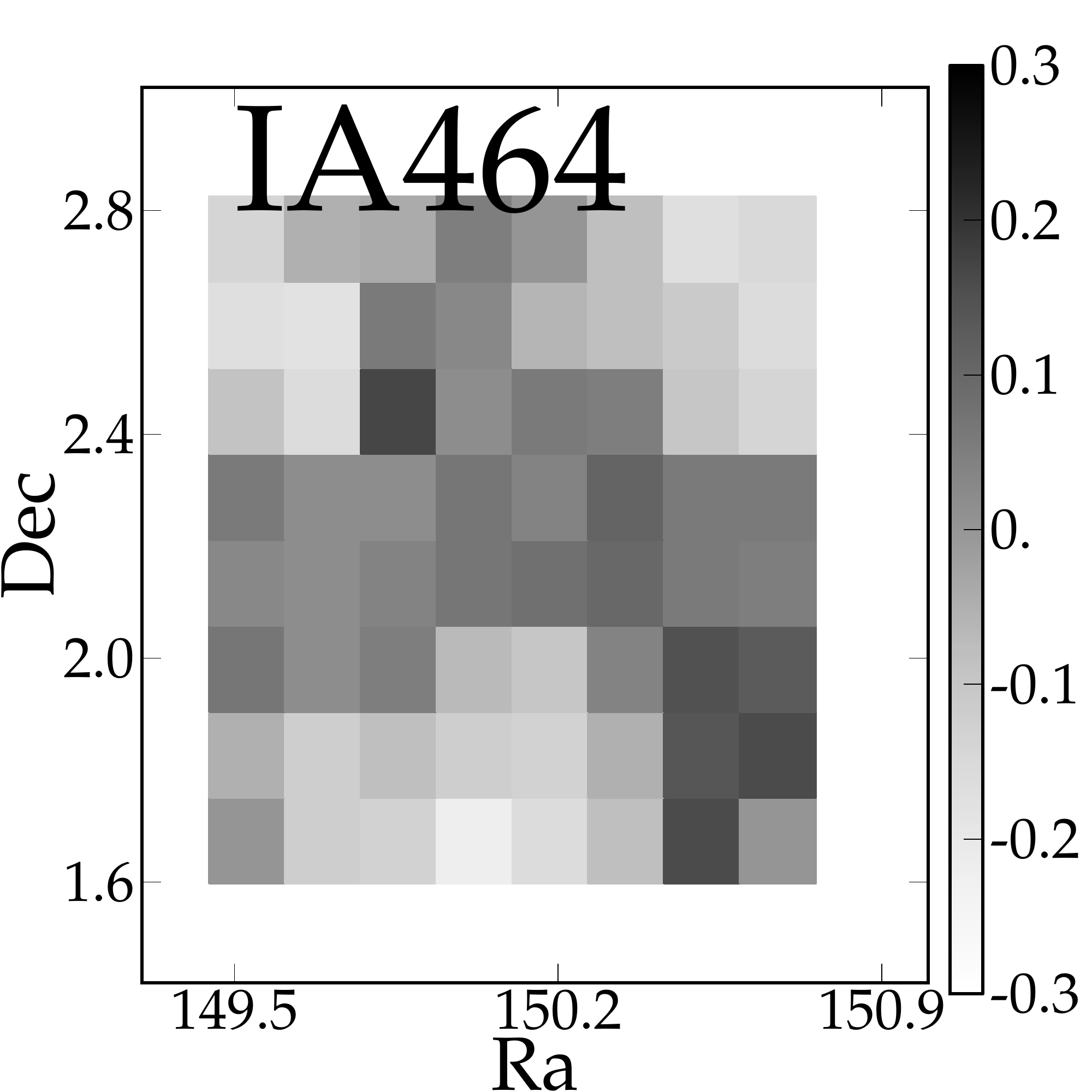}
\caption{Distribution of the difference local seeing and
    the median image seeing for the selected stars in the $u$ and
    $IA464$ bands as a function of the position. Note that for these
    seeing estimations, we did not fit each star \textit{individually}
    with a Moffat profile, but we used the Gaussian-profile-based {\rm
      FWHM\_WORLD} parameter from {\sc SExtractor}. While the $u$ band
    is relatively homogenous across the field, $IA464$ shows large
    positional variations. This is the most extreme case in our catalog.}
\label{Fig:VarPSF}
\end{center}
\end{figure*}

The flux obtained with the best-fitting PSF in each band is normalized
to the target profile and also plotted in Figure~\ref{Fig:PSF} (left
panel), before and after homogenization. For perfect homogenization,
this ratio should be one, independent of aperture. For the
3$\arcsecond$ diameter aperture, the relative photometric error for
point-sources objects after homogenization is below $5\%$ (or
  equivalently a difference of $\sim$0.05 in
  magnitude). Unfortunately, despite previous attempts at PSF
homogenization inside each field
\citep{2007ApJS..172...99C,McCracken:2012gd}, residual variations
remain across the field. These are shown in
  Figure~\ref{Fig:VarPSF}, which shows the distribution of the stellar
  FWHM and the median FWHM for two representative bands. While the PSF
  is relatively homogenous across the field for most of the bands (e.g
  $u$-band),there is larger scatter for some bands (e.g $IA464$). In
  Appendix~\ref{Ap:seeing}, we discuss the effect of these variations on
the aperture magnitude. 

Concerning the cosmetic quality of the image, the convolution
operation produces several undesirable effects. First, it induces a
covariance in the background noise which can lead photometric errors
being underestimated. Secondly, since the homogenization process acts
both on the FWHM and the profile slope ($\alpha$ and $\beta$
parameters), the convolution kernel may contains negative
components. In some bands it can lead to artefacts (such as rings)
around saturated objects. We mask these saturated objects in the final
catalog.  We deal with the correlation of the background by
multiplying in each band the photometric errors derived from {\sc
  SExtractor} by a correction factor (see section~\ref{Sec:corr} for
more details).

%%%%%%%%%%%%%%%
%%%%%%%%%%%%%%%
\section{Catalog extraction}
\label{section:prepIm}

\subsection{Photometric measurements}

\subsubsection{Optical and NIR data}
\label{SubSec:Extraction}

Object photometry is carried out using {\sc SExtractor} in ``dual
image'' mode. The $\chi^{2}$ $zYJHK_{\rm s}$ detection image
\citep{Szalay:1999p4804} is produced using {\sc SWarp}
\citep{Bertin:2002p5282} starting with the non-homogenized images.
Since the main objective of our new catalog is to probe the
high-redshift Universe and also to provide a catalog containing
UV-luminous sources at $z>2$, we create a detection image by combining
NIR images of UltraVISTA ($YJHK_{\rm s}$) with the optical $z^{++}$-band data from Subaru. We do not use $i^{+}$-band data since compact
objects in the $i^{+}$ image saturate around $i=21$.

We extract fluxes from 2 and 3$\arcsecond$ diameter apertures
  on PSF-homogenized images in each band. The well-known difficulty in
  source extraction from astronomical images is that objects have
  ill-defined, potentially overlapping boundaries, making flux
  measurements challenging. The two main parameters that control
  extraction are the deblending threshold and the flux threshold.
  Therefore, a reasonable balance must be found on one hand between
  deblending too much (splitting objects) and not
  deblending enough (leading to merging). Simliar problems occur with
  the choice of the detection threshold: a low detection threshold can
  create too many spurious objects, and one that is too high may miss
  objects. This can be mitigated in part by a judicious choice of
  detection threshold and the minimal number of contiguous pixels
  which constitute  an object. The solution we adopt is to set a low
deblending and detection threshold while increasing the number of
contiguous pixels to reject false detections. We validated this choice
through careful inspection of catalogs superimposed on the detection and
measurement images, which is feasible in the case of a single-field
survey like COSMOS.

The background is estimated locally within a rectangular annulus (30
pixels thick) around the objects, delimited by their isophotal
limits. Additionally, object mask flags indicating bad regions in
the optical and NIR bands were included and saturated pixels in
the optical bands were flagged by using the appropriate {\sc
  FLAG\_MAP}s. Our chosen parameters are given in
  Table~\ref{Tab:SEparam}.

In the last step, catalogs from each band are merged together into a
single FITS table and galactic extinction values are computed at each
object position using the \cite{1998ApJ...500..525S} values. These
reddening values have to be multiplied by a factor computed for each
band, derived from the filter response function and integrated
against the galactic extinction curve
\citep{2000A&A...363..476B,Allen1976}. These factors are shown in
Table~\ref{Tab:sysoff}.

\subsubsection{GALEX photometry}

As in \cite{2009ApJ...690.1236I}, GALEX photometry
  \citep{2007ApJS..172..468Z} for each object was derived by
  cross-matching our catalog with the publicly available photometric
  $i^{+}$-selected catalog described in
\cite{2007ApJS..172...99C}\footnote{\label{noteCap}The version of the
  catalog used is available at
  \url{http://irsa.ipac.caltech.edu/data/COSMOS/tables/photometry/}. This
  catalog supersedes that of \cite{2007ApJS..172...99C} with improved
  source detection and photometry.}. GALEX fluxes were measured
using a PSF fitting method with the $u^{*}$-band image used as a
prior.

\subsubsection{IRAC photometry}

The SPLASH data exceed the confusion limits in IRAC channel 1 and
channel 2; better deblending techniques are necessary to estimate
fluxes in these crowded, low-resolution images. We use {\sc IRACLEAN}
\citep{2012ApJS..203...23H} to derive the SPLASH photometry.This makes
use of the positional and morphological characteristics of objects
detected in a high-resolution image in a different waveband to deblend
objects and derive more accurate fluxes in low-resolution
images. Unlike other similar methods assuming that the intrinsic
morphology of an object is identical in the two wavebands (i.e., there
is no morphological \textit{k}-correction), {\sc IRACLEAN} is essentially the
same as {\sc CLEAN} deconvolution in radio imaging with nearly no
morphological restrictions, except for the locations where {\sc CLEAN}
can opperate. This scheme minimizes the effect of morphological
\textit{k}-correction when a high-resolution image (i.e., the prior) and its
low-resolution counterpart are taken in very different
wavebands. However, there is a limitation for this scheme. If the
separation of two objects is less than $\sim 1 $ FWHM, then the flux of the
brighter object can be overestimated while the flux of the fainter one
can be underestimated, as discussed in \cite{2012ApJS..203...23H}. To
solve this issue, we improve {\sc IRACLEAN} by taking the surface
brightness information in the prior into account. The new {\sc
  IRACLEAN} uses the surface brightness of the prior to weight where
{\sc CLEAN} occurs for each object. The weighting strength is
determined by the power of the surface brightness, i.e., (surface
brightness of the prior)$^{n}$, where $n$ is the weighting
parameter. If $n$ is zero, then the surface brightness of the prior is
ignored, and so the new {\sc IRACLEAN} behaves like the original {\sc
  IRACLEAN}. When $n$ is greater than zero, the higher the $n$ is, the
more heavily weighted the
surface brightness is. If the
wavebands of the prior and the target images are very different, then $n$
can be set to a lower value, e.g., 0.1 - 0.3.If the
wavebands of the prior and the target images are very similar, then $n$ can
be set to a higher value, e.g., 0.3 - 0.5. In general, $n=0.3$ is
sufficiently good for most cases.

In this paper, the UltraVISTA $zYJHK_{\rm s}$ chi-square image is used
as the prior for the SPLASH images in {\sc IRACLEAN}. To accelerate
the process, both the UltraVISTA $zYJHK_{\rm s}$ chi-square image and
the SPLASH images are broken up into the 144 tiles that are used for
the COSMOS Subaru/ACS data, making parallel processing easier. The
tiles overlap by 14.4$\arcsecond$ around the edges to avoid flux
underestimation for those objects close to the edges of the tiles. The
SPLASH PSFs in each tile are generated using point sources in that
tile. The aperture size used to measure the flux ratios between
sources and PSFs for the {\sc CLEAN} procedure is
$1.8 \arcsecond \times 1.8\arcsecond$, and the weighting parameter n
is 0.3. 
After the {\sc CLEAN} procedure is completed, a
residual map is generated which is used toe stimate the flux
errors. The flux error of each object is estimated based on the
fluctuations in the local area around that object in the residual
map. The {\sc IRACLEAN} procedure is described fully in
\cite{2012ApJS..203...23H}.

\subsubsection{X-ray photometry}
\label{Sec:XRay}

The \textit{Chandra COSMOS-Legacy} Survey
\citep{2016Civano,2016Marchesi} contains 4016 X-ray sources down to a
flux limit of $f_{X}\simeq$2 $\times$ 10$^{-16}$ erg s$^{-1}$ cm$^{-2}$ in
the 0.5-2 keV band: 3755 of these sources lie inside the UltraVISTA
field of view. The \textit{Chandra COSMOS-Legacy} catalog was matched
with the UltraVISTA catalog using the Likelihood Ratio (LR) ratio
technique \citep{1992Sutherland}. This method provides a much more
statistically accurate result than a simple positional match, taking
into account the following: (i) the separation between the X-ray source and the
candidate UltraVISTA counterpart; (ii) the counterpart $K$-band
magnitude with respect to the overall magnitude distribution of
sources in the field. Of the 3755 \textit{Chandra COSMOS-Legacy}
sources, 3459 ($\simeq$92\%) have an UltraVISTA counterpart.  
In the catalog we also added the match with X-ray detected sources from XMM-COSMOS
\citep{2007ApJS..172...29H,2007ApJS..172..341C,2010ApJ...716..348B}
and the previous Chandra COSMOS catalog
\citep{2009ApJS..184..158E,2012ApJS..201...30C}.

\subsubsection{Far-IR photometry}
Photometry at 24$\mu$m was obtained for a total of 42633 sources using
an updated version of the COSMOS MIPS-selected band-merged catalog
published by \cite{2009ApJ...703..222L}.  In this catalog, ~90\% of
the 24$\mu$m-selected sources were securely matched to their
$K_{\rm s}$-band counterpart using the WIRCAM COSMOS map
of~\cite{2010ApJ...708..202M}, assuming a matching radius of
2~$\arcsec$. Counterparts to another 5\% of the sample were found
using the IRAC-3.6$\mu$m COSMOS catalog of \cite{2007ApJS..172...86S},
while the rest of the 24$\mu$m source population remained unidentified
at shorter wavelengths. We thus considered the coordinates of the
WIRCAM K-band or IRAC counterparts (or the initial 24$\mu$m
coordinates for the unidentified MIPS sources), and cross-correlated
these positions with the VISTA catalog using a matching radius of
1$\arcsec$. VISTA counterparts were found for all of the 
previously-identified 24$\mu$m sources and for an additional set of
117 objects detected by MIPS which had no previous identification.

We also provide Far-IR photometry obtained at 100, 160, 250, 350, and
500$\mu$m using the PACS \citep{2010Pog} and SPIRE \citep{2010Griffin}
observations of the COSMOS field with the \textit{Herschel} Space Observatory.
The PACS data were obtained as part of the PEP guaranteed time program
\citep{2011A&A...532A..90L}, while the SPIRE observations were carried
out by the HERMES consortium \citep{2012MNRAS.424.1614O}. For each
band observed with \textit{Herschel}, source extraction was performed by a PSF
fitting algorithm and using the 24$\mu$m source catalog as
priors. Hence, far-IR matches to VISTA were unambiguously obtained
from the 24$\mu$m source counterparts described above, leading to a
total of 6608 sources with a PACS detection and 17923 sources detected
with SPIRE. Total uncertainties in the SPIRE bands include the
contribution from confusion. Flux density measurements with a signal to
noise smaller than 3 in the initial SPIRE COSMOS catalog published by
\cite{2012MNRAS.424.1614O} are not considered in our present work.

\subsection{Computation of photometric errors and upper limits}
\label{Sec:corr}

Precise photometric error measurements are essential for accurate
photometric redshifts. For each Subaru band, we use effective gain
values \citep{2007ApJS..172...99C} for the non-convolved data to
compute the magnitude errors. This is particularly important for
the Subaru bands because of the long exposure times used for each
individual exposure. However, because {\sc SExtractor} errors are
underestimated in data with correlated noise, we multiply the
magnitude and flux errors with a correction factor computed for each
band from empty apertures (based on the segmentation map,
  apertures that contain an object have been discarded). Following
\cite{2012A&A...545A..23B}, this factor is computed in each band for
the 2 and 3$\arcsec$ apertures and taken at the ratio between the
standard deviation of the flux extracted in empty apertures on the
field and the median of the {\sc SExtractor} errors. For UltraVISTA,
we compute separate values for the Ultra-deep (${\cal A}^{\rm UD}$)
and deep (${\cal A}^{\rm Deep}$) regions. The corrections are given in
Table~\ref{Tab:sysoff}.

In some bands, a source may be below the measurement threshold while at
the same time be detected in the combined $zYJHK_{\rm s}$ $\chi^{2}$
image. In this case, in the measurement band, {\sc SExtractor} may not
report consistent magnitudes or magnitude errors, and we report upper
limits on the source magnitudes in each band where they are too
faint to be detected. To compute the magnitude limits, we run {\sc
  SExtractor} on each individual image using the same detection
parameters. All of the pixels belonging to objects are flagged. Fluxes are
measured from PSF-homogenized images in empty apertures of 2 and
$3\arcsecond$, discarding all of the apertures containing an object. The
magnitude limit is then computed from the standard deviation of
fluxes in each aperture.

\begin{figure*}
\begin{center}
\includegraphics[scale=0.29]{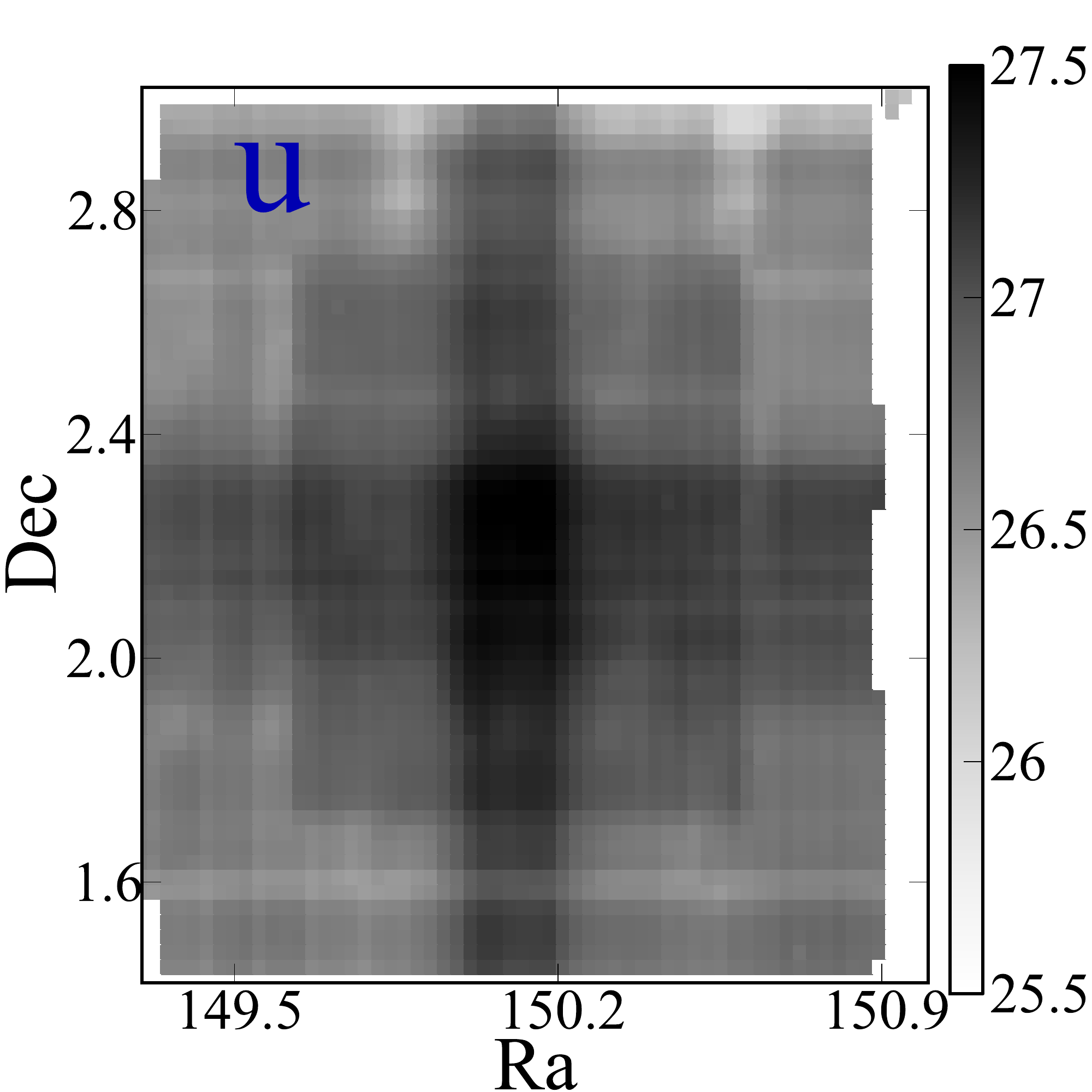}
\includegraphics[scale=0.29]{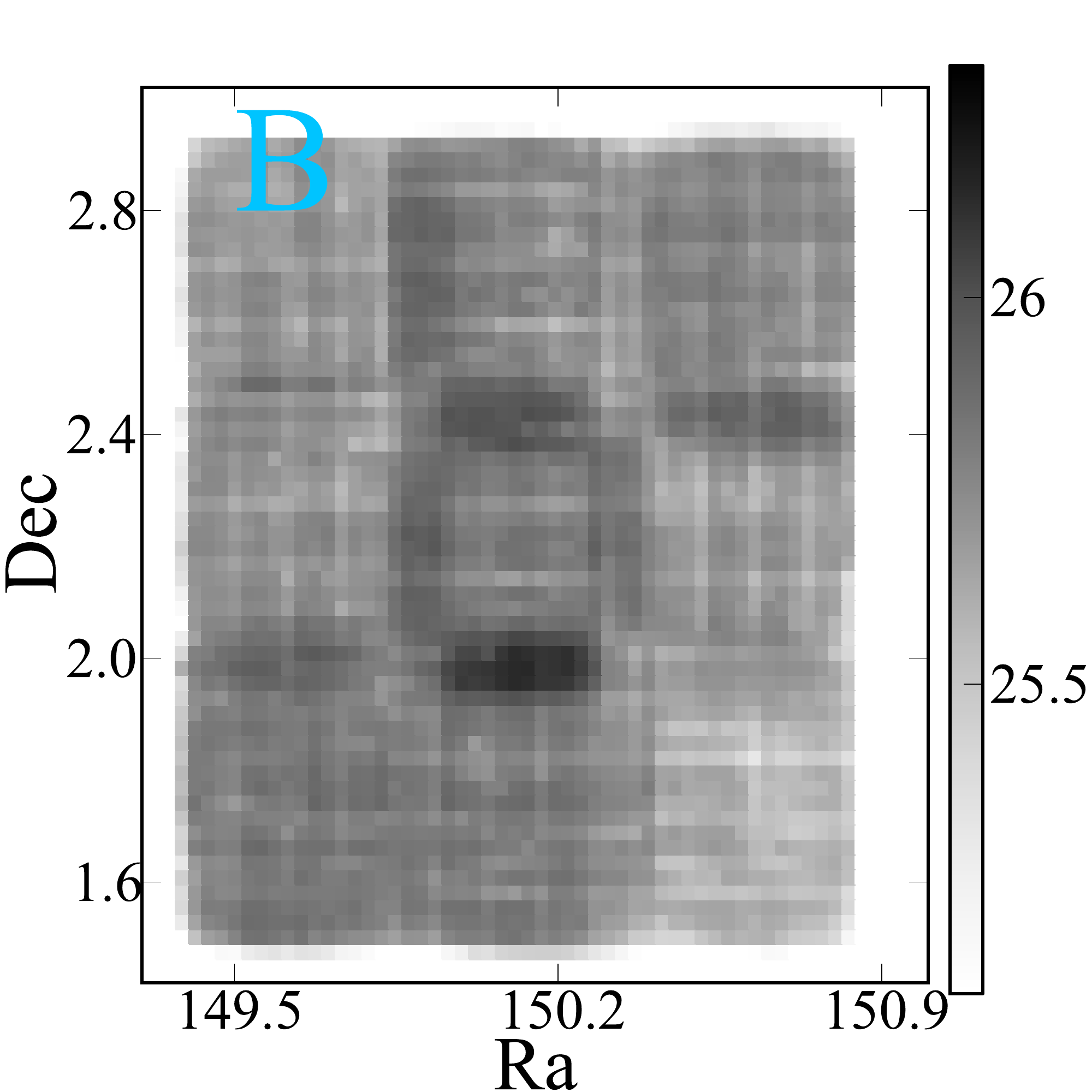}
\includegraphics[scale=0.29]{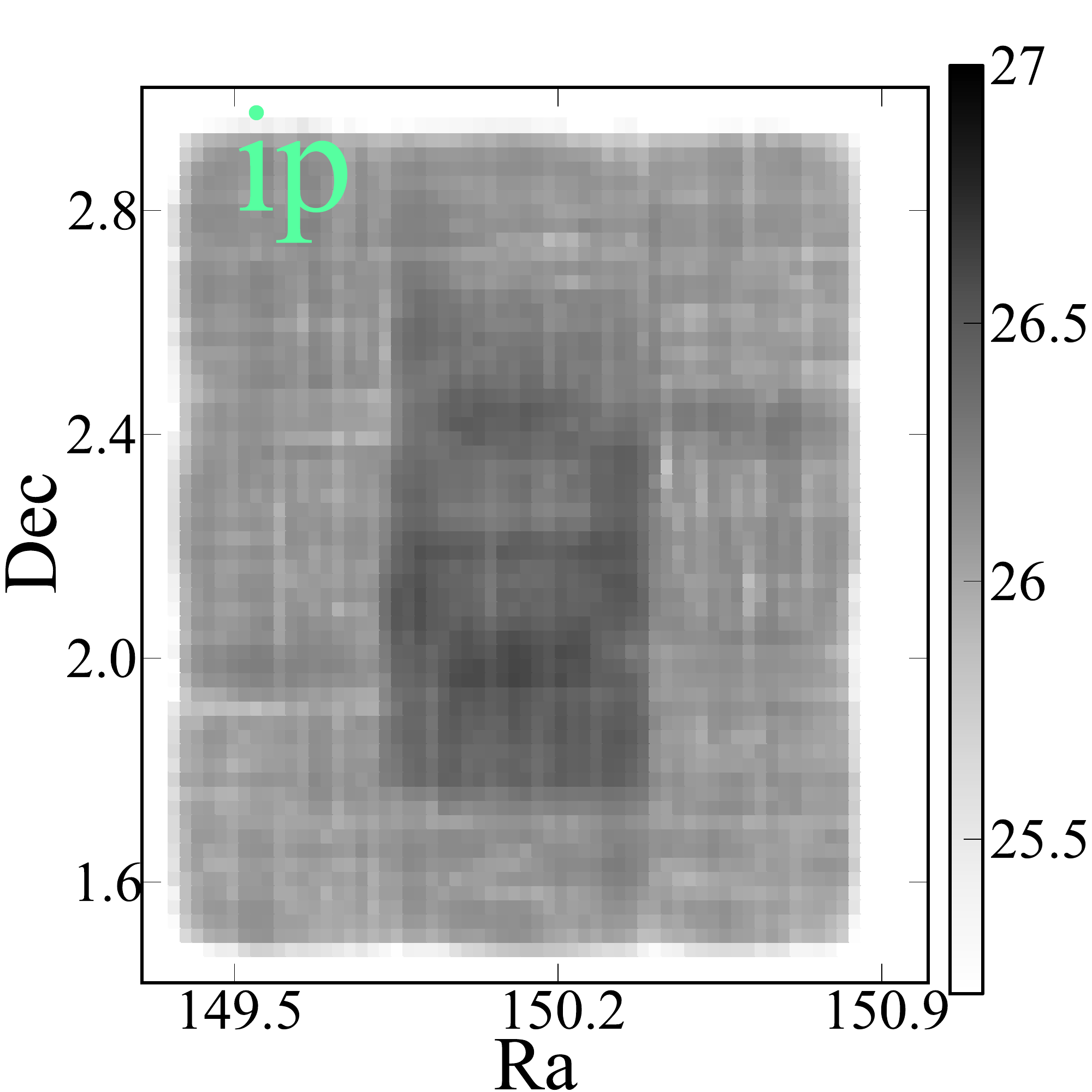}
\caption{The locally computed magnitude limits computed from the
  variance map as a function of right ascension and declination for
  $u$, $B$ and $i^{+}$ bands in a 2$\arcsec$ diameter aperture.}
\label{Fig:VarUp}
\end{center}
\end{figure*}

This method is not always appropriate since the values of the upper
limits may vary over the field, as shown in
Figure~\ref{Fig:VarUp}. This is why we use a local estimate for the
upper limits in the six broad bands of optical data ($u$, $B$, $V$, $r$,
$i^{+}$, $z^{++}$). In these
bands, upper limits are calculated for each object from the variance
map and are defined as being the square root of the variance per pixel
integrated over the aperture. The magnitude of the object is set to the 3$\sigma$
magnitude limit if the flux is below the 3$\sigma$ flux limit, or if the flux is
below the flux error. The averaged values of these upper limits are
consistent with the value computed with the first method and are displayed
in Table~\ref{Tab:limmag}. The upper limits in these bands are
important because young, star-forming objects at high redshift will
have apparent magnitudes in the optical bands of the order of the limiting
magnitude. The computation of the photometric redshift uses fluxes and so does not
use the upper limits which are only applied to the magnitudes, but it may be useful
when working with magnitudes to know whether or not the object is
within the upper limit. 

\subsection{Catalog validation}
\label{sec:catalog-validation}

\subsubsection{Number counts}
\label{sec:number-counts}

\begin{figure}
\begin{center}
\includegraphics[scale=0.44]{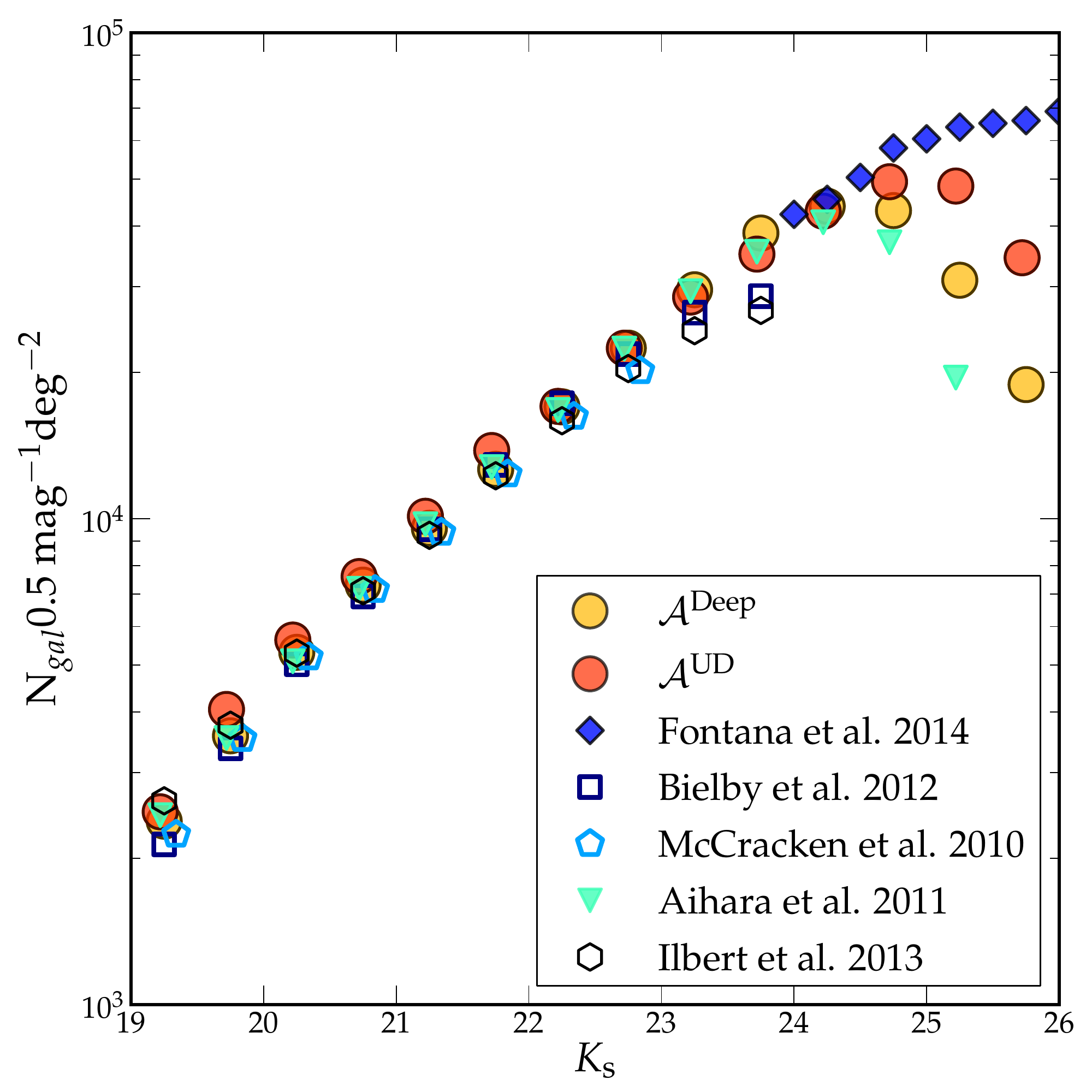}
\caption{$K_{\rm s}$-band-selected galaxy number counts of the
  $YJHK_{\rm s}z^{++}$-detected galaxies in ${\cal A}^{\rm UD}$
  (yellow circles) and ${\cal A}^{\rm Deep}$ (pink circles),
  compared to a selection of literature measurements. The
 \cite{Ilbert:2013dq} and \cite{2010ApJ...708..202M} points show
  previous measurements in COSMOS.}
\label{Fig:ncount}
\end{center}
\end{figure}

In Figure~\ref{Fig:ncount}, we plot the number of galaxies per square
degree per magnitude as a function of $K_{\rm s}$ magnitude for
objects in both the ${\cal A}^{\rm Deep}$ and ${\cal A}^{\rm UD}$ regions
(details of the stars-galaxy separations can be found in
Section~\ref{Sec:sep}). The corresponding values are presented in Table~\ref{Tab:count}.

Our counts are in excellent agreement with the literature. We reach
more than one magnitude deeper compared to the previous UltraVISTA-DR1
\citep{McCracken:2012gd}. In addition, our counts are in good
agreement with the much deeper Hawk-I survey \citep{2014Fontana}
up to at least $K_{\rm s}\sim24.5$.

At the $3\sigma$ limit in $K_{\rm s}$, we detect almost twice as many
objects per square degree in ${\cal A}^{\rm UD}$ than in
${\cal A}^{\rm Deep}$. Furthermore, our catalog contains
$\sim 1.5\times 10^{5}$ objects with $K_{\rm s}<24.7$ in the
${\cal A}^{\rm UD}$ compared to $\sim 0.8\times 10^{5}$ found with
UltraVista-DR1 \citep{McCracken:2012gd} in the same region at the
detection limit in $K_{\rm s}$.  In ${\cal A}^{\rm Deep}$, the
difference is less significant since the depths are comparable, with
$\sim 0.9\times 10^{5}$ objects compared to $\sim 0.7\times 10^{5}$
found in UltraVista-DR1.

\begin{figure}

% say something about star-galaxy separation. 

\begin{center}
\includegraphics[scale=0.44]{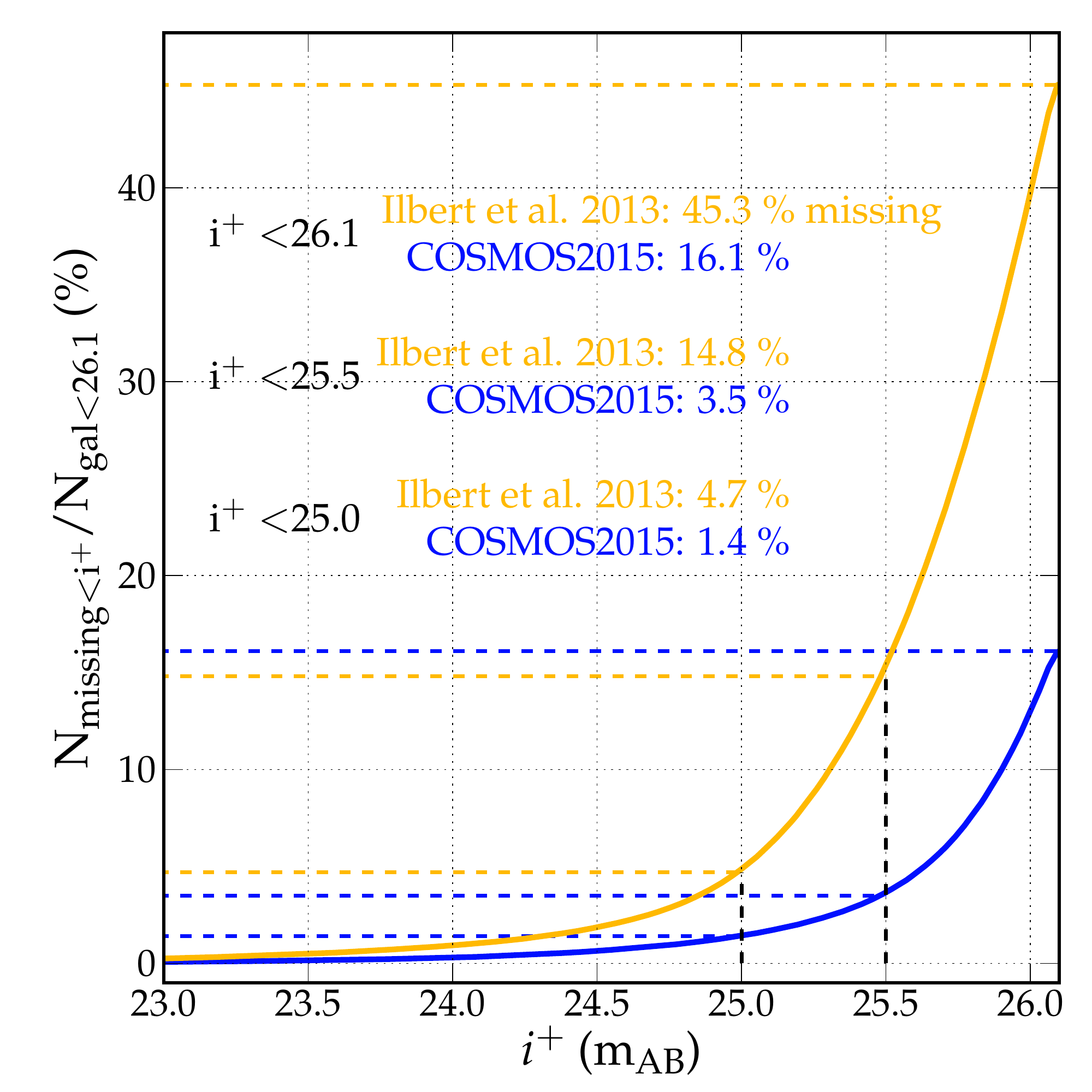}
\caption{Cumulative distribution of the number of galaxies over the
  total number of galaxies in a magnitude-limited sample ($i^{+}<26.1$) as a function of
  magnitude in the $i^{+}$-band from
  \cite{2007ApJS..172...99C} which are not recovered in
  \cite{Ilbert:2013dq} (orange line) and this catalog (blue line).}
\label{Fig:CheckIp}
\end{center}
\end{figure}

Compared with the previous publicly available photometric
$i^{+}$-selected catalog (described in footnote 4) at the detection
limit $i^{+}<26.1$ \citep[limiting magnitude at 5$\sigma$ in a 3
$\arcsec$ diameter aperture from][]{2007ApJS..172...99C}, we find
that 16.1\% of sources are not present in COSMOS2015, as shown on
figure~\ref{Fig:CheckIp}. Many of these missing sources are blue,
faint ($25.5<i^{+}<26.1$), low-mass, star forming galaxies. This
difference is to be expected, since an NIR-only selection and a pure
$i^{+}$-selection are not expected to sample the same galaxy
populations. However, we have mitigated this difference by including
the $z^{++}$ band in our detection image; this percentage is smaller
than in \cite{Ilbert:2013dq}, where the detection image was shallower
and did not include any optical bands. Furthermore, the previous
$i^{+}$-selected catalog also contained spurious objects near the
detection limit, and therefore the fraction of missed genuine objects
can be expected to be lower.

\nocite{2014Fontana} \nocite{2012A&A...545A..23B} 
\nocite{2010ApJ...708..202M}
\nocite{McCracken:2012gd}
\nocite{2011ApJS..193...29A}

\begin{table}
\begin{center}
  \caption{$K_{\rm s}$-band-selected galaxy number counts. Logarithmic
    galaxy number counts are in units of galaxies per 0.5 magnitude
    per square degree.}
\begin{tabular}{ccc}
\hline
 \textbf{Magnitude Bin} & \textbf{Deep Regions} & \textbf{Ultra-Deep Regions} \\
  & & \textbf{regions} \\\hline
$\left[16.0-16.5\right]$ & 2.24 & 2.19\\
$\left[16.5-17.0\right]$ & 2.28 & 2.33\\
$\left[17.0-17.5\right]$ & 2.58 & 2.60\\
$\left[17.5-18.0\right]$ & 2.72 &  2.80\\
$\left[18.0-18.5\right]$ & 2.94 &  3.00\\
$\left[18.5-19.0\right]$ & 3.19 &  3.24\\
$\left[19.0-19.5\right]$ & 3.39 &  3.42\\
$\left[19.5-20.0\right]$ & 3.57 &  3.62\\
$\left[20.0-20.5\right]$ & 3.73 &  3.76\\
$\left[20.5-21.0\right]$ & 3.87 &  3.89\\
$\left[21.0-21.5\right]$ & 3.99 &  4.01\\
$\left[21.5-22.0\right]$ & 4.11 &  4.14\\
$\left[22.0-22.5\right]$ & 4.23 &  4.23\\
$\left[22.5-23.0\right]$ & 4.35 &  4.36\\
$\left[23.0-23.5\right]$ & 4.48 &  4.46\\
$\left[23.5-24.0\right]$ & 4.59 &  4.55\\
$\left[24.0-24.5\right]$ & 4.65 &  4.64\\
$\left[24.5-25.0\right]$ & 4.63 &  4.69\\
$\left[25.0-25.5\right]$ & 4.49 &  4.68\\
$\left[25.5-26.0\right]$ & 4.27 &  4.54\\
\hline
\end{tabular}
\label{Tab:count}
\end{center}
\end{table}
%%%
\subsubsection{Astrometric accuracy}

\begin{figure}
\begin{center}
\includegraphics[scale=0.6]{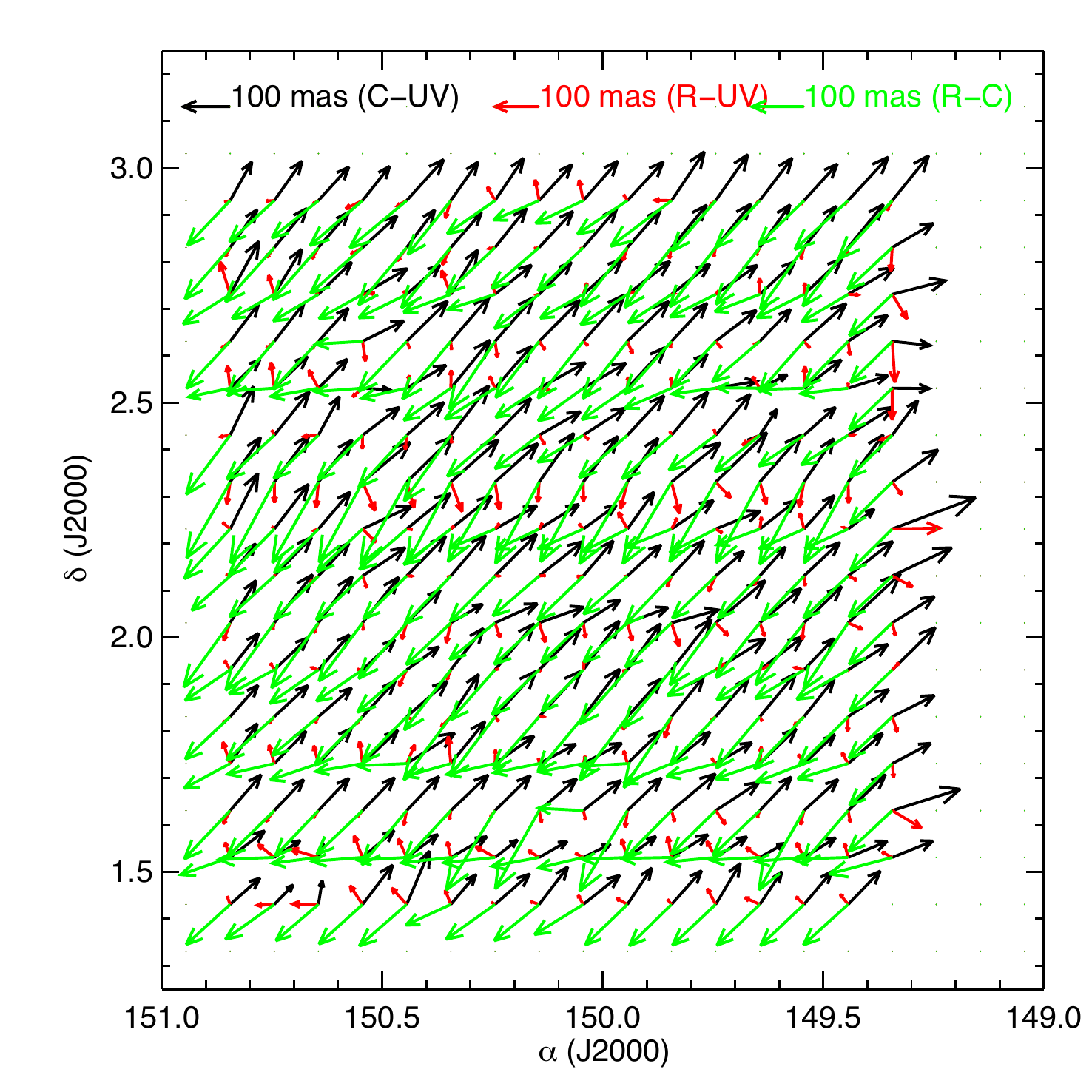}
\caption{Astrometric comparison for bright objects between our
  catalog, the catalog from \cite{2007ApJS..172..219L} and the
  publicly available COSMOS $i^{+}$-selected catalog
  \citep[]{2007ApJS..172...99C}. Black arrows show the shift between
  Capak et al. and COSMOS2015, red arrows between
  \cite{2007ApJS..172..219L} and COSMOS2015. Finally, green arrows
  show the shift between the two previous catalogs. All of these shifts
  occur below one pixel.}
\label{Fig:astrometrycheck}
\end{center}
\end{figure}

We compared the astrometric positions of bright, non-saturated objects
in COSMOS2015 with those in the COSMOS reference catalog from
\cite{2007ApJS..172..219L} and the publicly available $i^{+}$-selected
photometric catalog (footnote 4) described in
\cite{2007ApJS..172...99C}. This is illustrated in
Figure~\ref{Fig:astrometrycheck}. There is good agreement between
COSMOS2015 and the \cite{2007ApJS..172..219L} catalog. The shift
between the $i^{+}$-selected catalog and \cite{2007ApJS..172..219L} is
no longer present in COSMOS2015. This shift occurs below a pixel size of
0.15$\arcsec$. These comparisons show that our astrometry is accurate
to at least one pixel.

We note that the COSMOS astrometric reference catalog used in
\cite{2010ApJ...708..202M,McCracken:2012gd} and this paper is based
on a reference catalog extracted from a Megacam $i$-band (data taken
in 2004) image covering the full COSMOS field. The astrometric zero
point of this catalog was set using radio interferometric
observations \citep{Schinnerer:2004p12717}. Until now, at scale
smaller than the size of the resampled pixels, it has been challenging to test the astrometric
accuracy for our catalog given the lack of availability of
sufficiently dense astrometric catalogs. However, we have compared the
positions between our catalog and the catalogs extracted from the
independently reduced Hyper Suprime-Cam images described here, and this
has confirmed that our astrometric solutions are good at the level of
one pixel. For future data releases, we intend to improve our overall
astrometric precision by using densely sampled catalogs based on
either Hyper Suprime-Cam or Pann-Starrs data, which are tied to 2MASS. 
 
\section{Photometric Redshift and physical parameters}
\label{Sec:photoz}
\subsection{Input catalog}
\label{highz}

We use fluxes rather than magnitudes for our photometric measurements
to deal robustly with faint or non-detected objects. Faint objects
may have a physically meaningful flux measurement, whereas their
magnitudes and magnitude errors may be undetermined (for example, if
the flux is negative).  Consequently, when using magnitudes, we must
set an upper limit: for flux measurements with correct flux errors,
this is no longer necessary. There is no loss of information when using flux measurements. This leads to a better determination of
the photometric redshift and a lower number of catastrophic failures
at $z>2$.

Photometric redshifts are computed using 3$\arcsec$ aperture
fluxes. The fixed-aperture magnitude estimate is expected to be less
noisy for faint sources than the pseudo-total Kron
\citep{1980ApJS...43..305K} magnitudes {\sc mag\_auto}. This is
because {\sc mag\_auto}'s variable aperture is derived from the
detection image, which means that fainter objects have can potentially have
noisier colors \citep[][]{2012MNRAS.421.2355H,2016Moutard}. This magnitude measurement is also
susceptible to blended sources. We find that the 3$\arcsec$ aperture
photometry gives slightly better photometric redshifts than the
2$\arcsec$ aperture at low redshift (below $z\leq 1$) and we adopt
this aperture over the entire redshift range of our survey.  We
suspect that the photometric redshift precision is lower in the
2$\arcsec$ apertures due to small-scale residual astrometric
errors. This is being investigated for the upcoming DR3 UltraVISTA
release.

Photometric redshift computations use colors, and, consequently, should
not be sensitive to a systematic magnitude calibration
offset. However, in contrast to optical and NIR data, GALEX and IRAC
data provide total magnitudes or fluxes, which require an estimate of the
total flux from the corrected 3$\arcsecond$ aperture fluxes to be
consistent over the full wavelength range. This is also needed to
derive stellar masses. For each object, we compute a single
  offset $o$ (the same for all the bands) which allows for the conversion
  from aperture to total magnitude. The offset is computed following
\cite{2016Moutard}:

\begin{equation}
\begin{aligned}
o= & \frac{1}{\sum_{\rm filters \,i}{\rm w}_{i}}\times \\
& \sum_{\rm filters \,i}\left({\rm MAG}_{\rm AUTO}-
  {\rm MAG}_{\rm APER}\right)_{i}\times{\rm w}_{i}
\label{off}
\end{aligned}
\end{equation}
where we have:
\begin{equation}
{\rm w}_{i}=\frac{1}{\left({\sigma}^{2}_{\rm AUTO}+ {\sigma}^{2}_{\rm APER}\right)_{i}}
\end{equation}

This leads to the assumption that the PSF profile is the same in all of the
bands.  As it is averaged over all of the broad bands, i.e, $u$, $B$, $V$, $r$,
$i^{+}$, $z^{++}$, $Y$, $J$, $H$, and $K_{\rm s}$, this offset is more
robust than the one which would have been computed by band. These
offsets are given in the final catalog.

\subsection{Method}
\label{sec:Method}
To compute the photometric redshifts, we use {\sc LePhare} \citep{Arnouts:2002p10840, 2006A&A...457..841I} with the same
method as used in \cite{Ilbert:2013dq}.  
Our aim is to compute precise photometric redshifts over a wide
redshift range and for many object types with minimum
bias. Obviously, a single set of recipes will not perform as well as
several configurations, with each one tuned to optimize the fit at different redshifts.
That is why we use a set of 31 templates
including spiral and elliptical galaxies from
\cite{Polletta:2007p6857} and a set of 12 templates of young blue
star-forming galaxies using \cite{Bruzual:2003p963} models
(BC03). Extinction is added as a free parameter ($E(B-V)<0.5$) and
several extinction laws are considered: those of \cite{Calzetti:2000p6839},
\cite{Prevot:1984p6814} and a modified version of the Calzetti laws
including a ``bump'' at 2175\AA~\citep{1986ApJ...307..286F}.  Using a
spectroscopic sample of quiescent galaxies, \cite{2012ApJ...755...26O}
showed that the estimate of the photometric redshift for the quiescent
galaxies in \cite{2009ApJ...690.1236I} were underestimated at
$1.5<z<2$. Following \cite{Ilbert:2013dq}, we improved the photometric
redshift for this specific population by adding two new BC03 templates
assuming an exponentially declining SFR with a short timescale
$\tau=0.3\,\mathrm{Gyr}$ and extinction-free templates.

Finally, we compute the predicted fluxes in every band for each
template and following a redshift grid with a step of 0.01 and a
maximum redshift of 6. The computation of the fluxes also takes into
account the contribution of emission lines using an empirical relation
between the UV light and the emission line fluxes as described in
\cite{2009ApJ...690.1236I}.

The code performs the $\chi^{2}$ analysis between the fluxes
predicted by the templates and the observed fluxes of each galaxy.  
At
each redshift, $z_{\rm step}$, and for each template of the library,
the $\chi^{2}$ is computed as% template $T_{\rm min}$
%which minimizes the expression is found:
\begin{equation}
  \chi^{2}(z_{\rm step})=\sum_{\rm filters \, i}\frac{(F_{\rm obs\, i}-\alpha F_{\rm
      SED \, i}(z_{\rm step},T))^{2}}{\sigma^{2}_{\rm obs \, i}}
\end{equation}

where $ F_{\rm SED \, i}(z_{\rm step},T)$ is the flux predicted for a
template $T$ at $z_{\rm step}$ and $\alpha$ is the normalization
factor.  Then, the $\chi^{2}$ is converted to a probability of
$p=\exp^{-\chi^{2}/2}$. All of the probability values are summed up at
each redshift $z_{\rm step}$ to produce the Probability Distribution
Function (PDF).
We then determine then the
photometric redshift solution from the median of this
distribution. The 1$\sigma$ uncertainties given in the catalog are
derived directly from the PDF and enclose 68\% of the area around the
median.

An important aspect of the method is the computation of systematic
offsets which are applied to match the predicted magnitudes and the
observed ones \citep{2006A&A...457..841I}.  We measure these offsets
using the spectroscopic sample. For each object, we search for the
template which minimizes the $\chi^{2}$ at fixed redshift. Then, we
measure the systematic offset which minimizes the difference between
the predicted and observed magnitudes. This procedure iterates until
convergence.

\begin{figure*}
\begin{center}
\includegraphics[scale=0.44]{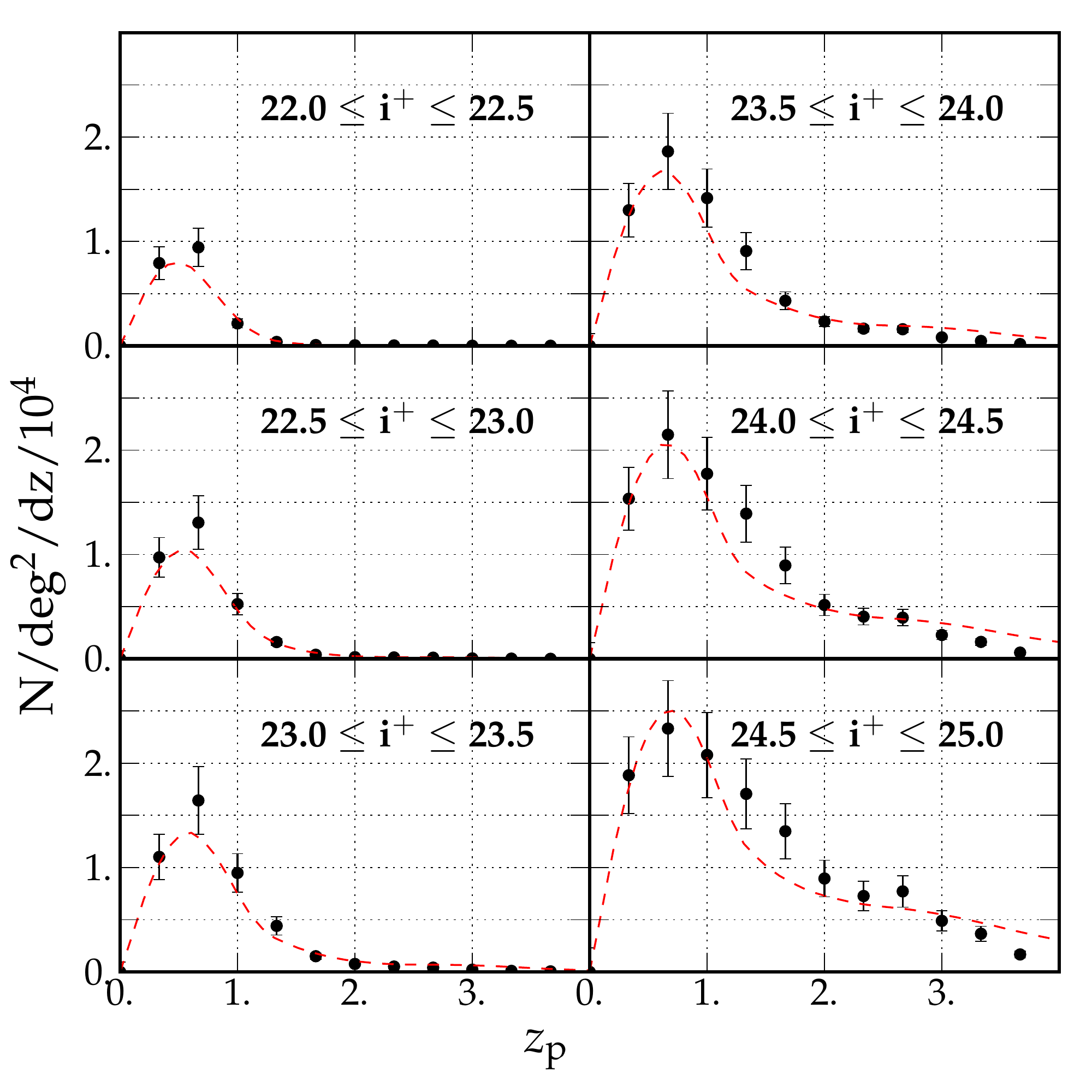}
\includegraphics[scale=0.44]{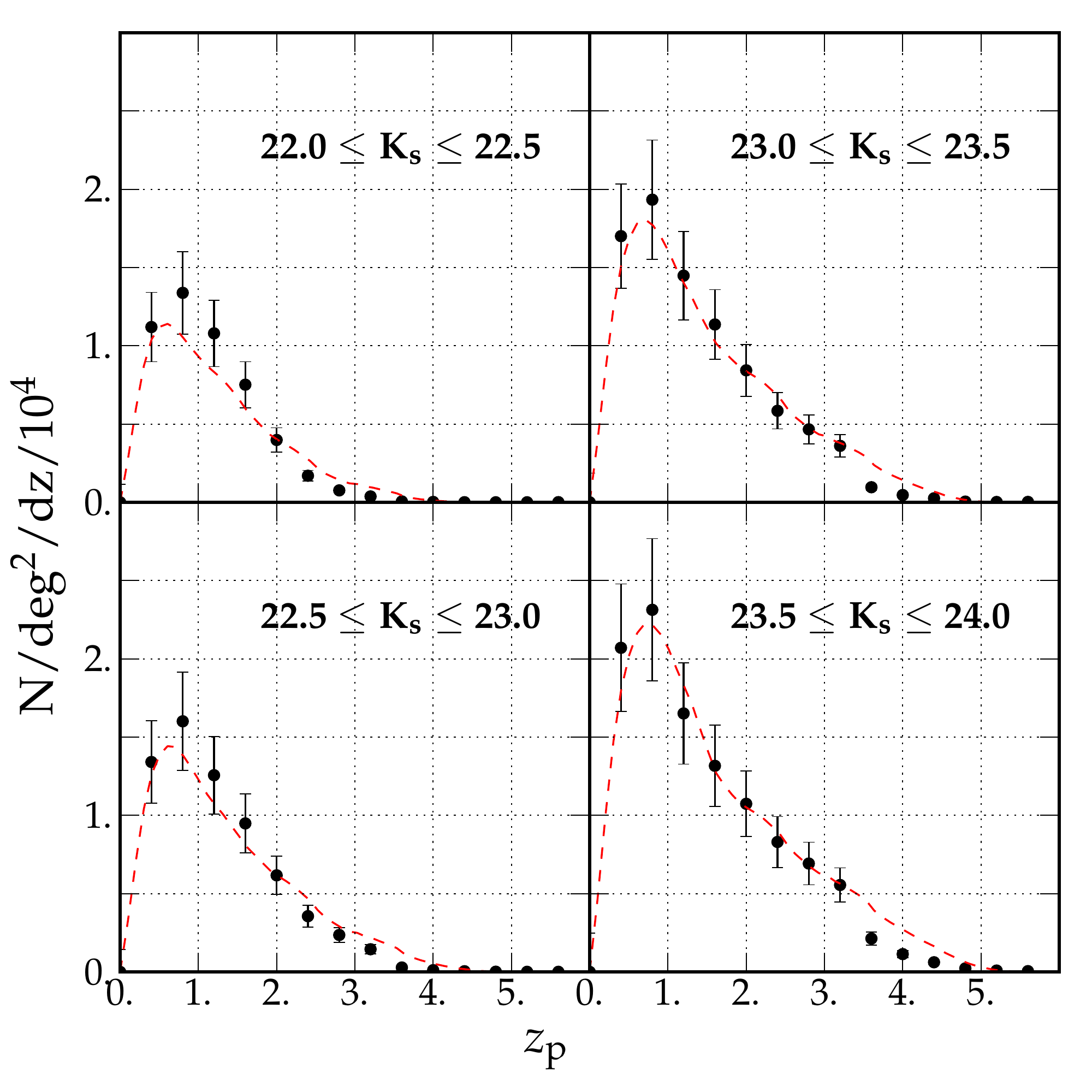}
\caption{ Photometric redshift distributions for $i^{+}-$
  (\textit{left}) and $K_{\rm s}$(\textit{right})-selected samples
  for the full sample, compared with a model prediction (red dashed
  line) from PEGASE.2
  \citep{1997A&A...326..950F,1999astro.ph.12179F}. Plotted errorbars
  are uncertainties estimated from jacknife errors, splitting the
  field into 25 sub-fields.}
\label{Fig:ZDistrib}
\end{center}
\end{figure*}

The photometric redshift distribution for the $i^{+}$- and $K_{\rm s}$-selected samples is given in Figure~\ref{Fig:ZDistrib}. Magnitudes are
measured in corrected 3'' aperture magnitudes with the derived
systematic offset applied. Several interesting trends are apparent. In
general, the median redshift of our $K_{\rm s}-$ sample is higher than
our $i^{+}$-selected samples. Also, the fraction of sources at higher
redshifts is greater for the NIR-selected samples. These
effects are largely due to the well-known positive
evolutionary corrections and $k-$ corrections for NIR-selected
samples. Optically selected samples at higher redshifts move
progressively to shorter rest-frame UV wavelengths, which are strongly
attenuated by dust and the intergalactic medium. We compare these
distributions with a simple three-components galaxy population model
generated with the PEGASE.2 code 
\citep{1997A&A...326..950F,1999astro.ph.12179F}.  Each population
starts forming at $z=8$ via the infall of pristine gas on a specific
timescale and gas is converted into stars at a specific rate.  The
corresponding star formation histories peak at $z\sim 4$, 2 and 0, and the
$z=0$ predicted optical-NIR colors correspond to those of local Sa,
Sbc, and Sd galaxies, respectively.  The total baryonic mass (gas,
stars, and hot halo-gas) of each galaxy is assumed to be constant, and
the mass function of each population is tuned so that the sum of the
three populations matches simultaneously the local luminosity function
in the $B$ band, the deep galaxy counts in the $B$, $V$, $I$, and $K_{\rm s}$
band, as well as the cosmic star formation rate density and the
stellar mass density observed at $z=0-6$.  The agreement between the
data and this simple three-component model is quite good. This success
lies in the differential contributions of the three galaxy populations to
the counts. Indeed, our modeled counts at $K_{\rm s}=24$ are the sums
of the 
almost equal contributions of Sd progenitors at $z\sim$0.7, of Sb
progenitors at $z\sim$1.2, and Sab progenitors at $z\sim$2. In
contrast, a simpler modeling of the galaxy populations using a single
scenario with an SFH proportional to SFRD($z$) (Star Formation Rate Density) and a
unique mass function leads to very good agreement between the integrated
counts in $i^{+}$ and $K_{\rm s}$ bands, as well as a good match between the
SFRD($z$) and $\rho_{*}(z)$, but it completely overshoots the mean
redshift of the $K_{\rm s}\sim$24 or $i^{+}\sim$24.5 sources
($z\sim$2, whereas the COSMOS data shows it is peaked at
$z\leq$1). Other choices of modeling that we explored also lead to a
high level of tension either in the SFRD($z$), $\rho_{*}(z)$ or in the
counts in the $i^{+}$, $K_{\rm s}$ or $B$ bands.

\begin{table}
\begin{center}
  \caption{Photometric corrections, including multiplicative error
    factors for {\sc SExtractor} (see Section~\ref{Sec:corr}),
    systematic offsets ($s_{f}$) derived from the spectroscopic sample (see
    Section~\ref{sec:Method}) and factors $F$ for the foreground
    extinction \citep{Allen1976}. $s_{f}$ values have to be subtracted to the apparent magnitudes.}

\begin{tabular}{ l  l l l l l  }
\hline
& \textbf{band}  & \textbf{error} & \textbf{error} & \textbf{$s_{f}$}   &
                                                                \textbf{$F$}
  \\
&   & \textbf{fact. (2$\arcsec$)} & \textbf{fact. (3$\arcsec$)} &   
                                                                 \\
\hline
HSC & $Y$ & 2.2 & 2.7& -0.014 & 1.298 \\
\hline
& $Y^{\rm UD}$ & 3.2  & 3.7 & 0.001 & 1.211 \\
 & $Y^{\rm Deep}$ & 2.8  &3.2  &0.001 & 1.211 \\
UVista & $J^{\rm UD}$ & 3.0 &3.3& 0.017 & 0.871 \\
UVista & $J^{\rm Deep}$ & 2.6& 2.9& 0.017 &  0.871 \\
& $H^{\rm UD}$ & 2.9 &3.1& 0.055  & 0.563 \\
& $H^{\rm Deep}$ & 2.4 &2.9& 0.055  & 0.563 \\
& $K_{\rm s}^{\rm UD}$ & 2.7 &3.1& -0.001 & 0.364\\
& $K_{\rm s}^{\rm Deep}$ & 2.3 &2.6 &-0.001 & 0.364 \\
 \hline
 WIRCam & $Ks$ & 2.1 &3.4& 0.068 & 0.364 \\
 & $H$ & 2.1 & 3.2 & -0.031 & 0.563 \\
 \hline
 CFHT & $u$ & 2.3 &3.3 & 0.010&  4.660\\
 \hline
 & $B$ & 1.6  & 1.8&0.146 & 4.020 \\
 & $V$ & 1.7 & 1.9 &-0.117 & 3.117 \\ 
 & $r$ & 1.4 & 1.7 & -0.012 & 2.660 \\
  & $i^{+}$ & 1.3 &1.7& 0.020 &  1.991 \\
  & $z^{++}$ & 2.0& 2.9&  -0.084 & 1.461 \\
 & $IA427$ & 1.7 & 2.5 & 0.050 & 4.260 \\
 & $IA464$ & 1.7 &2.4 &-0.014 & 3.843 \\
 & $IA484$ & 1.7 &2.5  &-0.002 & 3.621 \\
 & $IA505$ & 1.6 & 2.3& -0.013 & 3.425 \\
 & $IA527$ & 1.5 & 2.2&0.025 & 3.264 \\
 SUBARU & $IA574$&1.9 & 2.8 & 0.065 & 2.937 \\
 & $IA624$ & 1.4 & 2.0 &-0.010 & 2.694 \\
 & $IA679$ & 2.0 & 2.8 &-0.194  & 2.430 \\
 & $IA709$ & 1.7 & 2.4 &0.017  & 2.289 \\
 & $IA738$ & 1.5 &2.1 &0.020  & 2.150 \\
 & $IA767$ & 1.8 & 2.6&0.024 &  1.996 \\
 & $IA827$ & 2.2 &3.1 & -0.005  & 1.747  \\
 & $NB711$ & 1.2 & 1.8& 0.040 & 2.268  \\
 & $NB816$ & 2.5 & 3.5 &-0.035 & 1.787  \\
 \hline
 & ch1 &- & - &-0.025 & 0.162 \\
 IRAC & ch2 & -& -& -0.005 & 0.111 \\
  & ch3 & -& -&-0.061 &  0.075 \\
  & ch4 &- &- &-0.025 & 0.045 \\
 \hline
 GALEX & NUV &-& - & 0.128 & 8.621 \\
 \hline
 \end{tabular}
 \label{Tab:sysoff}
 \end{center}
 \end{table}
\subsection{Photometric redshift accuracy measured using spectroscopic samples}
\label{results_zs}

The COSMOS field is unique in its unparalleled spectroscopic data set.
These spectroscopic samples, derived from many hundreds of hours of
telescope time in many different observing programs, are a key
ingredient in allowing us to characterize the precision of our
photometric redshifts. 

From the COSMOS spectroscopic master catalog (Salvato M. et al. in
prep.), we retain only the highly reliable 97\% confidence-level
spectroscopic redshifts \citep[][]{2007ApJS..172...70L}. We estimate
the precision of the photometric redshift using the normalized median
absolute deviation (NMAD) \citep{1983ured.book.....H} defined as
\begin{math}
1.48 \;\times \; {\rm median}\left(| z_{\rm p} -z_{\rm s} |/(1+z_{\rm s})\right).
\end{math} 
This dispersion measurement, denoted by $\sigma$, is not affected by the fraction of catastrophic errors (denoted by $\eta$),
i.e. objects with $| z_{\rm p}-z_{\rm s} |/(1+z_{\rm s}) > 0.15$. 

\begin{figure*}
\begin{center}
\includegraphics[scale=0.61]{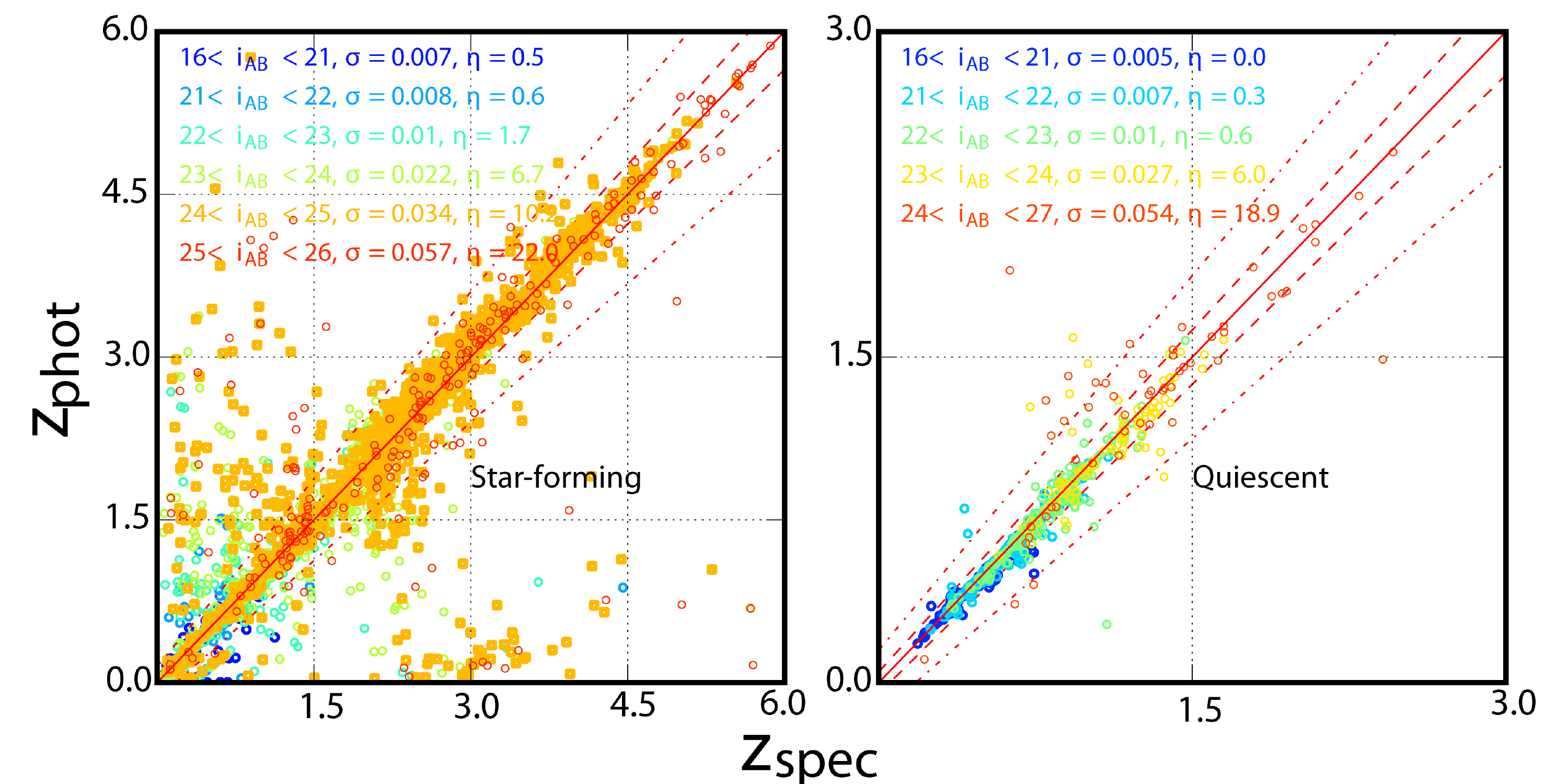}
\caption{Comparison between photometric and spectroscopic
  redshifts as a function of $i_{\rm AB}$ magnitude and type: star-forming galaxies(Left) and quiescent galaxies (Right), keeping only
 non-flagged galaxies. The dashed and dashed-dot lines show
  $z_{\rm p}=z_{\rm s}\pm 0.05(1+z_{\rm s})$ and
  $z_{\rm p}=z_{\rm s}\pm 0.15(1+z_{\rm s})$, respectively.}
\label{Fig:Zspec2}
\end{center}
\end{figure*}

\begin{figure*}
\begin{center}
\includegraphics[scale=0.91]{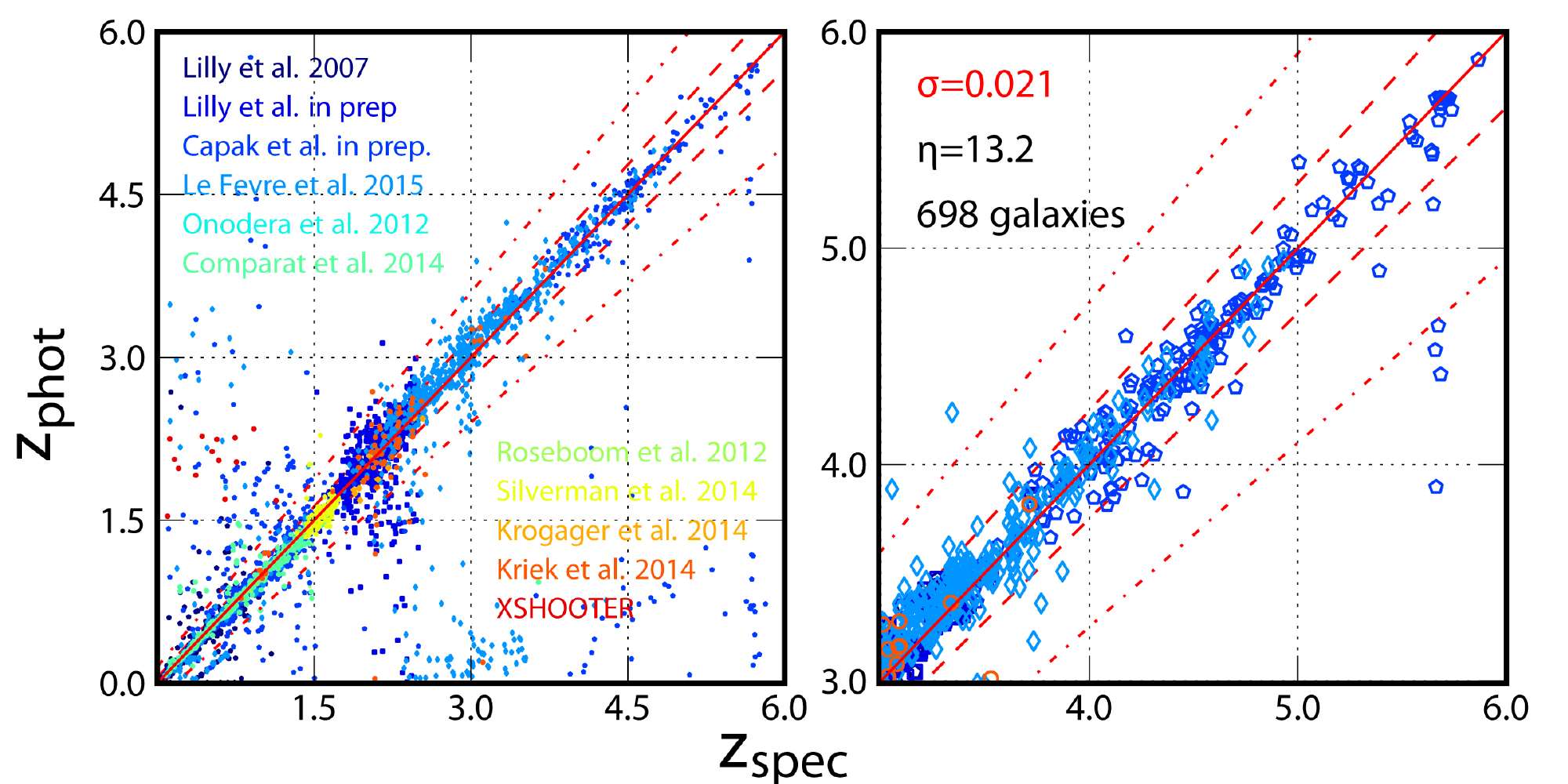}
\caption{ Left: comparison between photometric and spectroscopic
  redshifts for the different samples summarized in
  Table~\ref{acc}. Right: a magnified view of the high-redshift region. The
  number of galaxies, accuracy $\sigma$ and numbers of catastrophic
  failures $\eta$ and $\eta_{\rm lim}$ are computed from the zCOSMOS-faint, VUDS, DEIMOS,
  FMOS, and MOSDEF spectroscopic surveys taken together, keeping only
  non-flagged galaxies with a spectroscopic redshift greater than
  2.9. The dashed and dashed-dot lines show
  $z_{\rm p}=z_{\rm s}\pm 0.05(1+z_{\rm s})$ and
  $z_{\rm p}=z_{\rm s}\pm 0.15(1+z_{\rm s})$, respectively. Note that
  the given value for the precision and the percentage of catastrophic
  failures strongly depend on the spectroscopic sample. These values
  are detailed in
  Table~\ref{acc}. }
\label{Fig:Zspec}
\end{center}
\end{figure*}

The photometric redshift precision of the COSMOS2015 catalog is
described in Tables~\ref{acc2} and \ref{acc} as well as
Figures~\ref{Fig:Zspec2} and \ref{Fig:Zspec}.  In Table~\ref{acc2}, we
compare the photometric redshift precision in COSMOS2015 with that of
the catalog of \cite{Ilbert:2013dq} by cross-matching the two catalogs
and considering the same sources in both cases. Compared to
\cite{Ilbert:2013dq}, the number of catastrophic failures are reduced
and the photometric redshift precision is either increased or is
unchanged. It should be recalled, however, that the main gain of
COSMOS2015 is the considerable increase in catalog size compared
to \cite{Ilbert:2013dq}.

The left and right panels of Figure~\ref{Fig:Zspec2} show the photometric
redshift precision as a function of the $i$-band magnitude and for
star-forming and quiescent galaxies, respectively (classified using
NUV$-r$/$r-J$ diagram, Figure~\ref{Fig:classification}). Very bright,
low-redshift, star-forming galaxies have the most precise photometric
redshifts ($\sigma=0.007$, $\eta=0.5$\% for $16<i^{+}<21$). Moreover,
even at $z>3$, the accuracy is still very good (0.021), with only
13.2\% of catastrophic failures.\\

We now describe the photometric redshift precision and outlier fraction
for each spectroscopic sample. In all of the cases, the numbers correspond to
the fraction of secure spectroscopic redshifts not falling in masked
regions in our survey. These results are also summarized in
Table~\ref{acc} and plotted in Figure~\ref{Fig:Zspec}.

  \paragraph{zCosmos bright at $z<1.2$ \citep{2007ApJS..172...70L}.} 
  This sample from the zCOSMOS-bright survey includes 8608 galaxies
  selected with $i_{\rm AB}^+ \le 22.5$ (3$\sigma$,3$\arcsec$)
  observed with VIMOS at the VLT. We
  find $\sigma=0.007$ and $\eta=0.51$\%. 

  \paragraph{FORS2 sample at $z<3.7$ \citep{2015AA...575A..40C}.} This
  color-selected sample includes 788 objects and targets emission lines
  galaxies with 20 minute integration times with FORS2 at VLT. We find $\sigma=0.009$ and
  $\eta=2.03$\%. 

  \paragraph{The Keck follow-up reaching $z\sim 6$ \citep[][Capak et
    al. in prep]{2010ApJ...721...98K}.} This sample comprises
  spectroscopic redshifts of 2022 objects, some of which are $z>4$
  sub-populations selected in IR, and measured with DEIMOS at Keck II. We find $\sigma=0.014$ and
  $\eta=7.96$\%.

\paragraph{FMOS sample of IR luminous galaxies at $0.8<z<1.5$ \citep{2012MNRAS.426.1782R}.}  We compare our results
  with 26 \textit{Herschel} SPIRE and \textit{Spitzer} MIPS-selected galaxies observed with FMOS at Subaru. We find $\sigma=0.009$ and $\eta=7.69$\%.

\paragraph{A faint sample of quiescent galaxies at $1.2<z<2.1$
  \citep{2012ApJ...755...26O}.} This sample contains
10 faint,
  quiescent galaxies at $z<2$ obtained with MOIRCS at Subaru. We find
  $\sigma=0.017$, with no catastrophic failures.

    \paragraph{ FMOS-COSMOS survey at $1.4<z<1.8$ \citep{2015Silverman}.} 
These 178 FMOS at Subaru spectroscopic redshifts
were selected from the \cite{2009ApJ...690.1236I}
catalog, which implies that the fraction of catastrophic failures (1.12\%)
will be underestimated.  We find $\sigma=0.022$ and $\eta=1.12$\%.

 \paragraph{A faint sample of quiescent galaxies  at
 $1.9<z<2.5$ \cite{2014ApJ...797...17K}.} This sample contains 11 faint
quiescent galaxies obtained with the WFC3-grism observations from the
3D-HST survey. We find $\sigma=0.069$, with no catastrophic failures.

    \paragraph{zCosmos faint sample at $1.5<z<2.5$ (Lilly et al. in
      preparation).} This sample includes 767 galaxies color-selected
    to lie in the range $1.5 \lesssim z \lesssim 2.5$ and observed
    with VIMOS at the VLT. This
    redshift range is the least constrained in photometric
    redshift and the median magnitude $i_{\rm AB}^+$ is as faint as
    23.8 (3$\sigma$,3$\arcsec$). Nevertheless, we find $\sigma=0.032$
    and $\eta=7.95$\%.

\paragraph{MOSDEF survey \citep{2015Kriek}.} This sample includes 80
  galaxies observed with MOSFIRE at Keck I.  We find $\sigma=0.042$ and $\eta=10.0$\%.

\paragraph{ A sample of galaxies obtained with X-Shooter at VLT
  \citep[Stockmann et al. in prep,][]{phdthesisZabl15}.}
This sample contains eight massive quenched galaxies around $z\sim 2$ (Stockmann et al. in preparation) and  six narrow-band selected emission line galaxies at $z\sim 2.2$  \citep{phdthesisZabl15}:
five of the galaxies have been selected based on
[OII]$\lambda\lambda3726,3729$ emission in the VISTA $NB118$ data
\citep{2013A&A...560A..94M} using previous COSMOS photometric redshift,
and one of them through Ly-$\alpha$ emission from the sample of
\cite{2009A&A...498...13N}.  We find $\sigma=0.061$ and $\eta=7.14$\%.

  \paragraph{VUDS at $0.1<z<4$ \citep{2015AA...576A..79L}.} The VIMOS
  Ultra-Deep Survey targeted $z>2.4$ galaxies using color-color and
  photometric redshift selections. The VUDS sample includes extremely
  faint galaxies with a median magnitude of $i_{\rm AB}^+ \sim 24.6$
  (3$\sigma$, 3$\arcsec$) with a total exposure times of 40 hr per
  spectra. This sample contains a larger number of catastrophic
  failures, mostly because of the misidentification between the Lyman and Balmer
  break features.  This is because some of the objects do not have
  associated NIR data. Such data are extremely important at $z>1.5$.  We find
  $\sigma=0.028$ and $\eta=13.13$\%.

\begin{table}
\begin{center}
  \caption{Performance of the catalog as a function of magnitude and
    galaxy types for galaxies detected both in COSMOS2015 and
    \cite{Ilbert:2013dq} compared to spectroscopic samples. In almost all
    cases, photometric redshift precision ($\sigma_{13}$ and $\sigma_{15}$) increases and the number of
    catastrophic failures ( $\eta_{\rm 13}$ and  $\eta_{\rm 15}$) fall compared to \cite{Ilbert:2013dq} \textit{for a selection common to both
      catalogs. \label{acc2}}}
\begin{tabular}{ c | c c c c | cccc} \hline
 & \multicolumn{4}{c}{ Star-forming galaxies} & \multicolumn{4}{c}{ Quiescent galaxies} \\
     $i^{+}$  &   $\sigma_{15}$ &  $\eta_{\rm 15}$ &   $\sigma_{13}$ &  $\eta_{\rm 13}$ & $\sigma_{15}$ &  $\eta_{\rm 15}$ &   $\sigma_{13}$ &  $\eta_{\rm 13}$  \\
                         &       &    (\%)   & & (\%)   &       &    (\%) & & (\%)  \\ \hline
$\lbrack$16,21$\rbrack$ & 0.007 & 0.5 & 0.008&0.5 &0.005 &0.0 & 0.005& 0.0\\
$\lbrack$21,22$\rbrack$ & 0.008 & 0.6 &0.008 &0.6 &0.007 &0.3 &0.006&0.4\\
$\lbrack$22,23$\rbrack$ & 0.01 & 1.7 &0.01 &1.9 &0.01 &0.6 &0.011&0.6\\
$\lbrack$23,24$\rbrack$ & 0.022 & 6.7 &0.022 &7.2 &0.027 &6.0 &0.030&4.4\\
$\lbrack$24,25$\rbrack$ & 0.034 & 10.2 &0.037 &15.0 &0.054 &18.9 &0.062&16.7\\
$\lbrack$25,26$\rbrack$ & 0.057 & 22.0 &0.058 &24.2 & & &&\\
  \hline
\end{tabular}

\end{center}
\end{table}

\begin{table*}
\begin{center}
  \caption{Characteristics of the spectroscopic redshift samples and
    photometric redshift accuracy for the objects in a clean
    (non-flagged) regions. Only the most secure spectroscopic redshifts
    are considered (those with a flag between 3 and 4). The redshift
    range, median redshift
    and apparent magnitude in $i^{+}$-band are provided for each
    selected sample. \label{acc}}
\begin{tabular}{l c c c c c c c c c} \hline
    spectroscopic survey /reference  & Instrument/ &          Nb  & $z_{\rm med}$ & $z_{\rm range}$ &  $i^{+}_{\rm
                                                              med}$  &   $\sigma_{\Delta z/(1+z)}$ &  $\eta$  \\
                 &   Telescope    & spec-z &           &           &            &       &    (\%)  \vspace{0.2cm} \\ \hline

    zCOSMOS-bright \citep{2007ApJS..172...70L}  & VIMOS/VLT  & 8608  &  0.48  & [0.02,1.19]  &   21.6    &      0.007         &  0.51   \\ 
    \cite{2015AA...575A..40C}    &  FORS2/VLT   & 788   &  0.89  & [0.07,3.65]  &   22.6 & 0.009 & 2.03 \\ 
    Capak et al. in prep, \cite{2010ApJ...721...98K} & DEIMOS/Keck~II&~2022~&~0.93~&[0.02,5.87] &23.2  &   0.014  &  7.96    \\ 
 \cite{2012MNRAS.426.1782R} & FMOS/Subaru &26 & 1.21 & [0.82,1.50]& 22.5   &  0.009 & 7.69 \\ 
    \cite{2012ApJ...755...26O}   &   MOIRCS/Subaru &  10  &  1.41  & [1.24,2.09]  &   23.9    & 0.017  & 0.00  \\ 
FMOS-COSMOS  \citep{2015Silverman}   &       FMOS/Subaru   & 178  &  1.56  &  [1.34,1.73] &23.5  & 0.022 & 1.12    \\ 

WFC3-grism \citep{2014ApJ...797...17K}  &       WFC3/HST   & 11 &  2.03  &  [1.88,2.54] &25.1  & 0.069 & 0.00    \\ 

zCOSMOS-faint   (Lilly et al. in prep)    &    VIMOS/VLT     & 767& 2.11 & [1.50,2.50] & 23.8    &      0.032         &  7.95   \\ 
   MOSDEF \citep{2015Kriek} & MOSFIRE/Keck I & 80 & 2.15&[0.80,3.71] &24.2& 0.042 &10.0  \\
   Stockmann et al. in prep, \cite{phdthesisZabl15} & XSHOOTER/VLT & 14 & 2.19 &
                                                                 [1.98,2.48] & 22.2 & 0.061 & 7.14  \\
 VUDS     \citep{2015AA...576A..79L}     &   VIMOS/VLT  &  998 & 2.70&[0.10,4.93]&  24.6    &  0.028 &  13.13    \\                                                                 

\hline
\end{tabular}

\end{center}
\end{table*}

Note that the X-ray detected sources from XMM-COSMOS
\citep{2007ApJS..172...29H,2007ApJS..172..341C,2010ApJ...716..348B}
and Chandra COSMOS \citep{2009ApJS..184..158E,2012ApJS..201...30C} are
flagged and not used here. For those sources, the photometric redshift are computed with a
specific tuning and are presented in \cite{2011ApJ...742...61S}.

\subsection{Photometric redshift accuracy based on the Probability Distribution Function}
\label{results_pdf}

We also assess the photometric redshift accuracy using the
1$\sigma$ uncertainty derived from the photometric redshift
Probability Distribution Function (PDFz).  The advantage of this
method is that we can investigate the photometric redshift accuracy in
any redshift-magnitude range. However, it requires an accurate
estimate of the PDFz.

\begin{figure}
\begin{center}
\includegraphics[scale=0.44]{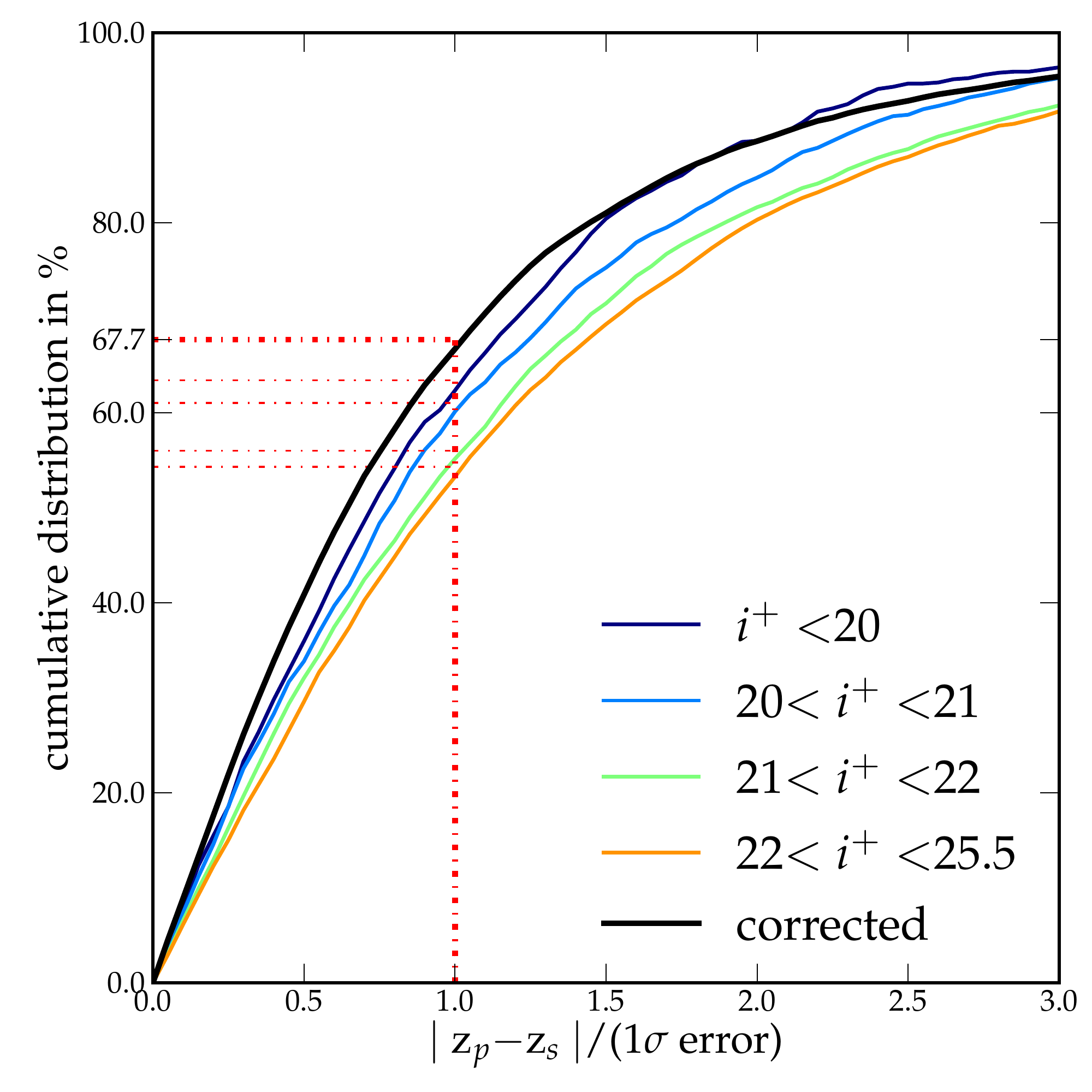}
\caption{Cumulative distribution of
  $\vert z_{\rm phot}-z_{\rm spec}\vert/1\sigma$. Of the
  spectroscopic redshifts, 58\%  have their photometric redshift within the
  1$\sigma$ error; this implies that photometric errors are slightly
  underestimated. This plot is made with the high-confidence
  spectroscopic redshift catalog.}
\label{Fig:PapCheckError}
\end{center}
\end{figure}

In Figure~\ref{Fig:PapCheckError}, we show the cumulative distribution
of the ratio $|z_p-z_s|/{{1\sigma}}$.The $1\sigma$ error
  given by {\sc LePhare} is defined as the value enclosing 68\% of the
  probability distribution function of the photometric
  redshift. Assuming that $z_{\rm s}$ is the true redshift, 68\% of the times it should
  fall within the $1\sigma$ error.  This comparison
  shows that the $1\sigma$-uncertainties enclose less than the 68\%
  of the expected value. This is confirmed when we split the spectroscopic
sample per magnitude and redshift bin. It appears that our errors on
photometric redshift are underestimated by a factor which depends on
the magnitude. We consequently chose to correct these errors by
applying the following magnitude-dependent correction: errors are
multiplied by a factor of 1.2 for bright objects ($i^{+}<20$) and by a
factor of $(0.1\times i^{+}-0.8)$ for faint objects ($i^{+}>20$).
This issue was already present in previous COSMOS photometric redshift
catalogs derived with {\sc LePhare}, and we have not been able to
determine why photometric redshift errors are underestimated; one
reason could be the lack of representativity of our set of templates,
while another reason could be that we do not include the intrinsic template
uncertainties. Another reason might be that the flux uncertainties in the
photometric catalog are still be underestimated. With this
magnitude-dependent correction, there is no consequence on the
computation of the physical parameters. However, the PDFz remains
systematically too peaky around the median values.

\begin{figure}
\begin{center}
\includegraphics[scale=0.44]{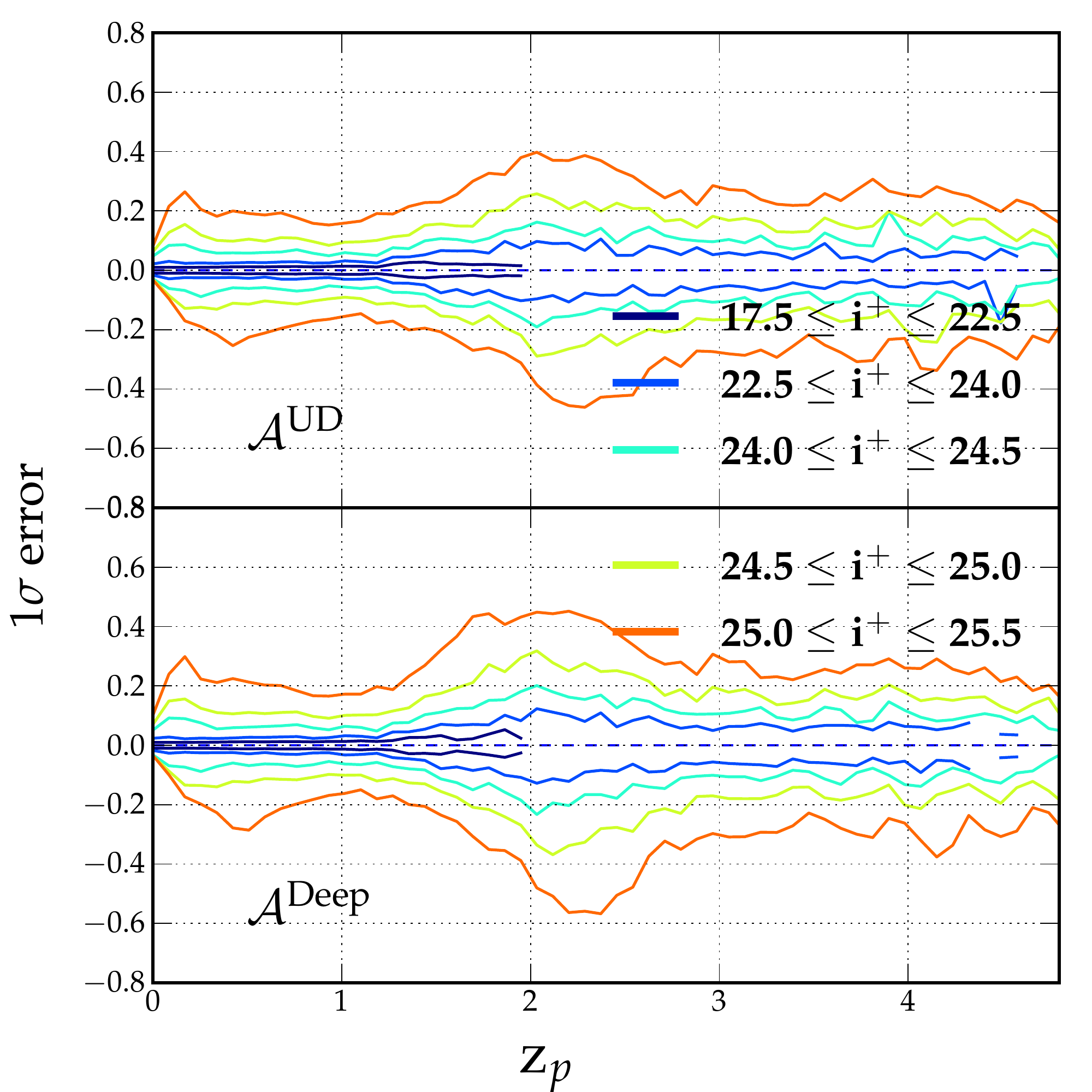}
\caption{Bottom and top panels: 1$\sigma$ photometric redshift error
  as a function of redshift for different magnitude bins on
  ${\cal A}^{\rm Deep}$ and on ${\cal A}^{\rm UD}$. }
\label{Fig:Paperror}
\end{center}
\end{figure}

Figure~\ref{Fig:Paperror} shows the 1$\sigma$ negative and positive
uncertainties as a function of redshift for different bins of apparent
magnitude. The magnitude-dependent correction described above has been
applied to this plot. Several clear conclusions emerge: first, the
photometric precision is lower for galaxies with fainter apparent
magnitudes at all redshifts; second, the photometric redshifts have
significantly lower uncertainties at $z\lesssim 1.4$. This is easy to
understand because here the Balmer break is redshifted within the
wavelength range covered by medium bands.  At
  $1.4\lesssim z \lesssim 2.5$, the redshift uncertainty increases by
  a factor of two. Such a trend is to be expected: the accuracy of the
  photometric redshift is mainly driven by an accurate knowledge of
  the Balmer break position.  Specifically, at $z>1.5$, the Balmer
  break moves outside the medium bands into the NIR range. Moreover,
  the absolute photometric precision is lower for a given
  signal-to-noise object in the near-infrared bands than in the
  optical. Additionally, the position of the Balmer break is less
  precisely determined using broad-band rather than medium-band
  photometry. This is reflected in the redshift uncertainty which
  rises at $z>1.5$.  For the same reason, we observe a difference in the
  redshift uncertainties which are lower in ${\cal A}^{\rm UD}$
  compared to ${\cal A}^{\rm Deep}$ regions, which is not the case at
  $z<1.4$: the photometric accuracy is higher in ${\cal A}^{\rm UD}$
  regions. At $z\sim2.5$, the Lyman-break enters into the optical
  bands and consequently the photometric redshift precision increases. In
  general, at bright magnitudes and lower redshifts, the dominant
  sources of error are probably related to photometric calibrations
  and spectral energy distribution (SED) fitting.
%Finally, the uncertainties are lower for a $K_{\rm s}$-selected sample

\subsection{Star/galaxy classification}
\label{Sec:sep}
We use {\sc LePhare} with both galaxy and stellar templates.  We
compare the best-fitting $\chi^{2}$ for the galaxy templates
$\chi^{2}_{\rm gal}$ and the ones derived for the stellar templates
$\chi^{2}_{\rm stars}$ to determine star-galaxy classification.
We flag as stars all objects for which
\begin{math}
\chi^{2}_{\rm gal}-\chi^{2}_{\rm stars}>0
\end{math}
but only if the object is detected in NIR or IRAC ($m_{3.6\mu m}\leq
25.5$ or $K_{\rm s}<24.7$) and is not too far from the $BzK$ stellar
sequence ($z^{++}-K_{\rm s}<(B-z^{++})*0.3-0.2$).

\begin{figure}
\begin{center}
\includegraphics[scale=0.45]{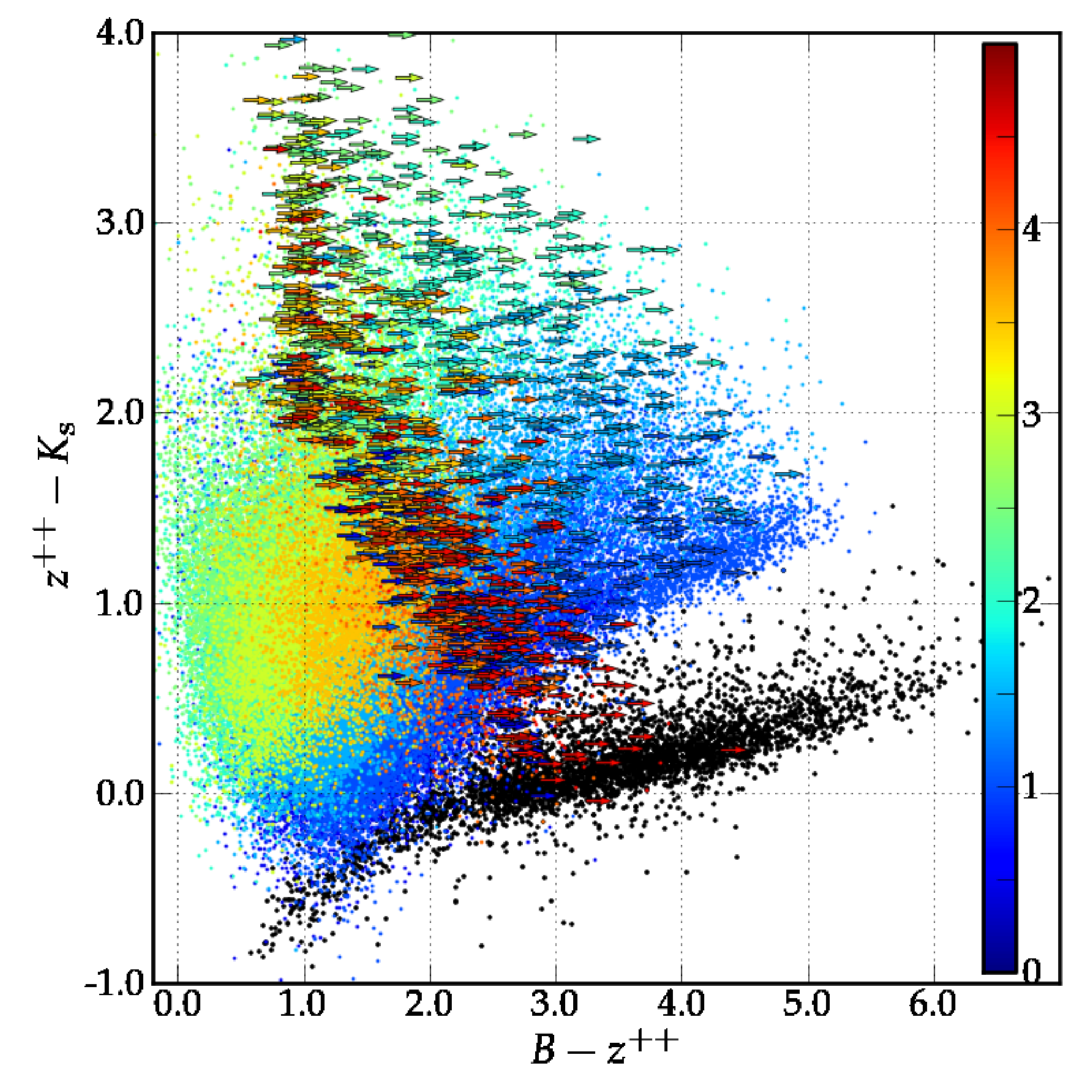}
\caption{Color-color diagram $z^{++}-K_{\rm s}$/$B-K_{\rm s}$ for all
  sources. Sources classified as stars are represented in
  black. Colors represent redshift. Right-pointing arrows are
  the upper limits in the $B$ band.}
\label{Fig:Papcolcol}
\end{center}
\end{figure}

Figure~\ref{Fig:Papcolcol} shows a $BzK$ color-color diagram for all
of the sources including stars and galaxies. Symbols are colored according
to their photometric redshifts. As expected, $B$-drop-outs occur
predominately at $z>4$, and galaxies with bluer $z^{++}-K$ color are
at lower redshifts. Stars selected using the above classification are
shown in black. In the
${\cal A}^{\rm UVISTA}$ region, 24,074 objects are classified as stars.  A cross-match with the ACS stellar
catalog \cite{2007ApJS..172..219L} shows that 77\% of the stars with
$i^{+}<24$ from ACS are classified as stars with this method. However,
15\% are misclassified as galaxies but are in masked areas. Finally,
0.6\% of the extended sources are misclassified as stars.

\subsection{Absolute magnitudes and stellar masses}

An estimate of the $k$-correction term \citep{1968ApJ...154...21O}
relies on the best-fitting template. This component is one of the main
sources of systematic error in the absolute magnitude and rest-frame
color estimate. To estimate these quantities, we follow the method
outlined in Appendix A of \cite{2005Ilbert}. In order to minimize the
\textit{k}-correction-induced uncertainties, the rest-frame luminosity at a
given wavelength $\lambda$ is derived from the apparent magnitude
$m_{obs}$ observed at the nearest filter to $\lambda (1 + z)$. Using
this procedure, the absolute magnitudes are less dependent on the
best-fit SED, but are more
dependent on any observational problem affecting $m_{\rm obs}$. Therefore,
we constrain the code to consider only the broad bands for $m_{\rm obs}$
and those bands with a systematic offset lower than 0.1 mag derived for
the photometric redshift.

We derive the stellar mass using {\sc LePhare} following exactly the
same method as in \cite{2015Ilbert}.  We derive the galaxy
stellar masses using a library of synthetic spectra generated using
the Stellar Population Synthesis (SPS) model of
\cite{Bruzual:2003p963}. We assume a \cite{2003PASP..115..763C}
Initial Mass Function (IMF). We combine the exponentially declining SFH
and delayed SFH ($\tau^{-2} t e^{-t/\tau}$). Two metallicities (solar
and half-solar) are considered. Emission lines are added following
\cite{2009ApJ...690.1236I}. We include two attenuation curves:
the
starburst curve of \cite{Calzetti:2000p6839} and a curve with a slope
$\lambda^{−0.9}$ \citep[Appendix A of
][]{2013A&A...558A..67A}. The $E(B-V)$ values are allowed to take values
as high as 0.7. We assign the mass using the median of the
marginalized probability distribution function (PDF). 
Given the
uncertainties on the SFR based on template
fitting \citep{2015Ilbert,2015ApJ...801...80L} we do not
include the SFRs estimated from template fitting in our distributed
catalogs.

\section{Characteristics of the global sample}
\label{sec:char-glob-sample}

\subsection{Galaxy classification}

\begin{figure}
\begin{center}
\includegraphics[scale=0.43]{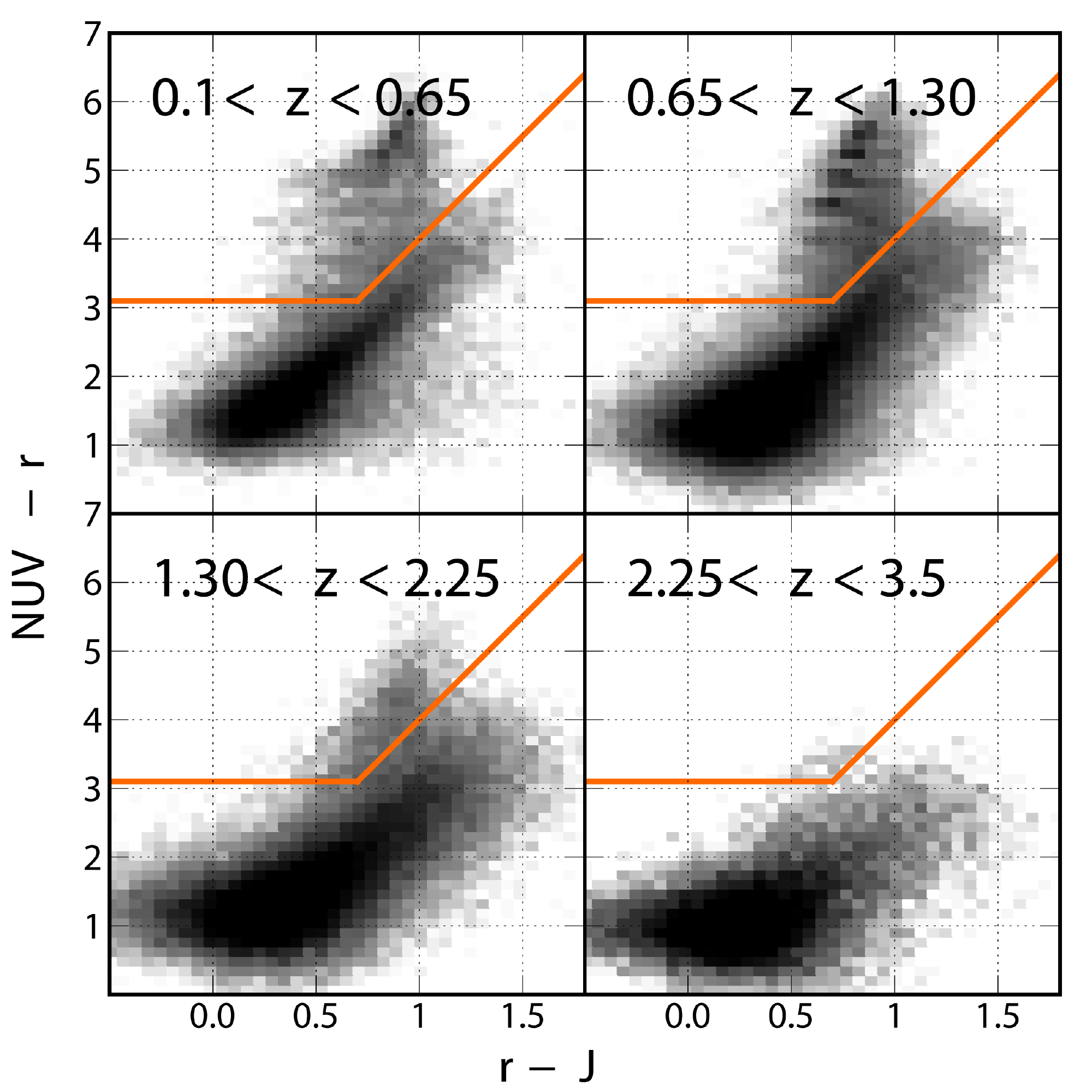}
\caption{ NUV$-r$/$r-J$ galaxy distributions.  Quiescent galaxies lie
  in the top-left corner. The objects fainter than limiting magnitudes
  are not used.}
\label{Fig:classification}
\end{center}
\end{figure}

Quiescent galaxies can be identified using the locations of galaxies in
the color-color plane NUV-$r$/$r$-$J$
\citep{Williams:2009p10339}. Quiescent objects are those with
$M_{NUV}-M_{r}>3(M_{r}-M_{J})+1$ and $M_{NUV}-M_{r}>3.1$.  This
technique is described in more detail in \cite{Ilbert:2013dq}; in
particular, this technique avoids mixing the red dusty galaxies and
quiescent galaxies. In our catalog, galaxies with a flag of 0 are
quiescent galaxies and the others are star-forming galaxies. The
redshift-dependent evolution of this distribution is presented in
figure~\ref{Fig:classification}. The rapid build-up of quiescent
galaxies at low redshift inside the box is evident, as is the
relative decrease in bright, star-forming galaxies outside the box. 

\subsection{Stellar mass Completeness}

We empirically estimate the stellar mass completeness
\citep{2010Pozetti,Ilbert:2013dq,2013Moustakas,2013Davidzon}. We first
determine the magnitude limit $K_{\rm s\,lim}$. For each galaxy, we
then determine the mass it would need to have to be observed, at that
redshift, at the magnitude limit:
\begin{equation}
\log M_{\rm lim}=\log M -0.4(K_{\rm s\,lim}-K_{\rm s}) . 
\end{equation} 

\begin{figure*}
\begin{center}
\includegraphics[scale=0.42]{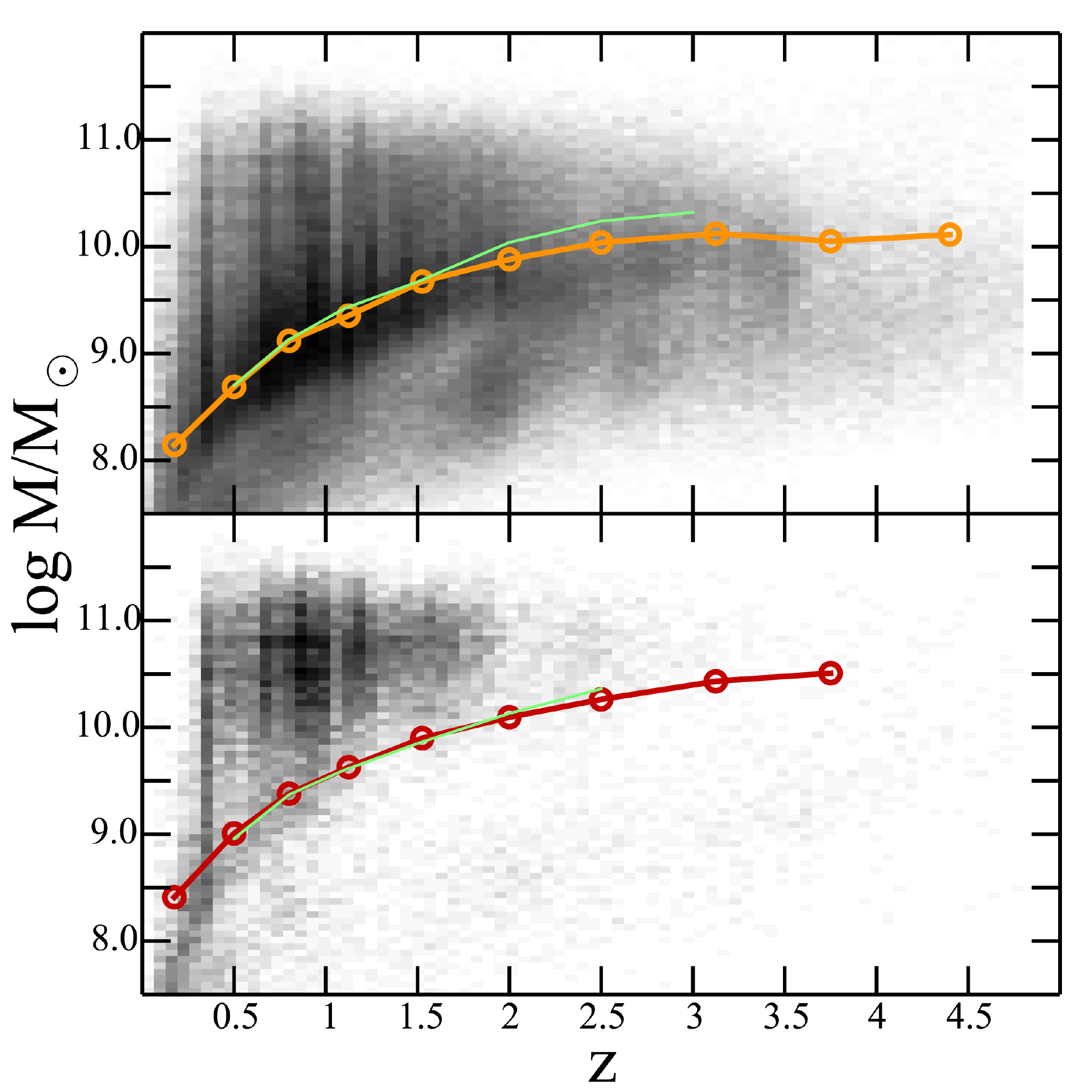}
\includegraphics[scale=0.42]{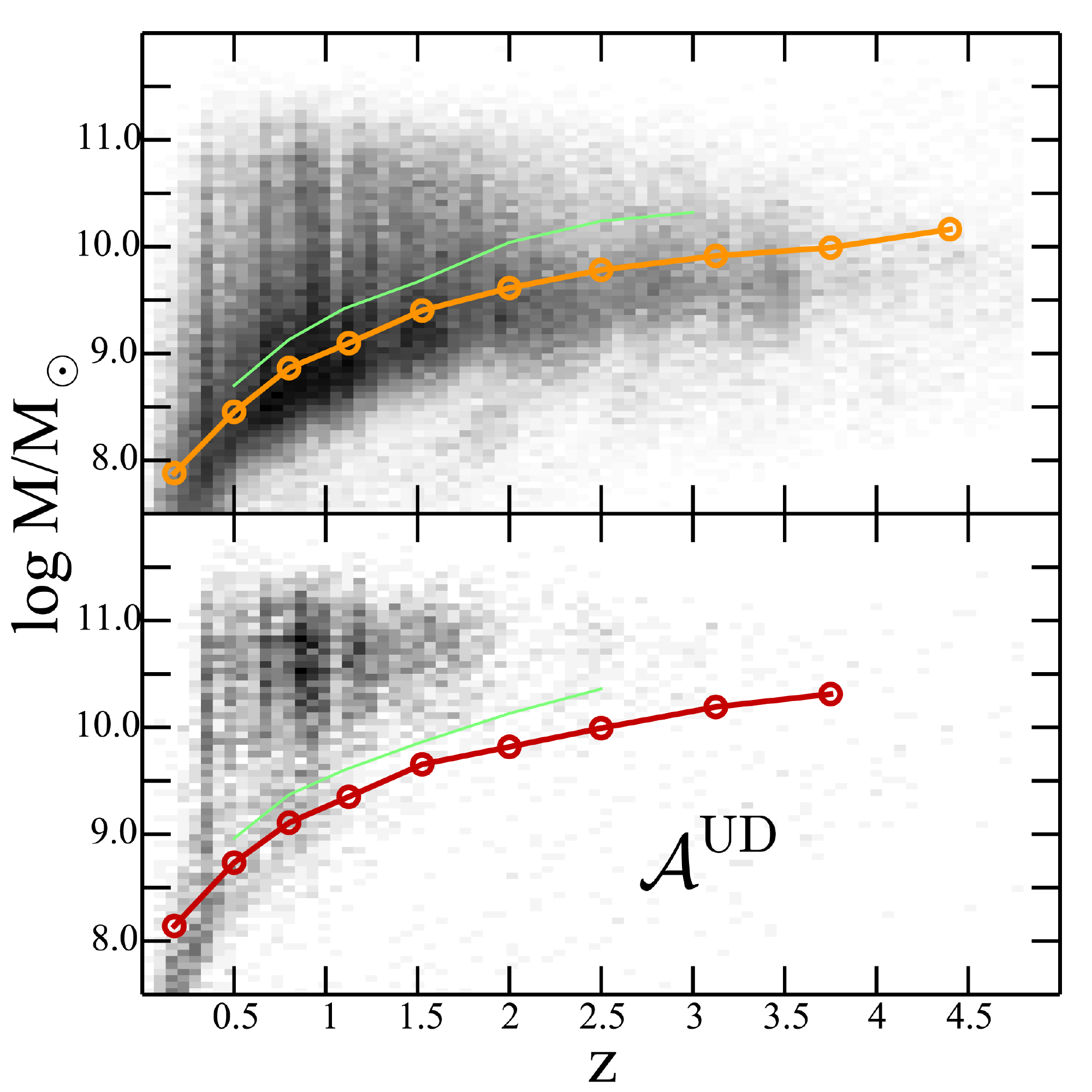}
\caption{Stellar mass-redshift histogram (the grayscale
  corresponds to the the number of galaxies in each cell) on
  ${\cal A}^{\rm Deep}$ (left) and
  ${\cal A}^{\rm UD}$ (right) regions for the full
  catalog (top) and for the quiescent sample (bottom). Orange lines
  shows the mass limit for the full catalog and red lines for the
  quiescent ones. These values are reported in
  Table~\ref{Tab:Mlim}. The solid green line is the mass limit in
  COSMOS as given in \cite{Ilbert:2013dq}.}
\label{Fig:Mlim}
\end{center}
\end{figure*}

Next, in each redshift bin, we independently estimate the stellar mass completeness
$M_{\rm lim}$ within which 90\% of the galaxies lie. We estimate
independently the mass limits on ${\cal A}^{\rm Deep}$ and
${\cal A}^{\rm UD}$. We compute these mass limits using the 3$\sigma$
limiting magnitude, which is 24.0 for ${\cal A}^{\rm Deep}$ and 24.7
for ${\cal A}^{\rm UD}$. These mass limits are given in
Table~\ref{Tab:Mlim} and are shown in Figure~\ref{Fig:Mlim}. In
  ${\cal A}^{\rm UD}$, the mass limits reach a factor of two lower in
  mass compared \cite{Ilbert:2013dq}. 
As expected, the mass limit
    is lower in  ${\cal A}^{\rm UD}$ compared to ${\cal A}^{\rm Deep}$,
    because  ${\cal A}^{\rm UD}$ reaches 0.7 magnitudes fainter in the $K_{\rm s}$ band. This
  estimate is robust to $z\sim4$ because the observed $K_{\rm s}$
  magnitude correlates well with stellar mass in this redshift
  range. However, these estimates should be treated cautiously at
  $z > 4$. Above this redshift, the rest-frame $K_{\rm s}$ band lies
  below the Balmer break and the $K_{\rm s}$ flux does not correspond
  precisely to the stellar mass. It is then better traced by mid-IR
  bands. We will estimate the mass limit at high redshift for an
  IRAC-selected sample in a future work (Davidzon et al. in preparation).

%%%%%%%%%%%%%%%%%%%%%%%%%
%% hjmcc stops here now %
%%%%%%%%%%%%%%%%%%%%%%%%%

\begin{table*}
\begin{center}
  \caption{Mass limits of COSMOS2015 for the full, quiescent and the
    star-forming samples in the ${\cal A}^{\rm Deep}$ and
    ${\cal A}^{\rm UD}$ regions. $N_{\rm gal}^{\rm
      full}$, $N_{\rm gal}^{\rm quies}$, $N_{\rm gal}^{\rm SF}$ are
    the percentage of galaxies in each redshift bins for the full,
    quiescent, and star forming populations. $M_{\rm lim}^{\rm full}$,
    $M_{\rm lim}^{\rm quies}$, $M_{\rm lim}^{\rm SF}$ are the
    logarithm of the  limiting mass in units of solar masses.
  }
\begin{tabular}{c|cccccc|cccccc}
\hline
        & \multicolumn{6}{c}{ ${\cal A}^{\rm Deep}$ } & \multicolumn{6}{c}{ ${\cal A}^{\rm UD}$ } \\
bin   & $N_{\rm gal}^{\rm full}$& $M_{\rm lim}^{\rm full}$   &$N_{\rm gal}^{\rm quies}$
      & $ M_{\rm lim}^{\rm quies}$ &$N_{\rm gal}^{\rm SF}$
      & $ M_{\rm lim}^{\rm SF}$ & $N_{\rm gal}^{\rm full}$
      & $ M_{\rm lim}^{\rm full}$  &$N_{\rm gal}^{\rm quies}$
      & $ M_{\rm lim}^{\rm quies}$&$N_{\rm gal}^{\rm SF}$
      & $ M_{\rm lim}^{\rm SF}$ \\
  \hline
0.00$< z <$0.35   &8.6 & 8.1 & 11.3 & 8.4 & 8.3 & 8.1 & 9.0  & 7.9 & 13.8   & 8.1& 8.7 & 7.8 \\
  0.35$< z <$0.65   & 14.1 & 8.7 & 18.4  & 9.0  & 13.7 & 8.6 & 13.5  & 8.4 &  19.2  & 8.7& 13.0& 8.4 \\
0.65$< z <$0.95  & 17.4 & 9.1 & 27.4  &  9.4  & 16.7 & 9.0 &  17.5   & 8.9 & 27.4  & 9.1 & 16.7 & 8.7\\
0.95$< z <$1.30  & 16.4 & 9.3  & 20.5  & 9.6 & 16.1 & 9.2 & 16.3  & 9.1 &18.9  & 9.3 & 16.1& 9.0\\
1.30$< z <$1.75   & 14.2 & 9.7 & 12.5  & 9.9  & 14.4 & 9.6 & 14.9  & 9.4 & 11.8  & 9.6 & 15.2& 9.3 \\
1.75$< z <$2.25 & 12.0 & 9.9 & 4.9  &  10.1  & 12.5 & 9.8 & 11.0  & 9.6 & 4.0  & 9.8 & 11.5& 9.6 \\
2.25$< z <$2.75   & 6.8  & 10.0 & 2.4 &  10.3  & 7.1 & 10.0 & 6.5  & 9.8 & 2.2  & 10.0 & 6.8 & 9.8\\
2.75$< z <$3.50  & 6.4 & 10.1 & 1.4 & 10.4  & 6.8 & 10.1 & 7.1  & 9.9 & 1.5  & 10.2  & 7.5  & 9.9 \\
3.50$< z <$4.00  & 1.9 & 10.1 & 0.5 &  10.5 & 2.0 & 10.5 &  2.0  & 10.0 & 0.4  & 10.3 & 2.1 & 10.0 \\
4.00$< z <$4.80  & 1.4  & 10.1 & -&  - & 1.5 & 10.8 &  1.5  & 10.2 & - & - & 1.6 & 10.1\\
     \hline
\end{tabular}

\label{Tab:Mlim}
\end{center}
\end{table*}

\subsection{Galaxy clustering measurements}
\label{sec:galaxy-clust-meas}

We estimate the projected galaxy clustering in our sample by computing
the angular two-point auto-correlation function $w(\theta)$. The
angular correlation function $w(\theta)$ measures the excess
probability of finding two objects separated by an angle $\theta$
compared to a random distribution in a series of angular bins. This
measurement is an excellent test of the uniformity of our photometric
catalog as $w$ is very sensitive to large-scale photometric
systematic errors. Adding cuts in stellar mass and photometric redshift
allows for an independent check of our photometric redshift procedures. We
$w$ to compute this using
ATHENA\footnote{\url{www.cosmostat.org/software/athena/}}, which uses
the usual \cite{Landy:1993p11082} estimator:

\begin{equation}
w(\theta)=\frac{1}{R R}\times\left(\frac{N_{r}(N_{r}-1)}{N_{d}(N_{d}-1)}DD-2\frac{N_{r}}{N_{d}}DR+RR\right)
\end{equation}

Where $N_{r}$ and $N_{d}$ are the number of points in the random and
the galaxy sample, and $RR$, $RD$ and $DD$ are the number of pairs in
the random catalog, between the random and galaxy catalog, and in the
galaxy catalog. Our random catalog contains 500,000 objects.  Our
measurements are corrected for the ``integral constraint''
\citep{Growth1977}, a systematic effect arising from using a
clustered sample to estimate the mean background density in a finite
area. 

\begin{figure*}
\begin{center}
\includegraphics[scale=0.58]{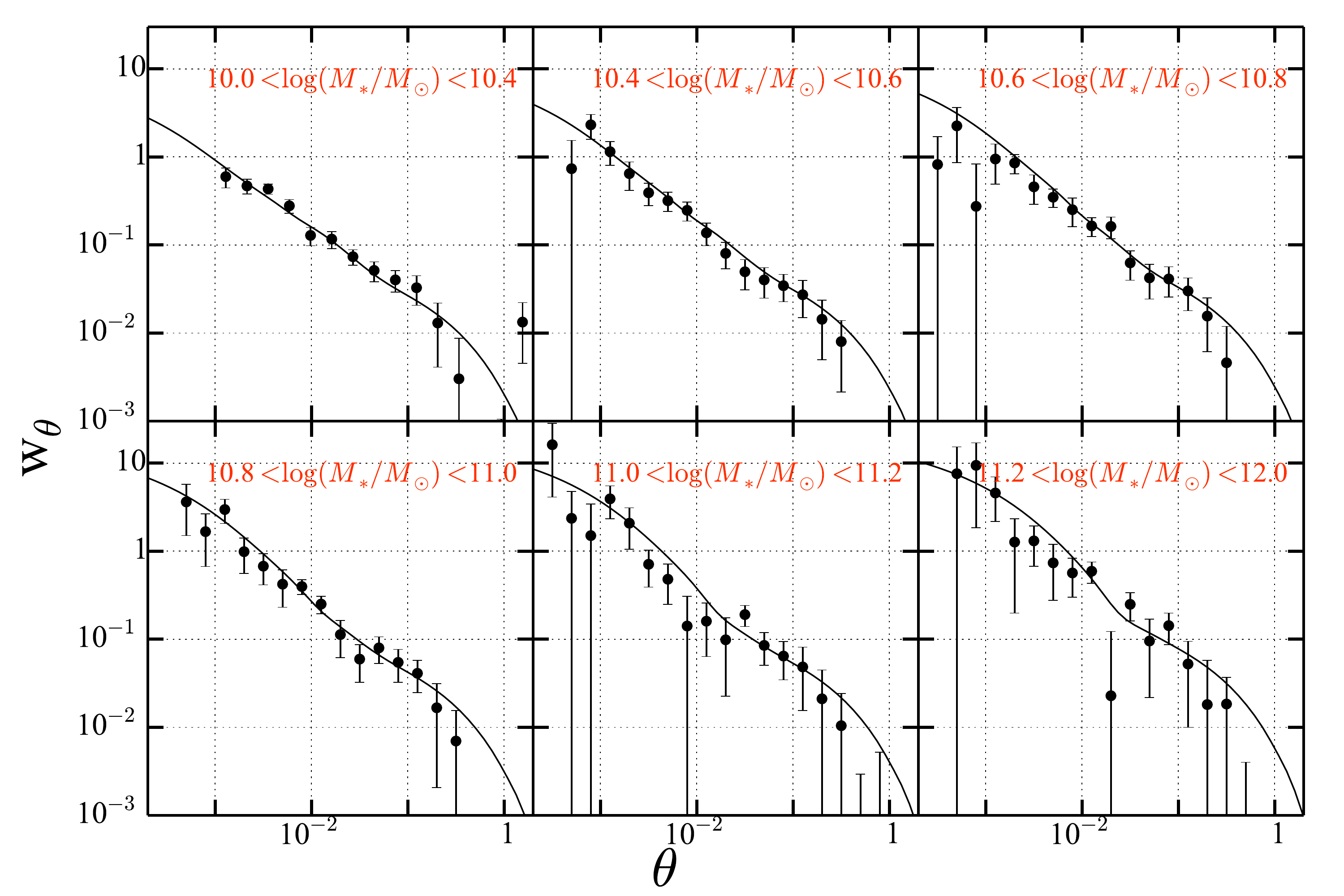}
\caption{Angular correlation function $w$ as a function of angular
  scale $\theta$ in degrees at $0.5<z<1$ for several bins in stellar
  mass. The solid line shows the analytic predictions of
  \cite{2015MNRASCoupon}.}
\label{Fig:clust}
\end{center}
\end{figure*}

Figure~\ref{Fig:clust} shows $w$ in $0.5<z<1$ in six mass bins
compared to the best-fitting occupation distribution (HOD) model
derived by \cite{2015MNRASCoupon} in the MIRACLES/CFHTLS field.
Our measurements are in excellent agreement with the
  predictions of \citeauthor{2015MNRASCoupon}'s best-fitting HOD
  model, computed from a larger 25deg~$^2$ field. This suggests
that, at these redshift ranges and masses, cosmic variance is not an
important issue in the COSMOS field. Only at high stellar masses and
small scales is there a systematic offset from the models, which may
indicate the limitations of the halo model in this mass regime. 

Finally, we note that, in contrast to this result, some works
  have noted that there is a clear excess in the number of galaxies in
  COSMOS compared to other fields \citep[see, e.g, Figure 33
  in][]{2014MNRAS.441.2891M} due to the presence of large structures
  at $z\sim1$ and below, which could influence our correlation
  function measurements \citep{2007ApJS..172..314M}. The measurements
  presented above cover quite a large redshift range and consequently
  probe a lare volume, and are therefore less susceptible to the
  effects of cosmic variance. In smaller redshift slices and at higher
  redshifts, the effect of cosmic variance becomes more pronounced,
  especially when these redshift ranges overlap with several of the large
  structures known to exist in the COSMOS field, for example, at
  $1<z<1.3$ \citep[see also the discussion in ][]{2015McCracken}.

\section{Conclusion}

\label{Sec:conclusion}

Using the unique combination of deep multi-wavelength data and
spectroscopic redshifts on the COSMOS field, we have computed a new
catalog containing precise photometric redshifts and 30-band
photometry. COSMOS2015 contains more than half a million secure
objects over two square degrees. Including new $YJHK_{\rm s}$ images
from the UltraVISTA-DR2 survey, $Y$-band images from Hyper Suprime-Cam and
IR data from the SPLASH \textit{Spitzer} legacy program, this NIR-selected catalog is highly optimized for the study of galaxy evolution
and environment in the early Universe. To maximize catalog
completeness to the highest redshifts, objects have been detected and
selected using an ultra-deep $\chi^{2}$ sum of the $YJHK_{\rm s}$ and
$z^{++}$ images.

The main improvements of the catalog compared with previous versions
are as follows.

\begin{itemize}
\item A great number of sources thanks to the combination of deeper data
  (UltraVISTA-DR2) and an improved extraction image. This image now
  contains the bluer $z^{++}$ band in addition to the redder NIR bands.
  There are now $\sim 6\times 10^5$ objects in the 1.5 deg$^{2}$
  UltraVISTA-DR2 area and $\sim 1.5\times 10^5$ in the ``ultra-deep
  stripes'' sub-region at the limiting magnitude in $K_{\rm s}$. This
  represents more than twice as many objects per square degree
  compared to \cite{Ilbert:2013dq}.
\item More precise photometric redshifts.  Based on comparisons with the unique
  spectroscopic redshift sample in the COSMOS field, we measure
  $\sigma_{\Delta z/(1+z_s)}$ = 0.021 for $3<z<6$ with 13.2\% of
  outliers. At lower
  redshifts, the precision is better than 0.01, only a few percent of
  catastrophic failures. The precision at low redshifts is consistent
  with \cite{Ilbert:2013dq}, while it improves significantly at high
  redshift.
\item The characteristic mass limits are much lower.  The deepest regions
  reach a completeness limit of 10$^{10}M_\odot$ to $z=4$, which
  is more than 0.3 dex better compared to \cite{Ilbert:2013dq} for the
  full sample.
\end{itemize}

Detailed comparisons of the color distributions, number counts and
clustering show good agreement with the literature in the mass
ranges where these previous studies overlap with ours. In particular,
our mass-selected clustering measurements at $0.5<z<1$ are in
excellent agreement with \citeauthor{2015MNRASCoupon}'s halo model
calibrated using $25~\deg^2$ of the CFHTLS. 

The COSMOS2015 catalog represents an invaluable resource which can be
used to
investigate the evolution of galaxies and structures back to the
earliest stages of the Universe. Sampling the galaxy
population out to $z\sim 4$ at degree scales it will allow us to study
the connection between galaxies, their host dark matter haloes, and
their large-scale environment, back to the earliest epochs of cosmic
time. 

\acknowledgments

C.L. is supported by the ILP LABEX (under reference ANR-10-LABX-63 and
ANR-11-IDEX-0004-02). This work is partially supported by the Spin(e)
grants ANR-13-BS05-0005 of the French Agence Nationale de la
Recherche. H.J.M.C.C. acknowledges financial support form the ``Programme
national cosmologie et galaxies'' (PNCG). O.I. acknowledges the funding
of the French Agence Nationale de la Recherche for the project
``SAGACE''. J.D.S. is supported by JSPS KAKENHI Grant Number 26400221,
the World Premier International Research Center Initiative (WPI),
MEXT, Japan and by CREST, JST. B.M.J. and J.P.U.F. acknowledge support from
the ERC-StG grant EGGS-278202. The Dark Cosmology Centre is funded by
the DNRF. S.T. and M.S. acknowledge support from the ERC Consolidator
Grant funding scheme (project ConTExt, grant number No. 648179). This
research is also partly supported by the Centre National d'Etudes
Spatiales (CNES). This work is based on data products from
observations made with ESO Telescopes at the La Silla Paranal
Observatory under ESO programme ID 179.A-2005 and on data products
produced by TERAPIX and the Cambridge Astronomy Survey Unit on behalf
of the UltraVISTA consortium. VS acknowledges the European Union's Seventh Framework pro-
gramme under grant agreement No. 337595.

\section{Appendix}

\subsection{Catalog description}
\label{Sec:Appendix1}

The details of the regions flagged in the catalog are presented in
Table~\ref{Tab:coordinates}. We perform COSMOS2015 quality checks only
in the inner part of the field covered by UltraVISTA-DR2. On the part
of the field not covered by UltraVISTA, the source extraction is
performed only on the $z^{++}$-band data and using the same
parameters. This part of the field has a higher fraction of spurious
sources and must be exploited carefully, particularly when selecting
a mass-selected sample. The area referred as ${\cal A}^{\rm Deep}$ above is
the region covered by ${\cal A}^{\rm UVISTA}$ not containing
${\cal A}^{\rm UD}$.

\begin{table*}
\begin{center}
\caption{Names, effective area, number of objects and coordinates of
  the regions flagged in the catalog and plotted in Figure~\ref{Fig:RegCos}. The region files are distributed with the catalog.}
\begin{threeparttable}
\begin{tabular}{llll}
\hline
 \textbf{Name} &
  \textbf{Area}   &\textbf{Nbr of} & \textbf{Coordinates
    or description} \\
    &
  \textbf{deg$^{2}$} &\textbf{objects} &  \\\hline
  & & & \\
  ${\cal A}^{\rm COSMOS}$ & 2  & 773118 & poly(148.70,0.79,151.53,3.620)\\
 ${\cal A}^{\rm!OPT}$ \&  ${\cal A}^{\rm COSMOS}$      &   1.77    & 694478 & not flagged regions in the
 optical bands inside the COSMOS
  2deg$^{2}$ field \\
 ${\cal A}^{\rm UD}$
                           & 0.62 & 247203 &poly(150.58,2.71,150.42,2.72,150.41,2.43,150.42,
  \\
  && &1.871,150.50,1.88,150.49,1.99,150.59,1.99) \\
 & & &poly(150.21,2.71,150.05,2.71,150.06,1.71,150.22,1.70,150.2,2.66) \\
 & && poly(149.84,2.71,149.68,2.71,149.68,1.71,149.85,1.70,149.84,2.66) \\
 & &&
  poly(149.48,2.71,149.33,2.72,149.32,1.71,149.49,1.71,149.48,2.66) \\
${\cal A}^{\rm UVISTA}$ &1.70&646939&
                                                     poly(150.77,2.81,149.31,2.81,149.32,1.61,150.41,1.61,150.41,1.66,\\
& & & 150.51,1.66,150.51,1.88,150.54,1.91,150.58,\\

& &&
      1.88,150.59,1.65,150.68,1.66,150.70,1.88,150.79,1.88)\\
 ${\cal A}^{\rm UD}$ \& ${\cal A}^{\rm COSMOS}$ & 0.53 &213716& Ultra-Deep area
                                                     inside the COSMOS
  2deg$^{2}$ field\\
${\cal A}^{\rm UD}$ \& ${\cal A}^{\rm !OPT}$ \& ${\cal A}^{\rm COSMOS} $& 0.46 &190650& Ultra-Deep area
                                                     inside the COSMOS 2deg$^{2}$ field, \\
& & & after removing
                                      flagged regions in the optical
                                                    bands\\

 ${\cal A}^{\rm UVISTA}$ \& ${\cal A}^{\rm COSMOS} $&1.58&604265& UVISTA area
                                                     inside the COSMOS
  2deg$^{2}$ field\\
 ${\cal A}^{\rm UVISTA}$ \& ${\cal A}^{\rm !OPT}$ \& ${\cal A}^{\rm COSMOS}$  &1.38& 536077 & UVISTA area inside the COSMOS 2deg$^{2}$ field, \\
 & & & after removing
                                      flagged regions in the optical
                                                    bands\\
                                                    & & & \\
%                    poly(150.01,2.76,149.89,2.76,149.89,1.71,150.01,1.71)
%  \\
%&UVISTA &&&
%  poly(150.76,2.76,150.62,2.76,150.61,1.73,150.67,1.73,150.76,1.89)
 % \\
%&&&&
%  poly(150.38,2.76,150.25,2.76,150.26,1.70,150.38,1.70) \\
% &&&&
%   poly(149.65,2.76,149.53,2.77,149.51,1.71,149.52,1.71,149.64,1.71,149.65,1.71)
%  \\
\hline
\end{tabular}
\label{Tab:coordinates}
\end{threeparttable}
\end{center}
\end{table*}

The parameters for the extraction of the photometry in dual mode with
{\sc SExtractor} are presented in Table~\ref{Tab:SEparam}.
%COSMOS2015 has been matched with 

Each column in the catalog is fully described by a README file
distributed with the catalog. We summarize the main content of our data
products in Table~\ref{Tab:summ}.
\subsection{From aperture magnitudes to total magnitudes}
\label{Sec:Appendix2}

Finally, we emphasize that to compute thr total magnitudes, one should
use 3$\arcsec$ diameter apertures, corrected for the photometric offsets
($o_{i}$, cf. Equation~\ref{off}) and systematic offsets ($s_{f}$,
cf. Table~\ref{Tab:sysoff}) according to the formula
\begin{equation}
{\rm MAG\_TOTAL}_{i,f}={\rm MAG\_APER3}_{i,f}+o_{i}-s_{f}
\end{equation}
where $i$ is the object identifier and $f$ the filter identifier.  A
similar procedure should be followed for the flux measurements.  Magnitudes
should also be corrected for foreground galactic extinction
using reddenning values $EBV$ given in the catalog and the extinction
factors ($F_{\rm f}$) mentioned in Table~\ref{Tab:sysoff} according
to
\begin{equation}
{\rm MAG\_TOTAL}_{i,f}={\rm MAG\_TOTAL}_{i,f}-EBV_{\rm i}*F_{\rm f}
\end{equation}
\begin{table*}
\begin{center}
  \caption{Summary of the main photometric and $z$-phot catalog
    columns. Matches with publicly available radio catalogs on COSMOS
    field are also provided. Please refer to the README file distributed with the
    catalog for more information.}
\begin{tabular}{ l l l}
\hline
%\textbf{Column Title} & \textbf{Unit} & \textbf{Description} \\\hline
\multicolumn{3}{c}{\textbf{General Parameters}}\\ 
%\hline 
 ID & - & identifiant \\
ALPHA\_2000, BETA\_2000& deg & Ra and Dec \\
X\_IMAGE, Y\_IMAGE & pix & pixel position \\
ERRX2\_IMAGE, ERRY2\_IMAGE, ERRXY\_IMAGE & pix & variances and
                                                 covariance on
                                                 positional
                                                  measurements\\
FLAGS\_\#$\left[{\rm Name\,of\,flag}\right]$ & - & flags as
                                                           described
                                                           in
                                                           Table~\ref{Tab:coordinates}\\
EBV& & Galactic extinction \citep{1998ApJ...500..525S} \\
%\hline 
\\
\multicolumn{3}{c}{ \textbf{optical and NIR photometry}}\\ 
%\hline 
 \#$\left[{\rm band}\right]$ \_FLUX\_APER2, \#$\left[{\rm band}\right]$\_FLUXERR\_APER2 & $\mu$Jy & flux and flux error measured
                              in a 2$\arcsecond$ aperture\\
 \#$\left[{\rm band}\right]$\_FLUX\_APER3, \#$\left[{\rm band}\right]$\_FLUXERR\_APER3 & $\mu$Jy & flux and flux error measured
                              in a 3$\arcsecond$ aperture\\
 \#$\left[{\rm band}\right]$\_MAG\_APER2, \#$\left[{\rm band}\right]$\_MAGERR\_APER2 & mag &  magnitude and magnitude error measured
                              in a 2$\arcsecond$ aperture\\
 \#$\left[{\rm band}\right]$\_MAG\_APER3, \#$\left[{\rm band}\right]$\_MAGERR\_APER3 & mag & magnitude and magnitude error measured
                              in a 3$\arcsecond$ aperture\\
\#$\left[{\rm band}\right]$\_MAG\_AUTO, \#$\left[{\rm band}\right]$\_MAGERR\_AUTO & mag & automatic aperture magnitude and magnitude
                                            error \\
\#$\left[{\rm band}\right]$\_MAG\_ISO, \#$\left[{\rm band}\right]$\_MAGERR\_ISO & mag & isophotal magnitude and
                                        magnitude error\\
\#$\left[{\rm band}\right]$\_FLAGS & & flags from {\sc SExtractor}\\
%\hline
\\
\multicolumn{3}{c}{\textbf{Match with the 24$\mu$m MIPS catalog \citep{2009ApJ...703..222L}}}\\ 
%\hline 
24\_FLUX, 24\_FLUXERR& $\mu$Jy & total flux and flux error\\
%\hline
\\
\multicolumn{3}{c}{\textbf{Match with the PACS/PEP catalog
  \citep{2011A&A...532A..90L}}}\\
%\hline 
100\_FLUX, 100\_FLUXERR& mJy & total 100$\mu$m flux and flux error\\
160\_FLUX, 160\_FLUXERR& mJy & total 160$\mu$m flux and flux error\\
\\
\multicolumn{3}{c}{\textbf{Match with the SPIRE/HerMES catalog
  \citep{2012MNRAS.424.1614O}}}\\
250\_FLUX, 250\_FLUXERR& mJy & total 250$\mu$m flux and flux error\\
350\_FLUX, 350\_FLUXERR& mJy & total 350$\mu$m flux and flux error\\
500\_FLUX, 500\_FLUXERR& mJy & total 500$\mu$m flux and flux error\\
%\hline 
\\
\multicolumn{3}{c}{\textbf{GALEX photometry \citep{2007ApJS..172...99C}}}\\
%\hline 
FLUX\_GALEX\_NUV, FLUXERR\_GALEX\_NUV & mJy & total flux and flux
                                                  error\\
FLUX\_GALEX\_FUV, FLUXERR\_GALEX\_FUV & mJy & total flux and flux error\\
 % \hline
\\
\multicolumn{3}{c}{\textbf{Match with the \textit{Chandra} COSMOS-Legacy survey
  \citep{2009ApJS..184..158E,2012ApJS..201...30C,2016Civano,2016Marchesi}}}\\ 
%\hline 
IDChandra & - & corresponding identifiant in the Chandra catalog\\
%\hline
\\
\multicolumn{3}{c}{\textbf{Match with ACS \citep{2007ApJS..172..219L}}}\\
 %\hline
814W\_FLUX, 814W\_FLUXERR &$\mu$Jy & flux and flux error for automatic
  aperture\\
%ACS\_STARS & - & \\
%ACS\_FWHM & & \\
%\hline 
\\
\multicolumn{3}{c}{\textbf{Match with previous multi-bands catalog}}\\
 %\hline
ID2006 & - & identifiant in the $1^{\rm st}$ version of the
             catalog from \cite{2007ApJS..172...99C}\\
ID2008 & - & identifiant in the $2^{\rm nd}$ version of the
             catalog from \cite{2007ApJS..172...99C}\\
ID2013 & - & corresponding identifiant in the catalog from \cite{Ilbert:2013dq}\\
 %\hline
\\
\multicolumn{3}{c}{\textbf{Parameters computed with \sc{LePhare}}}\\ 
%\hline 
total\_off & mag & weighted offset from MAG\_APER3 to total mag \\

type & - & 0 if galaxy, 1 if star, 2 if Xray source, -9 if failure in
         the fit \\
zPDF & - & median of the likelihood distribution \\
          zPDF\_l68, zPDF\_u68        &   -   &    lower and upper limits (68\% confidence level)\\
zMinChi2 & - & photo-z defines as the minimum of the $\chi^{2}$ distribution.\\
chi2best & - & reduced chi2 (-99 if less than 3 filters) for zMinChi2\\
zp\_2 & - &2$^{\rm nd}$ photo-z solution if a second peak is detected with P$>$5\% in the PDF\\
chi2\_2 & - & reduced chi2 for the second photo-z solution \\
NbFilt & - & number of filters used in the fit\\
zq, modq, chiq & - & z for the AGN library, best fit template and
                   associated reduced $\chi^{2}$\\
mods, chis & - & model for the star library and associated reduced $\chi^{2}$\\
model, age, extinction & - & best fit BC03 model at zPDF \\
M\_\#$\left[{\rm band}\right]$ & mag & absolute magnitudes in
                                   NUV,$u$,$B$,$r$,$i^{+}$,$z^{++}$,$Y$,$J$,$H$,$K_{\rm
                                   s}$\\
M\_NUV-M\_R& mag & color corrected from dust-extinction \\
 mass\_med & dex & log stellar mass from BC03 best-fit template (median)\\
mass\_med\_min68, mass\_med\_max68 & dex &  lower and upper limits (68\%
                                       confidence level)\\
mass\_best & dex & log stellar mass from BC03 best-fit template (minimum $\chi^{2}$)\\
L\_\#$\left[{\rm band}\right]$ & dex & luminosities in NUV,$r$,$K_{\rm s}$
  filters\\
\hline
  \end{tabular}
\label{Tab:summ}
\end{center}
\end{table*}

\begin{table}
\begin{center}
  \caption{{\sc SExtractor} parameters used for Dual-mode $\chi^{2}$
    detection and photometry.}
\begin{tabular}{ll}
\hline
 \textbf{Name} & \textbf{Value} \\\hline

ANALYS\_THRESH & 1.5 \\
FILTER\_NAME & gauss\_4.0\_7$\times$7.conv \\
CATALOG\_TYPE & FITS\_1.0 \\
DETECT\_TYPE & CCD \\
THRESH\_TYPE & ABSOLUTE \\
DETECT\_MINAREA & 10 \\
DETECT\_MAXAREA & 100 000\\
DETECT\_THRESH & 1.51 \\
FILTER & Y \\
DEBLEND\_NTHRESH & 32 \\
DEBLEND\_MINCONT & 0.00001 \\
CLEAN & Y \\
CLEAN\_PARAM & 1.0 \\
MASK\_TYPE & CORRECT \\
PHOT\_APERTURES & 13.33, 20.00, 47.33 \\
PHOT\_AUTOPARAMS & 2.5, 3.5\\
PHOT\_FLUXFRAC & 0.2, 0.5, 0.8\\
PHOT\_AUTOAPERS & 13.3, 13.3 \\
SATUR\_LEVEL & 30000. \\
MAG\_ZEROPOINT & depends on the band \\
GAIN & depends on the band \\
PIXEL\_SCALE & 4.16666$\times 10^{-5}$\\
BACK\_SIZE & 128 \\
BACK\_FILTERSIZE & 3 \\
BACKPHOTO\_TYPE & LOCAL \\
BACKPHOTO\_THICK & 30 \\
WEIGHT\_GAIN & N \\
RESCALE\_WEIGHTS & N  \\
WEIGHT\_TYPE & depends on the band \\
GAIN\_KEY & DUMMY \\
\hline
\end{tabular}

\label{Tab:SEparam}
\end{center}
\end{table}
%%%

\subsection{Effect of the seeing on the aperture magnitude}
\label{Ap:seeing}

\begin{figure*}
\begin{center}
\includegraphics[scale=0.43]{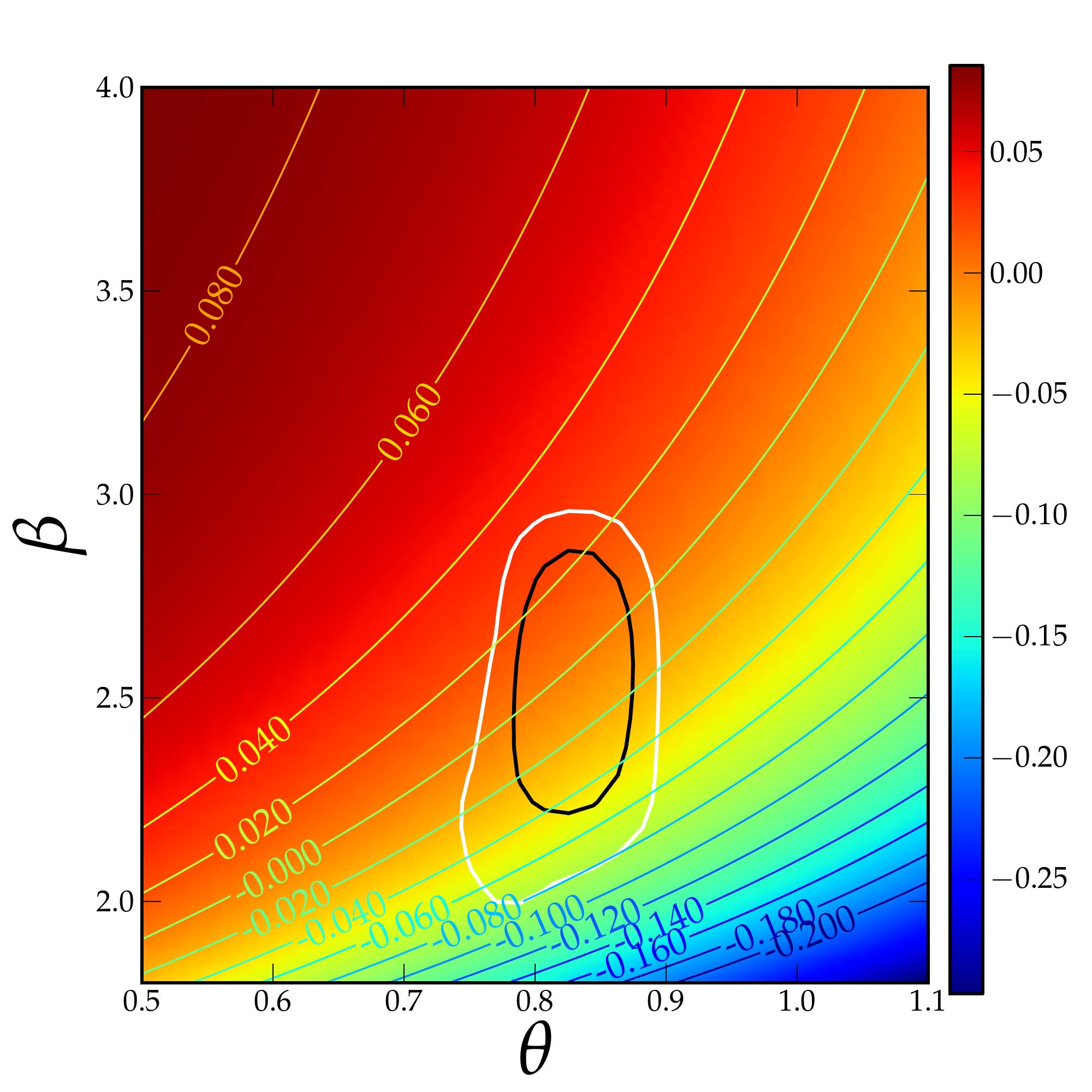}
\includegraphics[scale=0.43]{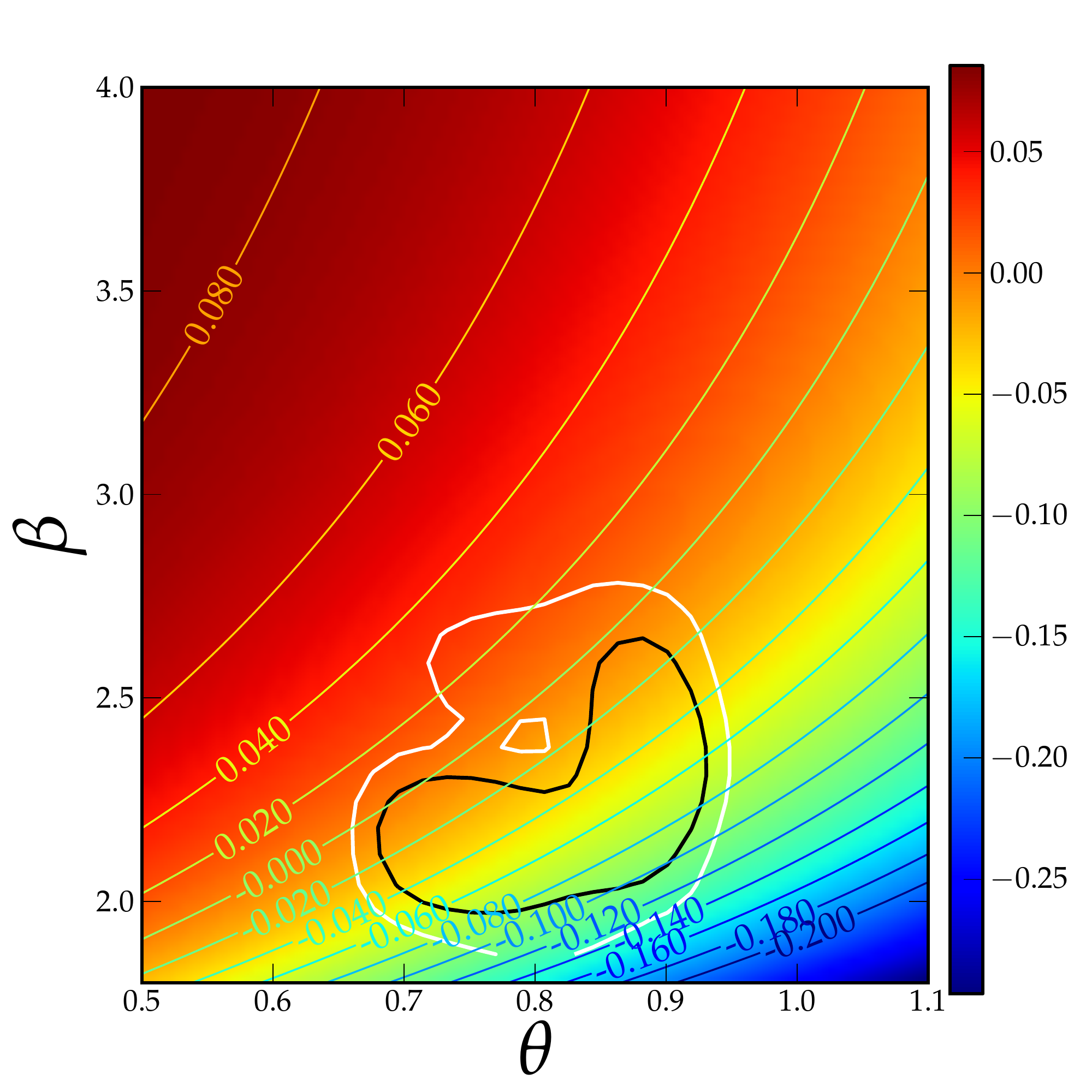}
\caption{Magnitude difference (measured in $3\arcsecond$ diameter
  apertures) between a point-like object convolved with the target PSF
  ${\cal M}\left[0.8\arcsecond,2.5\right]$ and with
  ${\cal M}\left[\theta,\beta\right]$ in the 2D parameter
  space $\left[\theta\arcsecond,\beta\right]$.  The black and white
  contours represent the regions which enclose 68\% and 95\% of the
  $\beta$-$\theta$ stellar distribution for two representative bands: $u$
  (Left), which is relatively homogenous across the field, and $IA464$
  which is not.}
\label{Fig:VarMag}
\end{center}
\end{figure*}

As discussed in section~\ref{Sec:preparation-images}, there is a
variation of the PSF within the field which is not taken into account
in our homogenization. For this reason, it is important to estimate the
magnitude differences arising from this variation.  To achieve this, we
present here a toy model to estimate the effect of the seeing
variation on the aperture magnitude for point-like objects.
We denote $D_{\rm stars}$($\theta$,$\beta$,$r$) as the difference
  of the aperture magnitudes for a PSF represented by a Moffat profile
  ${\cal M}\left[\theta,\beta \right]$ and with a PSF
  ${\cal M}\left[0.8\arcsec,2.5\right]$. $D_{\rm stars}$ is a function of
  $\theta$ and $\beta$, the two parameters which define the Moffat
  profile, and $r$ which is the aperture diameter.  We present in
  Figure~\ref{Fig:VarMag} $D_{\rm stars}$($\theta$,$\beta$,3$\arcsec$)
  in the two-dimensional (2D) parameter space $\left[\theta,\beta\right]$. We
  overplotted on this 2D distribution the contours which enclose 68\%
  and 95\% of the [$\theta$,$\beta$] distribution for the two bands
  $u$ and $IA464$. 
For the purpose of this figure, each star seeing is individually
computed from a fit with a Moffat profile on the PSF-homogenized star
profiles (reconstructed from the flux extracted at 14 fixed apertures,
logarithmically spaced between 0.25$\arcsec$ and 2.5$\arcsec$). Note
that since the Moffat Profile is fitted on individual stars from 14
discrete apertures and not on all of the point sources at the same time,
the precision of the fit is limited. However, this immediately provides a
qualitative insight as to the bias
  generated by internal PSF variation when extracting the star
  photometry within a 3$\arcsec$ aperture. For the worst band $IA464$,
  this bias is expected to remain below 0.1 mag. We also estimate
  that the median of the magnitude difference is below 0.05 mag, which
  is in agreement with Figure \ref{Fig:PSF}. 

\begin{figure}
\begin{center}
\includegraphics[scale=0.7]{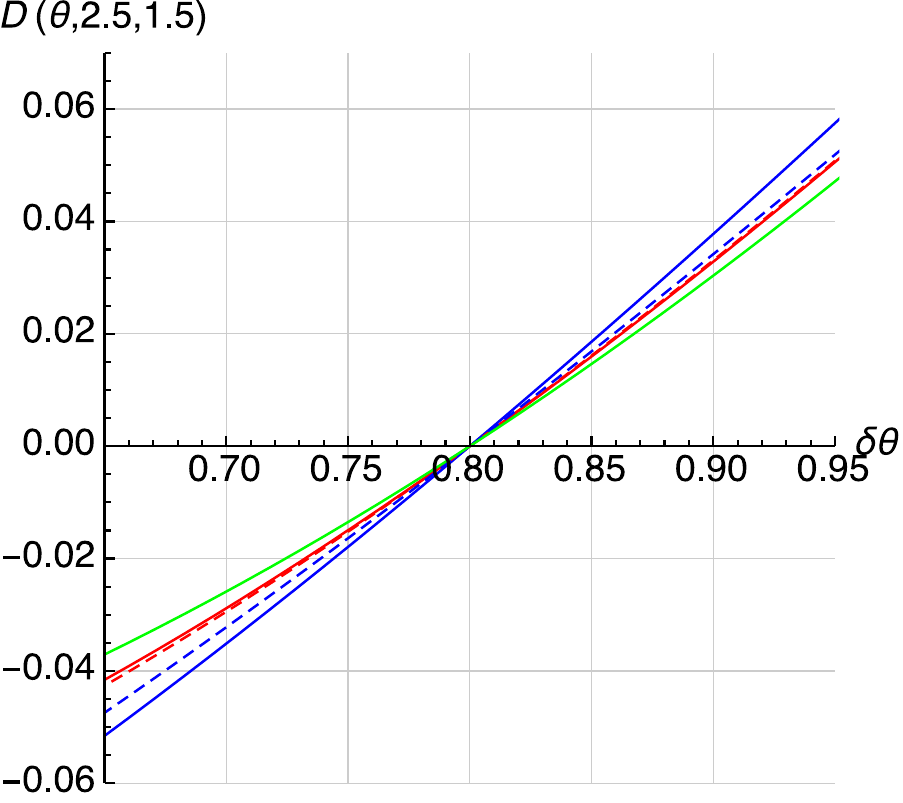}
\caption{Magnitude difference for a point-like object (green), an
  elliptical galaxy (red), and a spiral galaxy (blue) convolved with a
  PSF ${\cal M}\left[\theta,2.5\right]$ as a function of the seeing
  $\theta$ and for two different effective radii, 0.5 $\arcsec$ (solid
  line) and 0.8 $\arcsec$ (dashed line), when the extraction is
  performed in 3 $\arcsec$ diameter apertures.}
\label{Fig:2profiles}
\end{center}
\end{figure}

We then estimate this bias in the photometry for extended
  objects. We chose two different galaxy luminosity profiles, namely, a
  de Vaucouleurs profile (1948, 1959) to model a typical elliptical galaxy
  profile,
\begin{equation}
F_{\rm elliptical}(R_{e},r) \propto \exp \left[-7.67\left(\frac{r}{R_{e}}\right)^{\frac{1}{4}}\right]
\end{equation}
and an exponential profile to model a spiral galaxy profile,
\begin{equation}
F_{\rm spiral}(R_{e},r)\propto \exp\left[-\frac{r}{R_{e}}\right]
\end{equation}
Here, $R_{e}$ is the effective radius such that half of the
  total flux is within $R_{e}$. We then convolved the luminosity
  profiles with the Moffat profile, and integrate them in a circular
  aperture of 3$\arcsec$. For this exercise, we keep $\beta$ constant
  and equal to 2.5 and we allow $\theta$ to vary.  In
  Figure~\ref{Fig:2profiles}, we present the difference
  $D_{\rm spiral}$($\theta$,2.5,1.5$\arcsec$) and
  $D_{\rm elliptical}$($\theta$,2.5,1.5$\arcsec$) for two effective
  radii ($R_{e}=$ 0.5 and 0.8 $\arcsec$). We note that for FWHM
  differences below 0.1$\arcsec$ the induced magnitude discrepancies
  are always lower than 0.05, regardless of the galaxy profile.  

\bibliographystyle{apj}

\begin{thebibliography}{}
\expandafter\ifx\csname natexlab\endcsname\relax\def\natexlab#1{#1}\fi

\bibitem[{{Aihara} {et~al.}(2011){Aihara}, {Allende Prieto}, {An}, {Anderson},
  {Aubourg}, {Balbinot}, {Beers}, {Berlind}, {Bickerton}, {Bizyaev}, {Blanton},
  {Bochanski}, {Bolton}, {Bovy}, {Brandt}, {Brinkmann}, {Brown}, {Brownstein},
  {Busca}, {Campbell}, {Carr}, {Chen}, {Chiappini}, {Comparat}, {Connolly},
  {Cortes}, {Croft}, {Cuesta}, {da Costa}, {Davenport}, {Dawson}, {Dhital},
  {Ealet}, {Ebelke}, {Edmondson}, {Eisenstein}, {Escoffier}, {Esposito},
  {Evans}, {Fan}, {Femen{\'{\i}}a Castell{\'a}}, {Font-Ribera}, {Frinchaboy},
  {Ge}, {Gillespie}, {Gilmore}, {Gonz{\'a}lez Hern{\'a}ndez}, {Gott}, {Gould},
  {Grebel}, {Gunn}, {Hamilton}, {Harding}, {Harris}, {Hawley}, {Hearty}, {Ho},
  {Hogg}, {Holtzman}, {Honscheid}, {Inada}, {Ivans}, {Jiang}, {Johnson},
  {Jordan}, {Jordan}, {Kazin}, {Kirkby}, {Klaene}, {Knapp}, {Kneib},
  {Kochanek}, {Koesterke}, {Kollmeier}, {Kron}, {Lampeitl}, {Lang}, {Le Goff},
  {Lee}, {Lin}, {Long}, {Loomis}, {Lucatello}, {Lundgren}, {Lupton}, {Ma},
  {MacDonald}, {Mahadevan}, {Maia}, {Makler}, {Malanushenko}, {Malanushenko},
  {Mandelbaum}, {Maraston}, {Margala}, {Masters}, {McBride}, {McGehee},
  {McGreer}, {M{\'e}nard}, {Miralda-Escud{\'e}}, {Morrison}, {Mullally},
  {Muna}, {Munn}, {Murayama}, {Myers}, {Naugle}, {Neto}, {Nguyen}, {Nichol},
  {O'Connell}, {Ogando}, {Olmstead}, {Oravetz}, {Padmanabhan},
  {Palanque-Delabrouille}, {Pan}, {Pandey}, {P{\^a}ris}, {Percival},
  {Petitjean}, {Pfaffenberger}, {Pforr}, {Phleps}, {Pichon}, {Pieri}, {Prada},
  {Price-Whelan}, {Raddick}, {Ramos}, {Reyl{\'e}}, {Rich}, {Richards}, {Rix},
  {Robin}, {Rocha-Pinto}, {Rockosi}, {Roe}, {Rollinde}, {Ross}, {Ross},
  {Rossetto}, {S{\'a}nchez}, {Sayres}, {Schlegel}, {Schlesinger}, {Schmidt},
  {Schneider}, {Sheldon}, {Shu}, {Simmerer}, {Simmons}, {Sivarani}, {Snedden},
  {Sobeck}, {Steinmetz}, {Strauss}, {Szalay}, {Tanaka}, {Thakar}, {Thomas},
  {Tinker}, {Tofflemire}, {Tojeiro}, {Tremonti}, {Vandenberg}, {Vargas
  Maga{\~n}a}, {Verde}, {Vogt}, {Wake}, {Wang}, {Weaver}, {Weinberg}, {White},
  {White}, {Yanny}, {Yasuda}, {Yeche}, \& {Zehavi}}]{2011ApJS..193...29A}
{Aihara}, H., {Allende Prieto}, C., {An}, D., {et~al.} 2011, \apjs, 193, 29

\bibitem[{Allen(1976)}]{Allen1976}
Allen, C.~W. 1976, {Astrophysical Quantities}, 4th edn. (Addison-Wesley
  Professional)

\bibitem[{Arnouts {et~al.}(2002)Arnouts, Moscardini, Vanzella, Colombi,
  Cristiani, Fontana, Giallongo, Matarrese, \& Saracco}]{Arnouts:2002p10840}
Arnouts, S., Moscardini, L., Vanzella, E., {et~al.} 2002, \mnras, 329, 355

\bibitem[{Arnouts {et~al.}(2007)Arnouts, Walcher, Le~F{\`e}vre, Zamorani,
  Ilbert, Le~Brun, Pozzetti, Bardelli, Tresse, Zucca, Charlot, Lamareille,
  McCracken, Bolzonella, Iovino, Lonsdale, Polletta, Surace, Bottini, Garilli,
  Maccagni, Picat, Scaramella, Scodeggio, Vettolani, Zanichelli, Adami, Cappi,
  Ciliegi, Contini, de~la Torre, Foucaud, Franzetti, Gavignaud, Guzzo, Marano,
  Marinoni, Mazure, Meneux, Merighi, Paltani, Pell{\`o}, Pollo, Radovich,
  Temporin, \& Vergani}]{Arnouts:2007p3665}
Arnouts, S., Walcher, C.~J., Le~F{\`e}vre, O., {et~al.} 2007, \aap, 476, 137

\bibitem[{{Arnouts} {et~al.}(2013){Arnouts}, {Le Floc'h}, {Chevallard},
  {Johnson}, {Ilbert}, {Treyer}, {Aussel}, {Capak}, {Sanders}, {Scoville},
  {McCracken}, {Milliard}, {Pozzetti}, \& {Salvato}}]{2013A&A...558A..67A}
{Arnouts}, S., {Le Floc'h}, E., {Chevallard}, J., {et~al.} 2013, \aap, 558, A67

\bibitem[{{Bertin}(2013)}]{2013ascl.soft01001B}
{Bertin}, E. 2013, {PSFEx: Point Spread Function Extractor}, Astrophysics
  Source Code Library, ascl:1301.001

\bibitem[{Bertin \& Arnouts(1996)}]{Bertin:1996p13615}
Bertin, E., \& Arnouts, S. 1996, \apjs, 117, 393

\bibitem[{Bertin {et~al.}(2002)Bertin, Mellier, Radovich, Missonnier, Didelon,
  \& Morin}]{Bertin:2002p5282}
Bertin, E., Mellier, Y., Radovich, M., {et~al.} 2002, Astronomical Data
  Analysis Software and Systems XI, 281, 228

\bibitem[{B{\'e}thermin {et~al.}(2014)B{\'e}thermin, Kilbinger, Daddi, Gabor,
  Finoguenov, McCracken, Wolk, Aussel, Strazzulo, Le~Floc'h, Gobat, Rodighiero,
  Dickinson, Wang, Lutz, \& Heinis}]{Bethermin:2014dh}
B{\'e}thermin, M., Kilbinger, M., Daddi, E., {et~al.} 2014, \aap, 567, A103

\bibitem[{Bielby {et~al.}(2012)Bielby, Hudelot, McCracken, Ilbert, Daddi,
  Le~F{\`e}vre, Gonzalez-Perez, Kneib, Marmo, Mellier, Salvato, Sanders, \&
  Willott}]{2012A&A...545A..23B}
Bielby, R., Hudelot, P., McCracken, H.~J., {et~al.} 2012, \aap, 545, 23

\bibitem[{{Bieri} {et~al.}(2015){Bieri}, {Dubois}, {Silk}, \&
  {Mamon}}]{2015ApJ...812L..36B}
{Bieri}, R., {Dubois}, Y., {Silk}, J., \& {Mamon}, G.~A. 2015, \apjl, 812, L36

\bibitem[{{Bolzonella} {et~al.}(2000){Bolzonella}, {Miralles}, \&
  {Pell{\'o}}}]{2000A&A...363..476B}
{Bolzonella}, M., {Miralles}, J.-M., \& {Pell{\'o}}, R. 2000, \aap, 363, 476

\bibitem[{{Brusa} {et~al.}(2010){Brusa}, {Civano}, {Comastri}, {Miyaji},
  {Salvato}, {Zamorani}, {Cappelluti}, {Fiore}, {Hasinger}, {Mainieri},
  {Merloni}, {Bongiorno}, {Capak}, {Elvis}, {Gilli}, {Hao}, {Jahnke},
  {Koekemoer}, {Ilbert}, {Le Floc'h}, {Lusso}, {Mignoli}, {Schinnerer},
  {Silverman}, {Treister}, {Trump}, {Vignali}, {Zamojski}, {Aldcroft},
  {Aussel}, {Bardelli}, {Bolzonella}, {Cappi}, {Caputi}, {Contini},
  {Finoguenov}, {Fruscione}, {Garilli}, {Impey}, {Iovino}, {Iwasawa},
  {Kampczyk}, {Kartaltepe}, {Kneib}, {Knobel}, {Kovac}, {Lamareille},
  {Leborgne}, {Le Brun}, {Le Fevre}, {Lilly}, {Maier}, {McCracken}, {Pello},
  {Peng}, {Perez-Montero}, {de Ravel}, {Sanders}, {Scodeggio}, {Scoville},
  {Tanaka}, {Taniguchi}, {Tasca}, {de la Torre}, {Tresse}, {Vergani}, \&
  {Zucca}}]{2010ApJ...716..348B}
{Brusa}, M., {Civano}, F., {Comastri}, A., {et~al.} 2010, \apj, 716, 348

\bibitem[{Bruzual \& Charlot(2003)}]{Bruzual:2003p963}
Bruzual, G., \& Charlot, S. 2003, \mnras, 344, 1000

\bibitem[{Calzetti {et~al.}(2000)Calzetti, Armus, Bohlin, Kinney, Koornneef, \&
  Storchi-Bergmann}]{Calzetti:2000p6839}
Calzetti, D., Armus, L., Bohlin, R.~C., {et~al.} 2000, \apj, 533, 682

\bibitem[{Capak {et~al.}(2007)Capak, Aussel, Ajiki, McCracken, Mobasher,
  Scoville, Shopbell, Taniguchi, Thompson, Tribiano, Sasaki, Blain, Brusa,
  Carilli, Comastri, Carollo, Cassata, Colbert, Ellis, Elvis, Giavalisco,
  Green, Guzzo, Hasinger, Ilbert, Impey, Jahnke, Kartaltepe, Kneib, Koda,
  Koekemoer, Komiyama, Leauthaud, Le~F{\`e}vre, Lilly, Liu, Massey, Miyazaki,
  Murayama, Nagao, Peacock, Pickles, Porciani, Renzini, Rhodes, Rich, Salvato,
  Sanders, Scarlata, Schiminovich, Schinnerer, Scodeggio, Sheth, Shioya, Tasca,
  Taylor, Yan, \& Zamorani}]{2007ApJS..172...99C}
Capak, P., Aussel, H., Ajiki, M., {et~al.} 2007, \apjs, 172, 99

\bibitem[{{Cappelluti} {et~al.}(2007){Cappelluti}, {Hasinger}, {Brusa},
  {Comastri}, {Zamorani}, {B{\"o}hringer}, {Brunner}, {Civano}, {Finoguenov},
  {Fiore}, {Gilli}, {Griffiths}, {Mainieri}, {Matute}, {Miyaji}, \&
  {Silverman}}]{2007ApJS..172..341C}
{Cappelluti}, N., {Hasinger}, G., {Brusa}, M., {et~al.} 2007, \apjs, 172, 341

\bibitem[{{Chabrier}(2003)}]{2003PASP..115..763C}
{Chabrier}, G. 2003, \pasp, 115, 763

\bibitem[{{Civano} {et~al.}(2012){Civano}, {Elvis}, {Brusa}, {Comastri},
  {Salvato}, {Zamorani}, {Aldcroft}, {Bongiorno}, {Capak}, {Cappelluti},
  {Cisternas}, {Fiore}, {Fruscione}, {Hao}, {Kartaltepe}, {Koekemoer}, {Gilli},
  {Impey}, {Lanzuisi}, {Lusso}, {Mainieri}, {Miyaji}, {Lilly}, {Masters},
  {Puccetti}, {Schawinski}, {Scoville}, {Silverman}, {Trump}, {Urry},
  {Vignali}, \& {Wright}}]{2012ApJS..201...30C}
{Civano}, F., {Elvis}, M., {Brusa}, M., {et~al.} 2012, \apjs, 201, 30

\bibitem[{{Civano} {et~al.}(2016){Civano}, {Marchesi}, {Comastri}, {Urry},
  {Elvis}, {Cappelluti}, {Puccetti}, {Brusa}, {Zamorani}, {Hasinger},
  {Aldcroft}, {Alexander}, {Allevato}, {Brunner}, {Capak}, {Finoguenov},
  {Fiore}, {Fruscione}, {Gilli}, {Glotfelty}, {Griffiths}, {Hao}, {Harrison},
  {Jahnke}, {Kartaltepe}, {Karim}, {LaMassa}, {Lanzuisi}, {Miyaji}, {Ranalli},
  {Salvato}, {Sargent}, {Scoville}, {Schawinski}, {Schinnerer}, {Silverman},
  {Smolcic}, {Stern}, {Toft}, {Trakhenbrot}, {Treister}, \&
  {Vignali}}]{2016Civano}
{Civano}, F., {Marchesi}, S., {Comastri}, A., {et~al.} 2016, ArXiv e-prints,
  arXiv:1601.00941



\bibitem[{{Codis} {et~al.}(2012){Codis}, {Pichon}, {Devriendt}, {Slyz},
  {Pogosyan}, {Dubois}, \& {Sousbie}}]{2012MNRAS.427.3320C}
{Codis}, S., {Pichon}, C., {Devriendt}, J., {et~al.} 2012, \mnras, 427, 3320

\bibitem[{{Codis} {et~al.}(2015){Codis}, {Pichon}, \&
  {Pogosyan}}]{2015MNRASCodis}
{Codis}, S., {Pichon}, C., \& {Pogosyan}, D. 2015, \mnras, 452, 3369

\bibitem[{{Comparat} {et~al.}(2015){Comparat}, {Richard}, {Kneib}, {Ilbert},
  {Gonzalez-Perez}, {Tresse}, {Zoubian}, {Arnouts}, {Brownstein}, {Baugh},
  {Delubac}, {Ealet}, {Escoffier}, {Ge}, {Jullo}, {Lacey}, {Ross}, {Schlegel},
  {Schneider}, {Steele}, {Tasca}, {Yeche}, {Lesser}, {Jiang}, {Jing}, {Fan},
  {Fan}, {Ma}, {Nie}, {Wang}, {Wu}, {Zhang}, {Zhou}, {Zhou}, \&
  {Zou}}]{2015AA...575A..40C}
{Comparat}, J., {Richard}, J., {Kneib}, J.-P., {et~al.} 2015, \aap, 575, A40

\bibitem[{Cooray \& Sheth(2002)}]{Cooray:2002p846}
Cooray, A., \& Sheth, R. 2002, \physrep, 372, 1

\bibitem[{{Coupon} {et~al.}(2015){Coupon}, {Arnouts}, {van Waerbeke},
  {Moutard}, {Ilbert}, {van Uitert}, {Erben}, {Garilli}, {Guzzo}, {Heymans},
  {Hildebrandt}, {Hoekstra}, {Kilbinger}, {Kitching}, {Mellier}, {Miller},
  {Scodeggio}, {Bonnett}, {Branchini}, {Davidzon}, {De Lucia}, {Fritz}, {Fu},
  {Hudelot}, {Hudson}, {Kuijken}, {Leauthaud}, {Le F{\`e}vre}, {McCracken},
  {Moscardini}, {Rowe}, {Schrabback}, {Semboloni}, \&
  {Velander}}]{2015MNRASCoupon}
{Coupon}, J., {Arnouts}, S., {van Waerbeke}, L., {et~al.} 2015, \mnras, 449,
  1352

\bibitem[{Cowie {et~al.}(1996)Cowie, Songaila, Hu, \& Cohen}]{Cowie:1996p8471}
Cowie, L.~L., Songaila, A., Hu, E.~M., \& Cohen, J.~G. 1996, Astronomical
  Journal v.112, 112, 839

\bibitem[{{Croton} {et~al.}(2006){Croton}, {Springel}, {White}, {De Lucia},
  {Frenk}, {Gao}, {Jenkins}, {Kauffmann}, {Navarro}, \&
  {Yoshida}}]{2006MNRAS.367..864C}
{Croton}, D.~J., {Springel}, V., {White}, S.~D.~M., {et~al.} 2006, \mnras, 367,
  864

\bibitem[{Daddi {et~al.}(2007)Daddi, Dickinson, Morrison, Chary, Cimatti,
  Elbaz, Frayer, Renzini, Pope, Alexander, Bauer, Giavalisco, Huynh, Kurk, \&
  Mignoli}]{Daddi:2007p2924}
Daddi, E., Dickinson, M., Morrison, G., {et~al.} 2007, \apj, 670, 156

\bibitem[{{Darvish} {et~al.}(2014){Darvish}, {Sobral}, {Mobasher}, {Scoville},
  {Best}, {Sales}, \& {Smail}}]{2014ApJ...796...51D}
{Darvish}, B., {Sobral}, D., {Mobasher}, B., {et~al.} 2014, \apj, 796, 51

\bibitem[{{Davidzon} {et~al.}(2013){Davidzon}, {Bolzonella}, {Coupon},
  {Ilbert}, {Arnouts}, {de la Torre}, {Fritz}, {De Lucia}, {Iovino}, {Granett},
  {Zamorani}, {Guzzo}, {Abbas}, {Adami}, {Bel}, {Bottini}, {Branchini},
  {Cappi}, {Cucciati}, {Franzetti}, {Fumana}, {Garilli}, {Krywult}, {Le Brun},
  {Le F{\`e}vre}, {Maccagni}, {Ma{\l}ek}, {Marulli}, {McCracken}, {Paioro},
  {Peacock}, {Polletta}, {Pollo}, {Schlagenhaufer}, {Scodeggio}, {Tasca},
  {Tojeiro}, {Vergani}, {Zanichelli}, {Burden}, {Di Porto}, {Marchetti},
  {Marinoni}, {Mellier}, {Moscardini}, {Moutard}, {Nichol}, {Percival},
  {Phleps}, \& {Wolk}}]{2013Davidzon}
{Davidzon}, I., {Bolzonella}, M., {Coupon}, J., {et~al.} 2013, \aap, 558, A23

\bibitem[{{Dekel} {et~al.}(2009){Dekel}, {Birnboim}, {Engel}, {Freundlich},
  {Goerdt}, {Mumcuoglu}, {Neistein}, {Pichon}, {Teyssier}, \&
  {Zinger}}]{2009Natur.457..451D}
{Dekel}, A., {Birnboim}, Y., {Engel}, G., {et~al.} 2009, \nat, 457, 451

\bibitem[{Dressler(1980)}]{1980ApJ...236..351D}
Dressler, A. 1980, \apj, 236, 351

\bibitem[{{Dubois} {et~al.}(2014){Dubois}, {Pichon}, {Welker}, {Le Borgne},
  {Devriendt}, {Laigle}, {Codis}, {Pogosyan}, {Arnouts}, {Benabed}, {Bertin},
  {Blaizot}, {Bouchet}, {Cardoso}, {Colombi}, {de Lapparent}, {Desjacques},
  {Gavazzi}, {Kassin}, {Kimm}, {McCracken}, {Milliard}, {Peirani}, {Prunet},
  {Rouberol}, {Silk}, {Slyz}, {Sousbie}, {Teyssier}, {Tresse}, {Treyer},
  {Vibert}, \& {Volonteri}}]{2014MNRAS.444.1453D}
{Dubois}, Y., {Pichon}, C., {Welker}, C., {et~al.} 2014, \mnras, 444, 1453

\bibitem[{Elbaz {et~al.}(2007)Elbaz, Daddi, Le~Borgne, Dickinson, Alexander,
  Chary, Starck, Brandt, Kitzbichler, MacDonald, Nonino, Popesso, Stern, \&
  Vanzella}]{Elbaz:2007p11978}
Elbaz, D., Daddi, E., Le~Borgne, D., {et~al.} 2007, \aap, 468, 33

\bibitem[{{Elvis} {et~al.}(2009){Elvis}, {Civano}, {Vignali}, {Puccetti},
  {Fiore}, {Cappelluti}, {Aldcroft}, {Fruscione}, {Zamorani}, {Comastri},
  {Brusa}, {Gilli}, {Miyaji}, {Damiani}, {Koekemoer}, {Finoguenov}, {Brunner},
  {Urry}, {Silverman}, {Mainieri}, {Hasinger}, {Griffiths}, {Carollo}, {Hao},
  {Guzzo}, {Blain}, {Calzetti}, {Carilli}, {Capak}, {Ettori}, {Fabbiano},
  {Impey}, {Lilly}, {Mobasher}, {Rich}, {Salvato}, {Sanders}, {Schinnerer},
  {Scoville}, {Shopbell}, {Taylor}, {Taniguchi}, \&
  {Volonteri}}]{2009ApJS..184..158E}
{Elvis}, M., {Civano}, F., {Vignali}, C., {et~al.} 2009, \apjs, 184, 158

\bibitem[{{Fioc} \& {Rocca-Volmerange}(1997)}]{1997A&A...326..950F}
{Fioc}, M., \& {Rocca-Volmerange}, B. 1997, \aap, 326, 950

\bibitem[{{Fioc} \& {Rocca-Volmerange}(1999)}]{1999astro.ph.12179F}
---. 1999, ArXiv Astrophysics e-prints, astro-ph/9912179

\bibitem[{{Fitzpatrick} \& {Massa}(1986)}]{1986ApJ...307..286F}
{Fitzpatrick}, E.~L., \& {Massa}, D. 1986, \apj, 307, 286

\bibitem[{{Fontana} {et~al.}(2014){Fontana}, {Dunlop}, {Paris}, {Targett},
  {Boutsia}, {Castellano}, {Galametz}, {Grazian}, {McLure}, {Merlin},
  {Pentericci}, {Wuyts}, {Almaini}, {Caputi}, {Chary}, {Cirasuolo},
  {Conselice}, {Cooray}, {Daddi}, {Dickinson}, {Faber}, {Fazio}, {Ferguson},
  {Giallongo}, {Giavalisco}, {Grogin}, {Hathi}, {Koekemoer}, {Koo}, {Lucas},
  {Nonino}, {Rix}, {Renzini}, {Rosario}, {Santini}, {Scarlata}, {Sommariva},
  {Stark}, {van der Wel}, {Vanzella}, {Wild}, {Yan}, \&
  {Zibetti}}]{2014Fontana}
{Fontana}, A., {Dunlop}, J.~S., {Paris}, D., {et~al.} 2014, \aap, 570, A11

\bibitem[{{Gaibler} {et~al.}(2012){Gaibler}, {Khochfar}, {Krause}, \&
  {Silk}}]{2012MNRAS.425..438G}
{Gaibler}, V., {Khochfar}, S., {Krause}, M., \& {Silk}, J. 2012, \mnras, 425,
  438

\bibitem[{{Griffin} {et~al.}(2010){Griffin}, {Abergel}, {Abreu}, {Ade},
  {Andr{\'e}}, {Augueres}, {Babbedge}, {Bae}, {Baillie}, {Baluteau}, {Barlow},
  {Bendo}, {Benielli}, {Bock}, {Bonhomme}, {Brisbin}, {Brockley-Blatt},
  {Caldwell}, {Cara}, {Castro-Rodriguez}, {Cerulli}, {Chanial}, {Chen},
  {Clark}, {Clements}, {Clerc}, {Coker}, {Communal}, {Conversi}, {Cox},
  {Crumb}, {Cunningham}, {Daly}, {Davis}, {de Antoni}, {Delderfield}, {Devin},
  {di Giorgio}, {Didschuns}, {Dohlen}, {Donati}, {Dowell}, {Dowell}, {Duband},
  {Dumaye}, {Emery}, {Ferlet}, {Ferrand}, {Fontignie}, {Fox}, {Franceschini},
  {Frerking}, {Fulton}, {Garcia}, {Gastaud}, {Gear}, {Glenn}, {Goizel},
  {Griffin}, {Grundy}, {Guest}, {Guillemet}, {Hargrave}, {Harwit}, {Hastings},
  {Hatziminaoglou}, {Herman}, {Hinde}, {Hristov}, {Huang}, {Imhof}, {Isaak},
  {Israelsson}, {Ivison}, {Jennings}, {Kiernan}, {King}, {Lange}, {Latter},
  {Laurent}, {Laurent}, {Leeks}, {Lellouch}, {Levenson}, {Li}, {Li},
  {Lilienthal}, {Lim}, {Liu}, {Lu}, {Madden}, {Mainetti}, {Marliani}, {McKay},
  {Mercier}, {Molinari}, {Morris}, {Moseley}, {Mulder}, {Mur}, {Naylor},
  {Nguyen}, {O'Halloran}, {Oliver}, {Olofsson}, {Olofsson}, {Orfei}, {Page},
  {Pain}, {Panuzzo}, {Papageorgiou}, {Parks}, {Parr-Burman}, {Pearce},
  {Pearson}, {P{\'e}rez-Fournon}, {Pinsard}, {Pisano}, {Podosek}, {Pohlen},
  {Polehampton}, {Pouliquen}, {Rigopoulou}, {Rizzo}, {Roseboom}, {Roussel},
  {Rowan-Robinson}, {Rownd}, {Saraceno}, {Sauvage}, {Savage}, {Savini},
  {Sawyer}, {Scharmberg}, {Schmitt}, {Schneider}, {Schulz}, {Schwartz},
  {Shafer}, {Shupe}, {Sibthorpe}, {Sidher}, {Smith}, {Smith}, {Smith},
  {Spencer}, {Stobie}, {Sudiwala}, {Sukhatme}, {Surace}, {Stevens}, {Swinyard},
  {Trichas}, {Tourette}, {Triou}, {Tseng}, {Tucker}, {Turner}, {Vaccari},
  {Valtchanov}, {Vigroux}, {Virique}, {Voellmer}, {Walker}, {Ward}, {Waskett},
  {Weilert}, {Wesson}, {White}, {Whitehouse}, {Wilson}, {Winter}, {Woodcraft},
  {Wright}, {Xu}, {Zavagno}, {Zemcov}, {Zhang}, \& {Zonca}}]{2010Griffin}
{Griffin}, M.~J., {Abergel}, A., {Abreu}, A., {et~al.} 2010, \aap, 518, L3


\bibitem[{{Growth} \& {Peebles}(1977)}]{Growth1977}
{Growth}, E.~J., \& {Peebles}, P.~J.~E. 1977, \apj, 217, 385


\bibitem[{{Hasinger} {et~al.}(2007){Hasinger}, {Cappelluti}, {Brunner},
  {Brusa}, {Comastri}, {Elvis}, {Finoguenov}, {Fiore}, {Franceschini}, {Gilli},
  {Griffiths}, {Lehmann}, {Mainieri}, {Matt}, {Matute}, {Miyaji}, {Molendi},
  {Paltani}, {Sanders}, {Scoville}, {Tresse}, {Urry}, {Vettolani}, \&
  {Zamorani}}]{2007ApJS..172...29H}
{Hasinger}, G., {Cappelluti}, N., {Brunner}, H., {et~al.} 2007, \apjs, 172, 29

\bibitem[{{Heinis} {et~al.}(2007){Heinis}, {Milliard}, {Arnouts}, {Blaizot},
  {Schiminovich}, {Budav{\'a}ri}, {Ilbert}, {Donas}, {Treyer}, {Wyder},
  {McCracken}, {Barlow}, {Forster}, {Friedman}, {Martin}, {Morrissey}, {Neff},
  {Seibert}, {Small}, {Bianchi}, {Heckman}, {Lee}, {Madore}, {Rich}, {Szalay},
  {Welsh}, {Yi}, \& {Xu}}]{2007ApJS..173..503H}
{Heinis}, S., {Milliard}, B., {Arnouts}, S., {et~al.} 2007, \apjs, 173, 503

\bibitem[{{Hildebrandt} {et~al.}(2012){Hildebrandt}, {Erben}, {Kuijken}, {van
  Waerbeke}, {Heymans}, {Coupon}, {Benjamin}, {Bonnett}, {Fu}, {Hoekstra},
  {Kitching}, {Mellier}, {Miller}, {Velander}, {Hudson}, {Rowe}, {Schrabback},
  {Semboloni}, \& {Ben{\'{\i}}tez}}]{2012MNRAS.421.2355H}
{Hildebrandt}, H., {Erben}, T., {Kuijken}, K., {et~al.} 2012, \mnras, 421, 2355

\bibitem[{{Hoaglin} {et~al.}(1983){Hoaglin}, {Mosteller}, \&
  {Tukey}}]{1983ured.book.....H}
{Hoaglin}, D.~C., {Mosteller}, F., \& {Tukey}, J.~W. 1983, {Understanding
  robust and exploratory data anlysis}

\bibitem[{{Hopkins} {et~al.}(2006){Hopkins}, {Hernquist}, {Cox}, {Di Matteo},
  {Robertson}, \& {Springel}}]{2006ApJS..163....1H}
{Hopkins}, P.~F., {Hernquist}, L., {Cox}, T.~J., {et~al.} 2006, \apjs, 163, 1

\bibitem[{{Hsieh} {et~al.}(2012){Hsieh}, {Wang}, {Hsieh}, {Lin}, {Yan}, {Lim},
  \& {Ho}}]{2012ApJS..203...23H}
{Hsieh}, B.-C., {Wang}, W.-H., {Hsieh}, C.-C., {et~al.} 2012, \apjs, 203, 23

\bibitem[{{Ilbert} {et~al.}(2005){Ilbert}, {Tresse}, {Zucca}, {Bardelli},
  {Arnouts}, {Zamorani}, {Pozzetti}, {Bottini}, {Garilli}, {Le Brun}, {Le
  F{\`e}vre}, {Maccagni}, {Picat}, {Scaramella}, {Scodeggio}, {Vettolani},
  {Zanichelli}, {Adami}, {Arnaboldi}, {Bolzonella}, {Cappi}, {Charlot},
  {Contini}, {Foucaud}, {Franzetti}, {Gavignaud}, {Guzzo}, {Iovino},
  {McCracken}, {Marano}, {Marinoni}, {Mathez}, {Mazure}, {Meneux}, {Merighi},
  {Paltani}, {Pello}, {Pollo}, {Radovich}, {Bondi}, {Bongiorno}, {Busarello},
  {Ciliegi}, {Lamareille}, {Mellier}, {Merluzzi}, {Ripepi}, \&
  {Rizzo}}]{2005Ilbert}
{Ilbert}, O., {Tresse}, L., {Zucca}, E., {et~al.} 2005, \aap, 439, 863

\bibitem[{Ilbert {et~al.}(2006)Ilbert, Arnouts, McCracken, Bolzonella, Bertin,
  Le~F{\`e}vre, Mellier, Zamorani, Pell{\`o}, Iovino, Tresse, Le~Brun, Bottini,
  Garilli, Maccagni, Picat, Scaramella, Scodeggio, Vettolani, Zanichelli,
  Adami, Bardelli, Cappi, Charlot, Ciliegi, Contini, Cucciati, Foucaud,
  Franzetti, Gavignaud, Guzzo, Marano, Marinoni, Mazure, Meneux, Merighi,
  Paltani, Pollo, Pozzetti, Radovich, Zucca, Bondi, Bongiorno, Busarello, de~la
  Torre, Gregorini, Lamareille, Mathez, Merluzzi, Ripepi, Rizzo, \&
  Vergani}]{2006A&A...457..841I}
Ilbert, O., Arnouts, S., McCracken, H.~J., {et~al.} 2006, \aap, 457, 841

\bibitem[{Ilbert {et~al.}(2009)Ilbert, Capak, Salvato, Aussel, McCracken,
  Sanders, Scoville, Kartaltepe, Arnouts, Le~Floc'h, Mobasher, Taniguchi,
  Lamareille, Leauthaud, Sasaki, Thompson, Zamojski, Zamorani, Bardelli,
  Bolzonella, Bongiorno, Brusa, Caputi, Carollo, Contini, Cook, Coppa,
  Cucciati, de~la Torre, De~Ravel, Franzetti, Garilli, Hasinger, Iovino,
  Kampczyk, Kneib, Knobel, Kovac, Le~Borgne, Le~Brun, F{\`e}vre, Lilly, Looper,
  Maier, Mainieri, Mellier, Mignoli, Murayama, Pell{\`o}, Peng, Perez-Montero,
  Renzini, Ricciardelli, Schiminovich, Scodeggio, Shioya, Silverman, Surace,
  Tanaka, Tasca, Tresse, Vergani, \& Zucca}]{2009ApJ...690.1236I}
Ilbert, O., Capak, P., Salvato, M., {et~al.} 2009, \apj, 690, 1236

\bibitem[{Ilbert {et~al.}(2013)Ilbert, McCracken, Le~F{\`e}vre, Capak, Dunlop,
  Karim, Renzini, Caputi, Boissier, Arnouts, Aussel, Comparat, Guo, Hudelot,
  Kartaltepe, Kneib, Krogager, Le~Floc'h, Lilly, Mellier, Milvang-Jensen,
  Moutard, Onodera, Richard, Salvato, Sanders, Scoville, Silverman, Taniguchi,
  Tasca, Thomas, Toft, Tresse, Vergani, Wolk, \& Zirm}]{Ilbert:2013dq}
Ilbert, O., McCracken, H.~J., Le~F{\`e}vre, O., {et~al.} 2013, \aap, 556, A55

\bibitem[{{Ilbert} {et~al.}(2015){Ilbert}, {Arnouts}, {Le Floc'h}, {Aussel},
  {Bethermin}, {Capak}, {Hsieh}, {Kajisawa}, {Karim}, {Le F{\`e}vre}, {Lee},
  {Lilly}, {McCracken}, {Michel-Dansac}, {Moutard}, {Renzini}, {Salvato},
  {Sanders}, {Scoville}, {Sheth}, {Silverman}, {Smol{\v c}i{\'c}}, {Taniguchi},
  \& {Tresse}}]{2015Ilbert}
{Ilbert}, O., {Arnouts}, S., {Le Floc'h}, E., {et~al.} 2015, \aap, 579, A2

\bibitem[{{Kartaltepe} {et~al.}(2010){Kartaltepe}, {Sanders}, {Le Floc'h},
  {Frayer}, {Aussel}, {Arnouts}, {Ilbert}, {Salvato}, {Scoville}, {Surace},
  {Yan}, {Capak}, {Caputi}, {Carollo}, {Cassata}, {Civano}, {Hasinger},
  {Koekemoer}, {Le F{\`e}vre}, {Lilly}, {Liu}, {McCracken}, {Schinnerer},
  {Smol{\v c}i{\'c}}, {Taniguchi}, {Thompson}, {Trump}, {Baldassare}, \&
  {Fiorenza}}]{2010ApJ...721...98K}
{Kartaltepe}, J.~S., {Sanders}, D.~B., {Le Floc'h}, E., {et~al.} 2010, \apj,
  721, 98

\bibitem[{{Katz} {et~al.}(2003){Katz}, {Keres}, {Dave}, \&
  {Weinberg}}]{2003ASSL..281..185K}
{Katz}, N., {Keres}, D., {Dave}, R., \& {Weinberg}, D.~H. 2003, in Astrophysics
  and Space Science Library, Vol. 281, The IGM/Galaxy Connection. The
  Distribution of Baryons at z=0, ed. J.~L. {Rosenberg} \& M.~E. {Putman}, 185

\bibitem[{{Kauffmann} {et~al.}(2004){Kauffmann}, {White}, {Heckman},
  {M{\'e}nard}, {Brinchmann}, {Charlot}, {Tremonti}, \&
  {Brinkmann}}]{2004MNRAS.353..713K}
{Kauffmann}, G., {White}, S.~D.~M., {Heckman}, T.~M., {et~al.} 2004, \mnras,
  353, 713

\bibitem[{{Kere{\v s}} {et~al.}(2005){Kere{\v s}}, {Katz}, {Weinberg}, \&
  {Dav{\'e}}}]{2005MNRAS.363....2K}
{Kere{\v s}}, D., {Katz}, N., {Weinberg}, D.~H., \& {Dav{\'e}}, R. 2005,
  \mnras, 363, 2

\bibitem[{{Khandai} {et~al.}(2015){Khandai}, {Di Matteo}, {Croft}, {Wilkins},
  {Feng}, {Tucker}, {DeGraf}, \& {Liu}}]{2015Khandai}
{Khandai}, N., {Di Matteo}, T., {Croft}, R., {et~al.} 2015, \mnras, 450, 1349

\bibitem[{{Kochiashvili} {et~al.}(2015){Kochiashvili}, {M{\o}ller},
  {Milvang-Jensen}, {Christensen}, {Fynbo}, {Freudling}, {Cl{\'e}ment}, {Cuby},
  {Zabl}, \& {Zibetti}}]{2015Koch}
{Kochiashvili}, I., {M{\o}ller}, P., {Milvang-Jensen}, B., {et~al.} 2015, \aap,
  580, A42

\bibitem[{Koekemoer {et~al.}(2007)Koekemoer, Aussel, Calzetti, Capak,
  Giavalisco, Kneib, Leauthaud, Le~F{\`e}vre, McCracken, Massey, Mobasher,
  Rhodes, Scoville, \& Shopbell}]{2007ApJS..172..196K}
Koekemoer, A.~M., Aussel, H., Calzetti, D., {et~al.} 2007, \apjs, 172, 196

\bibitem[{{Kriek} {et~al.}(2015){Kriek}, {Shapley}, {Reddy}, {Siana}, {Coil},
  {Mobasher}, {Freeman}, {de Groot}, {Price}, {Sanders}, {Shivaei}, {Brammer},
  {Momcheva}, {Skelton}, {van Dokkum}, {Whitaker}, {Aird}, {Azadi}, {Kassis},
  {Bullock}, {Conroy}, {Dav{\'e}}, {Kere{\v s}}, \& {Krumholz}}]{2015Kriek}
{Kriek}, M., {Shapley}, A.~E., {Reddy}, N.~A., {et~al.} 2015, \apjs, 218, 15

\bibitem[{{Krogager} {et~al.}(2014){Krogager}, {Zirm}, {Toft}, {Man}, \&
  {Brammer}}]{2014ApJ...797...17K}
{Krogager}, J.-K., {Zirm}, A.~W., {Toft}, S., {Man}, A., \& {Brammer}, G. 2014,
  \apj, 797, 17

\bibitem[{Kron(1980)}]{1980ApJS...43..305K}
Kron, R.~G. 1980, \apjs, 43, 305

\bibitem[{{Laigle} {et~al.}(2015){Laigle}, {Pichon}, {Codis}, {Dubois}, {Le
  Borgne}, {Pogosyan}, {Devriendt}, {Peirani}, {Prunet}, {Rouberol}, {Slyz}, \&
  {Sousbie}}]{2015MNRAS.446.2744L}
{Laigle}, C., {Pichon}, C., {Codis}, S., {et~al.} 2015, \mnras, 446, 2744

\bibitem[{Landy \& Szalay(1993)}]{Landy:1993p11082}
Landy, S.~D., \& Szalay, A.~S. 1993, \apj, 412, 64

\bibitem[{{Le Fevre} {et~al.}(2015){Le Fevre}, {Tasca}, {Cassata}, {Garilli},
  {Le Brun}, {Maccagni}, {Pentericci}, {Thomas}, {Vanzella}, {Zamorani},
  {Zucca}, {Amorin}, {Bardelli}, {Capak}, {Cassar{\`a}}, {Castellano},
  {Cimatti}, {Cuby}, {Cucciati}, {de la Torre}, {Durkalec}, {Fontana},
  {Giavalisco}, {Grazian}, {Hathi}, {Ilbert}, {Lemaux}, {Moreau}, {Paltani},
  {Ribeiro}, {Salvato}, {Schaerer}, {Scodeggio}, {Sommariva}, {Talia},
  {Taniguchi}, {Tresse}, {Vergani}, {Wang}, {Charlot}, {Contini}, {Fotopoulou},
  {L{\'o}pez-Sanjuan}, {Mellier}, \& {Scoville}}]{2015AA...576A..79L}
{Le Fevre}, O., {Tasca}, L.~A.~M., {Cassata}, P., {et~al.} 2015, \aap, 576, A79

\bibitem[{{Le Floc'h} {et~al.}(2009){Le Floc'h}, {Aussel}, {Ilbert},
  {Riguccini}, {Frayer}, {Salvato}, {Arnouts}, {Surace}, {Feruglio},
  {Rodighiero}, {Capak}, {Kartaltepe}, {Heinis}, {Sheth}, {Yan}, {McCracken},
  {Thompson}, {Sanders}, {Scoville}, \& {Koekemoer}}]{2009ApJ...703..222L}
{Le Floc'h}, E., {Aussel}, H., {Ilbert}, O., {et~al.} 2009, \apj, 703, 222

\bibitem[{Leauthaud {et~al.}(2007)Leauthaud, Massey, Kneib, Rhodes, Johnston,
  Capak, Heymans, Ellis, Koekemoer, Le~F{\`e}vre, Mellier, Refregier, Robin,
  Scoville, Tasca, Taylor, \& Van~Waerbeke}]{2007ApJS..172..219L}
Leauthaud, A., Massey, R., Kneib, J.-P., {et~al.} 2007, \apjs, 172, 219

\bibitem[{{Lee} {et~al.}(2012){Lee}, {Alberts}, {Atlee}, {Dey}, {Pope},
  {Jannuzi}, {Reddy}, \& {Brown}}]{2012ApJ...758L..31L}
{Lee}, K.-S., {Alberts}, S., {Atlee}, D., {et~al.} 2012, \apjl, 758, L31

\bibitem[{{Lee} {et~al.}(2015){Lee}, {Sanders}, {Casey}, {Toft}, {Scoville},
  {Hung}, {Le Floc'h}, {Ilbert}, {Zahid}, {Aussel}, {Capak}, {Kartaltepe},
  {Kewley}, {Li}, {Schawinski}, {Sheth}, \& {Xiao}}]{2015ApJ...801...80L}
{Lee}, N., {Sanders}, D.~B., {Casey}, C.~M., {et~al.} 2015, \apj, 801, 80

\bibitem[{{Libeskind} {et~al.}(2013){Libeskind}, {Hoffman}, {Steinmetz},
  {Gottl{\"o}ber}, {Knebe}, \& {Hess}}]{2013ApJ...766L..15L}
{Libeskind}, N.~I., {Hoffman}, Y., {Steinmetz}, M., {et~al.} 2013, \apjl, 766,
  L15

\bibitem[{Lilly {et~al.}(2007)Lilly, Le~F{\`e}vre, Renzini, Zamorani,
  Scodeggio, Contini, Carollo, Hasinger, Kneib, Iovino, Le~Brun, Maier,
  Mainieri, Mignoli, Silverman, Tasca, Bolzonella, Bongiorno, Bottini, Capak,
  Caputi, Cimatti, Cucciati, Daddi, Feldmann, Franzetti, Garilli, Guzzo,
  Ilbert, Kampczyk, Kovac, Lamareille, Leauthaud, Borgne, McCracken, Marinoni,
  Pell{\`o}, Ricciardelli, Scarlata, Vergani, Sanders, Schinnerer, Scoville,
  Taniguchi, Arnouts, Aussel, Bardelli, Brusa, Cappi, Ciliegi, Finoguenov,
  Foucaud, Franceschini, Halliday, Impey, Knobel, Koekemoer, Kurk, Maccagni,
  Maddox, Marano, Marconi, Meneux, Mobasher, Moreau, Peacock, Porciani,
  Pozzetti, Scaramella, Schiminovich, Shopbell, Smail, Thompson, Tresse,
  Vettolani, Zanichelli, \& Zucca}]{2007ApJS..172...70L}
Lilly, S.~J., Le~F{\`e}vre, O., Renzini, A., {et~al.} 2007, \apjs, 172, 70

\bibitem[{{Lin} {et~al.}(2014){Lin}, {Jian}, {Foucaud}, {Norberg}, {Bower},
  {Cole}, {Arnalte-Mur}, {Chen}, {Coupon}, {Hsieh}, {Heinis}, {Phleps}, {Chen},
  {Lee}, {Burgett}, {Chambers}, {Denneau}, {Draper}, {Flewelling}, {Hodapp},
  {Huber}, {Kaiser}, {Kudritzki}, {Magnier}, {Metcalfe}, {Price}, {Tonry},
  {Wainscoat}, \& {Waters}}]{2014ApJ...782...33L}
{Lin}, L., {Jian}, H.-Y., {Foucaud}, S., {et~al.} 2014, \apj, 782, 33

\bibitem[{{Lutz} {et~al.}(2011){Lutz}, {Poglitsch}, {Altieri}, {Andreani},
  {Aussel}, {Berta}, {Bongiovanni}, {Brisbin}, {Cava}, {Cepa}, {Cimatti},
  {Daddi}, {Dominguez-Sanchez}, {Elbaz}, {F{\"o}rster Schreiber}, {Genzel},
  {Grazian}, {Gruppioni}, {Harwit}, {Le Floc'h}, {Magdis}, {Magnelli},
  {Maiolino}, {Nordon}, {P{\'e}rez Garc{\'{\i}}a}, {Popesso}, {Pozzi},
  {Riguccini}, {Rodighiero}, {Saintonge}, {Sanchez Portal}, {Santini}, {Shao},
  {Sturm}, {Tacconi}, {Valtchanov}, {Wetzstein}, \&
  {Wieprecht}}]{2011A&A...532A..90L}
{Lutz}, D., {Poglitsch}, A., {Altieri}, B., {et~al.} 2011, \aap, 532, A90

\bibitem[{{Marchesi} {et~al.}(2016){Marchesi}, {Civano}, {Elvis}, {Salvato},
  {Brusa}, {Comastri}, {Gilli}, {Hasinger}, {Lanzuisi}, {Miyaji}, {Treister},
  {Urry}, {Vignali}, {Zamorani}, {Allevato}, {Cappelluti}, {Cardamone},
  {Finoguenov}, {Griffiths}, {Karim}, {Laigle}, {LaMassa}, {Jahnke}, {Ranalli},
  {Schawinski}, {Schinnerer}, {Silverman}, {Smolcic}, {Suh}, \&
  {Trakhtenbrot}}]{2016Marchesi}
{Marchesi}, S., {Civano}, F., {Elvis}, M., {et~al.} 2016, \apj, 817, 34

\bibitem[{{Massey} \& {Refregier}(2005)}]{2005MNRAS.363..197M}
{Massey}, R., \& {Refregier}, A. 2005, \mnras, 363, 197

\bibitem[{McCracken {et~al.}(2007)McCracken, Peacock, Guzzo, Capak, Porciani,
  Scoville, Aussel, Finoguenov, James, Kitzbichler, Koekemoer, Leauthaud,
  Le~F{\`e}vre, Massey, Mellier, Mobasher, Norberg, Rhodes, Sanders, Sasaki,
  Taniguchi, Thompson, White, \& El-Zant}]{2007ApJS..172..314M}
McCracken, H.~J., Peacock, J.~A., Guzzo, L., {et~al.} 2007, \apjs, 172, 314

\bibitem[{McCracken {et~al.}(2010)McCracken, Capak, Salvato, Aussel, Thompson,
  Daddi, Sanders, Kneib, Willott, Mancini, Renzini, Cook, Le~F{\`e}vre, Ilbert,
  Kartaltepe, Koekemoer, Mellier, Murayama, Scoville, Shioya, \&
  Tanaguchi}]{2010ApJ...708..202M}
McCracken, H.~J., Capak, P., Salvato, M., {et~al.} 2010, \apj, 708, 202

\bibitem[{McCracken {et~al.}(2012)McCracken, Milvang-Jensen, Dunlop, Franx,
  Fynbo, Le~F{\`e}vre, Holt, Caputi, Goranova, Buitrago, Emerson, Freudling,
  Hudelot, L{\'o}pez-Sanjuan, Magnard, Mellier, M{\o}ller, Nilsson, Sutherland,
  Tasca, \& Zabl}]{McCracken:2012gd}
McCracken, H.~J., Milvang-Jensen, B., Dunlop, J., {et~al.} 2012, \aap, 544,
  A156

\bibitem[{{McCracken} {et~al.}(2015){McCracken}, {Wolk}, {Colombi},
  {Kilbinger}, {Ilbert}, {Peirani}, {Coupon}, {Dunlop}, {Milvang-Jensen},
  {Caputi}, {Aussel}, {B{\'e}thermin}, \& {Le F{\`e}vre}}]{2015McCracken}
{McCracken}, H.~J., {Wolk}, M., {Colombi}, S., {et~al.} 2015, \mnras, 449, 901

\bibitem[{{Milliard} {et~al.}(2007){Milliard}, {Heinis}, {Blaizot}, {Arnouts},
  {Schiminovich}, {Budav{\'a}ri}, {Donas}, {Treyer}, {Laget}, {Viton}, {Wyder},
  {Szalay}, {Barlow}, {Forster}, {Friedman}, {Martin}, {Morrissey}, {Neff},
  {Seibert}, {Small}, {Bianchi}, {Heckman}, {Lee}, {Madore}, {Rich}, {Welsh},
  {Yi}, \& {Xu}}]{2007ApJS..173..494M}
{Milliard}, B., {Heinis}, S., {Blaizot}, J., {et~al.} 2007, \apjs, 173, 494

\bibitem[{{Milvang-Jensen} {et~al.}(2013){Milvang-Jensen}, {Freudling}, {Zabl},
  {Fynbo}, {M{\o}ller}, {Nilsson}, {McCracken}, {Hjorth}, {Le F{\`e}vre},
  {Tasca}, {Dunlop}, \& {Sobral}}]{2013A&A...560A..94M}
{Milvang-Jensen}, B., {Freudling}, W., {Zabl}, J., {et~al.} 2013, \aap, 560,
  A94

\bibitem[{{Miyazaki} {et~al.}(2012){Miyazaki}, {Komiyama}, {Nakaya}, {Kamata},
  {Doi}, {Hamana}, {Karoji}, {Furusawa}, {Kawanomoto}, {Morokuma}, {Ishizuka},
  {Nariai}, {Tanaka}, {Uraguchi}, {Utsumi}, {Obuchi}, {Okura}, {Oguri},
  {Takata}, {Tomono}, {Kurakami}, {Namikawa}, {Usuda}, {Yamanoi}, {Terai},
  {Uekiyo}, {Yamada}, {Koike}, {Aihara}, {Fujimori}, {Mineo}, {Miyatake},
  {Yasuda}, {Nishizawa}, {Saito}, {Tanaka}, {Uchida}, {Katayama}, {Wang},
  {Chen}, {Lupton}, {Loomis}, {Bickerton}, {Price}, {Gunn}, {Suzuki},
  {Miyazaki}, {Muramatsu}, {Yamamoto}, {Endo}, {Ezaki}, {Itoh}, {Miwa},
  {Yokota}, {Matsuda}, {Ebinuma}, \& {Takeshi}}]{2012SPIE.8446E..0ZM}
{Miyazaki}, S., {Komiyama}, Y., {Nakaya}, H., {et~al.} 2012, in Society of
  Photo-Optical Instrumentation Engineers (SPIE) Conference Series, Vol. 8446,
  Society of Photo-Optical Instrumentation Engineers (SPIE) Conference Series,
  0

\bibitem[{Moffat(1969)}]{Moffat:1969p12721}
Moffat, A. F.~J. 1969, \aap, 3, 455

\bibitem[{{Molino} {et~al.}(2014){Molino}, {Ben{\'{\i}}tez}, {Moles},
  {Fern{\'a}ndez-Soto}, {Crist{\'o}bal-Hornillos}, {Ascaso},
  {Jim{\'e}nez-Teja}, {Schoenell}, {Arnalte-Mur}, {Povi{\'c}}, {Coe},
  {L{\'o}pez-Sanjuan}, {D{\'{\i}}az-Garc{\'{\i}}a}, {Varela}, {Stefanon},
  {Cenarro}, {Matute}, {Masegosa}, {M{\'a}rquez}, {Perea}, {Del Olmo},
  {Husillos}, {Alfaro}, {Aparicio-Villegas}, {Cervi{\~n}o}, {Huertas-Company},
  {Aguerri}, {Broadhurst}, {Cabrera-Ca{\~n}o}, {Cepa}, {Gonz{\'a}lez},
  {Infante}, {Mart{\'{\i}}nez}, {Prada}, \& {Quintana}}]{2014MNRAS.441.2891M}
{Molino}, A., {Ben{\'{\i}}tez}, N., {Moles}, M., {et~al.} 2014, \mnras, 441,
  2891

\bibitem[{{Moustakas} {et~al.}(2013){Moustakas}, {Coil}, {Aird}, {Blanton},
  {Cool}, {Eisenstein}, {Mendez}, {Wong}, {Zhu}, \& {Arnouts}}]{2013Moustakas}
{Moustakas}, J., {Coil}, A.~L., {Aird}, J., {et~al.} 2013, \apj, 767, 50


\bibitem[{{Moutard} {et~al.}(2016){Moutard}, {Arnouts}, {Ilbert}}]{2016Moutard}
{Moutard}, T., {Arnouts}, S., {Ilbert}, O., {et~al.} 2016, ArXiv e-prints,
  arXiv:1602.05915

\bibitem[{{Muzzin} {et~al.}(2013){Muzzin}, {Marchesini}, {Stefanon}, {Franx},
  {Milvang-Jensen}, {Dunlop}, {Fynbo}, {Brammer}, {Labb{\'e}}, \& {van
  Dokkum}}]{2013ApJS..206....8M}
{Muzzin}, A., {Marchesini}, D., {Stefanon}, M., {et~al.} 2013, \apjs, 206, 8

\bibitem[{{Nilsson} {et~al.}(2009){Nilsson}, {Tapken}, {M{\o}ller},
  {Freudling}, {Fynbo}, {Meisenheimer}, {Laursen}, \&
  {{\"O}stlin}}]{2009A&A...498...13N}
{Nilsson}, K.~K., {Tapken}, C., {M{\o}ller}, P., {et~al.} 2009, \aap, 498, 13

\bibitem[{Noeske {et~al.}(2007)Noeske, Weiner, Faber, Papovich, Koo,
  Somerville, Bundy, Conselice, Newman, Schiminovich, Floc'h, Coil, Rieke,
  Lotz, Primack, Barmby, Cooper, Davis, Ellis, Fazio, Guhathakurta, Huang,
  Kassin, Martin, Phillips, Rich, Small, Willmer, \& Wilson}]{Noeske:2007p5802}
Noeske, K.~G., Weiner, B.~J., Faber, S.~M., {et~al.} 2007, \apj, 660, L43

\bibitem[{{Ocvirk} {et~al.}(2008){Ocvirk}, {Pichon}, \&
  {Teyssier}}]{2008MNRAS.390.1326O}
{Ocvirk}, P., {Pichon}, C., \& {Teyssier}, R. 2008, \mnras, 390, 1326

\bibitem[{Oke(1974)}]{Oke:1974p12716}
Oke, J.~B. 1974, \apjs, 27, 21

\bibitem[{{Oke} \& {Sandage}(1968)}]{1968ApJ...154...21O}
{Oke}, J.~B., \& {Sandage}, A. 1968, \apj, 154, 21

\bibitem[{{Oliver} {et~al.}(2012){Oliver}, {Bock}, {Altieri}, {Amblard},
  {Arumugam}, {Aussel}, {Babbedge}, {Beelen}, {B{\'e}thermin}, {Blain},
  {Boselli}, {Bridge}, {Brisbin}, {Buat}, {Burgarella},
  {Castro-Rodr{\'{\i}}guez}, {Cava}, {Chanial}, {Cirasuolo}, {Clements},
  {Conley}, {Conversi}, {Cooray}, {Dowell}, {Dubois}, {Dwek}, {Dye}, {Eales},
  {Elbaz}, {Farrah}, {Feltre}, {Ferrero}, {Fiolet}, {Fox}, {Franceschini},
  {Gear}, {Giovannoli}, {Glenn}, {Gong}, {Gonz{\'a}lez Solares}, {Griffin},
  {Halpern}, {Harwit}, {Hatziminaoglou}, {Heinis}, {Hurley}, {Hwang}, {Hyde},
  {Ibar}, {Ilbert}, {Isaak}, {Ivison}, {Lagache}, {Le Floc'h}, {Levenson},
  {Faro}, {Lu}, {Madden}, {Maffei}, {Magdis}, {Mainetti}, {Marchetti},
  {Marsden}, {Marshall}, {Mortier}, {Nguyen}, {O'Halloran}, {Omont}, {Page},
  {Panuzzo}, {Papageorgiou}, {Patel}, {Pearson}, {P{\'e}rez-Fournon}, {Pohlen},
  {Rawlings}, {Raymond}, {Rigopoulou}, {Riguccini}, {Rizzo}, {Rodighiero},
  {Roseboom}, {Rowan-Robinson}, {S{\'a}nchez Portal}, {Schulz}, {Scott},
  {Seymour}, {Shupe}, {Smith}, {Stevens}, {Symeonidis}, {Trichas}, {Tugwell},
  {Vaccari}, {Valtchanov}, {Vieira}, {Viero}, {Vigroux}, {Wang}, {Ward},
  {Wardlow}, {Wright}, {Xu}, \& {Zemcov}}]{2012MNRAS.424.1614O}
{Oliver}, S.~J., {Bock}, J., {Altieri}, B., {et~al.} 2012, \mnras, 424, 1614

\bibitem[{{Onodera} {et~al.}(2012){Onodera}, {Renzini}, {Carollo},
  {Cappellari}, {Mancini}, {Strazzullo}, {Daddi}, {Arimoto}, {Gobat}, {Yamada},
  {McCracken}, {Ilbert}, {Capak}, {Cimatti}, {Giavalisco}, {Koekemoer}, {Kong},
  {Lilly}, {Motohara}, {Ohta}, {Sanders}, {Scoville}, {Tamura}, \&
  {Taniguchi}}]{2012ApJ...755...26O}
{Onodera}, M., {Renzini}, A., {Carollo}, M., {et~al.} 2012, \apj, 755, 26

%\bibitem[{Peebles(1980)}]{Peebles:1980p5506}
%Peebles, P. J.~E. 1980, Research supported by the National Science Foundation.
%  Princeton


\bibitem[{Peng {et~al.}(2010)Peng, Lilly, Kova{\v c}, Bolzonella, Pozzetti,
  Renzini, Zamorani, Ilbert, Knobel, Iovino, Maier, Cucciati, Tasca, Carollo,
  Silverman, Kampczyk, de~Ravel, Sanders, Scoville, Contini, Mainieri,
  Scodeggio, Kneib, Le~F{\`e}vre, Bardelli, Bongiorno, Caputi, Coppa, de~la
  Torre, Franzetti, Garilli, Lamareille, Le~Borgne, Le~Brun, Mignoli, Montero,
  Pello, Ricciardelli, Tanaka, Tresse, Vergani, Welikala, Zucca, Oesch, Abbas,
  Barnes, Bordoloi, Bottini, Cappi, Cassata, Cimatti, Fumana, Hasinger,
  Koekemoer, Leauthaud, Maccagni, Marinoni, McCracken, Memeo, Meneux, Nair,
  Porciani, Presotto, \& Scaramella}]{Peng:2010p11940}
Peng, Y.-j., Lilly, S.~J., Kova{\v c}, K., {et~al.} 2010, \apj, 721, 193

\bibitem[{{Poglitsch} {et~al.}(2010){Poglitsch}, {Waelkens}, {Geis},
  {Feuchtgruber}, {Vandenbussche}, {Rodriguez}, {Krause}, {Renotte}, {van
  Hoof}, {Saraceno}, {Cepa}, {Kerschbaum}, {Agn{\`e}se}, {Ali}, {Altieri},
  {Andreani}, {Augueres}, {Balog}, {Barl}, {Bauer}, {Belbachir}, {Benedettini},
  {Billot}, {Boulade}, {Bischof}, {Blommaert}, {Callut}, {Cara}, {Cerulli},
  {Cesarsky}, {Contursi}, {Creten}, {De Meester}, {Doublier}, {Doumayrou},
  {Duband}, {Exter}, {Genzel}, {Gillis}, {Gr{\"o}zinger}, {Henning},
  {Herreros}, {Huygen}, {Inguscio}, {Jakob}, {Jamar}, {Jean}, {de Jong},
  {Katterloher}, {Kiss}, {Klaas}, {Lemke}, {Lutz}, {Madden}, {Marquet},
  {Martignac}, {Mazy}, {Merken}, {Montfort}, {Morbidelli}, {M{\"u}ller},
  {Nielbock}, {Okumura}, {Orfei}, {Ottensamer}, {Pezzuto}, {Popesso},
  {Putzeys}, {Regibo}, {Reveret}, {Royer}, {Sauvage}, {Schreiber}, {Stegmaier},
  {Schmitt}, {Schubert}, {Sturm}, {Thiel}, {Tofani}, {Vavrek}, {Wetzstein},
  {Wieprecht}, \& {Wiezorrek}}]{2010Pog}
{Poglitsch}, A., {Waelkens}, C., {Geis}, N., {et~al.} 2010, \aap, 518, L2

\bibitem[{{Pogosyan} {et~al.}(1996){Pogosyan}, {Bond}, {Kofman}, \&
  {Wadsley}}]{1996AAS...189.1303P}
{Pogosyan}, D., {Bond}, J.~R., {Kofman}, L., \& {Wadsley}, J. 1996, in Bulletin
  of the American Astronomical Society, Vol.~28, American Astronomical Society
  Meeting Abstracts, 1289

\bibitem[{Polletta {et~al.}(2007)Polletta, Tajer, Maraschi, Trinchieri,
  Lonsdale, Chiappetti, Andreon, Pierre, Le~F{\`e}vre, Zamorani, Maccagni,
  Garcet, Surdej, Franceschini, Alloin, Shupe, Surace, Fang, Rowan-Robinson,
  Smith, \& Tresse}]{Polletta:2007p6857}
Polletta, M., Tajer, M., Maraschi, L., {et~al.} 2007, \apj, 663, 81

\bibitem[{Pozzetti {et~al.}(2007)Pozzetti, Bolzonella, Lamareille, Zamorani,
  Franzetti, Le~F{\`e}vre, Iovino, Temporin, Ilbert, Arnouts, Charlot,
  Brinchmann, Zucca, Tresse, Scodeggio, Guzzo, Bottini, Garilli, Le~Brun,
  Maccagni, Picat, Scaramella, Vettolani, Zanichelli, Adami, Bardelli, Cappi,
  Ciliegi, Contini, Foucaud, Gavignaud, McCracken, Marano, Marinoni, Mazure,
  Meneux, Merighi, Paltani, Pell{\`o}, Pollo, Radovich, Bondi, Bongiorno,
  Cucciati, de~la Torre, Gregorini, Mellier, Merluzzi, Vergani, \&
  Walcher}]{2007A&A...474..443P}
Pozzetti, L., Bolzonella, M., Lamareille, F., {et~al.} 2007, \aap, 474, 443

\bibitem[{{Pozzetti} {et~al.}(2010){Pozzetti}, {Bolzonella}, {Zucca},
  {Zamorani}, {Lilly}, {Renzini}, {Moresco}, {Mignoli}, {Cassata}, {Tasca},
  {Lamareille}, {Maier}, {Meneux}, {Halliday}, {Oesch}, {Vergani}, {Caputi},
  {Kova{\v c}}, {Cimatti}, {Cucciati}, {Iovino}, {Peng}, {Carollo}, {Contini},
  {Kneib}, {Le F{\'e}vre}, {Mainieri}, {Scodeggio}, {Bardelli}, {Bongiorno},
  {Coppa}, {de la Torre}, {de Ravel}, {Franzetti}, {Garilli}, {Kampczyk},
  {Knobel}, {Le Borgne}, {Le Brun}, {Pell{\`o}}, {Perez Montero},
  {Ricciardelli}, {Silverman}, {Tanaka}, {Tresse}, {Abbas}, {Bottini}, {Cappi},
  {Guzzo}, {Koekemoer}, {Leauthaud}, {Maccagni}, {Marinoni}, {McCracken},
  {Memeo}, {Porciani}, {Scaramella}, {Scarlata}, \& {Scoville}}]{2010Pozetti}
{Pozzetti}, L., {Bolzonella}, M., {Zucca}, E., {et~al.} 2010, \aap, 523, A13

\bibitem[{Prevot {et~al.}(1984)Prevot, Lequeux, Prevot, Maurice, \&
  Rocca-Volmerange}]{Prevot:1984p6814}
Prevot, M.~L., Lequeux, J., Prevot, L., Maurice, E., \& Rocca-Volmerange, B.
  1984, Astronomy and Astrophysics (ISSN 0004-6361), 132, 389

\bibitem[{{Rees} \& {Ostriker}(1977)}]{1977MNRAS.179..541R}
{Rees}, M.~J., \& {Ostriker}, J.~P. 1977, \mnras, 179, 541

\bibitem[{{Roseboom} {et~al.}(2012){Roseboom}, {Bunker}, {Sumiyoshi}, {Wang},
  {Dalton}, {Akiyama}, {Bock}, {Bonfield}, {Buat}, {Casey}, {Chapin},
  {Clements}, {Conley}, {Curtis-Lake}, {Cooray}, {Dunlop}, {Farrah}, {Ham},
  {Ibar}, {Iwamuro}, {Kimura}, {Lewis}, {Macaulay}, {Magdis}, {Maihara},
  {Marsden}, {Mauch}, {Moritani}, {Ohta}, {Oliver}, {Page}, {Schulz}, {Scott},
  {Symeonidis}, {Takato}, {Tamura}, {Totani}, {Yabe}, \&
  {Zemcov}}]{2012MNRAS.426.1782R}
{Roseboom}, I.~G., {Bunker}, A., {Sumiyoshi}, M., {et~al.} 2012, \mnras, 426,
  1782

\bibitem[{{Salmon} {et~al.}(2015){Salmon}, {Papovich}, {Finkelstein}, {Tilvi},
  {Finlator}, {Behroozi}, {Dahlen}, {Dav{\'e}}, {Dekel}, {Dickinson},
  {Ferguson}, {Giavalisco}, {Long}, {Lu}, {Mobasher}, {Reddy}, {Somerville}, \&
  {Wechsler}}]{2015ApJ...799..183S}
{Salmon}, B., {Papovich}, C., {Finkelstein}, S.~L., {et~al.} 2015, \apj, 799,
  183

\bibitem[{{Salvato} {et~al.}(2011){Salvato}, {Ilbert}, {Hasinger}, {Rau},
  {Civano}, {Zamorani}, {Brusa}, {Elvis}, {Vignali}, {Aussel}, {Comastri},
  {Fiore}, {Le Floc'h}, {Mainieri}, {Bardelli}, {Bolzonella}, {Bongiorno},
  {Capak}, {Caputi}, {Cappelluti}, {Carollo}, {Contini}, {Garilli}, {Iovino},
  {Fotopoulou}, {Fruscione}, {Gilli}, {Halliday}, {Kneib}, {Kakazu},
  {Kartaltepe}, {Koekemoer}, {Kovac}, {Ideue}, {Ikeda}, {Impey}, {Le Fevre},
  {Lamareille}, {Lanzuisi}, {Le Borgne}, {Le Brun}, {Lilly}, {Maier},
  {Manohar}, {Masters}, {McCracken}, {Messias}, {Mignoli}, {Mobasher}, {Nagao},
  {Pello}, {Puccetti}, {Perez-Montero}, {Renzini}, {Sargent}, {Sanders},
  {Scodeggio}, {Scoville}, {Shopbell}, {Silvermann}, {Taniguchi}, {Tasca},
  {Tresse}, {Trump}, \& {Zucca}}]{2011ApJ...742...61S}
{Salvato}, M., {Ilbert}, O., {Hasinger}, G., {et~al.} 2011, \apj, 742, 61

\bibitem[{Sanders {et~al.}(2007)Sanders, Salvato, Aussel, Ilbert, Scoville,
  Surace, Frayer, Sheth, Helou, Brooke, Bhattacharya, Yan, Kartaltepe, Barnes,
  Blain, Calzetti, Capak, Carilli, Carollo, Comastri, Daddi, Ellis, Elvis,
  Fall, Franceschini, Giavalisco, Hasinger, Impey, Koekemoer, Le~F{\`e}vre,
  Lilly, Liu, McCracken, Mobasher, Renzini, Rich, Schinnerer, Shopbell,
  Taniguchi, Thompson, Urry, \& Williams}]{2007ApJS..172...86S}
Sanders, D.~B., Salvato, M., Aussel, H., {et~al.} 2007, \apjs, 172, 86

\bibitem[{{Schaye} {et~al.}(2015){Schaye}, {Crain}, {Bower}, {Furlong},
  {Schaller}, {Theuns}, {Dalla Vecchia}, {Frenk}, {McCarthy}, {Helly},
  {Jenkins}, {Rosas-Guevara}, {White}, {Baes}, {Booth}, {Camps}, {Navarro},
  {Qu}, {Rahmati}, {Sawala}, {Thomas}, \& {Trayford}}]{2015MNRAS.446..521S}
{Schaye}, J., {Crain}, R.~A., {Bower}, R.~G., {et~al.} 2015, \mnras, 446, 521

\bibitem[{Schinnerer {et~al.}(2004)Schinnerer, Carilli, Scoville, Bondi,
  Ciliegi, Vettolani, Le~F{\`e}vre, Koekemoer, Bertoldi, \&
  Impey}]{Schinnerer:2004p12717}
Schinnerer, E., Carilli, C.~L., Scoville, N.~Z., {et~al.} 2004, \aj, 128, 1974

\bibitem[{Schlegel {et~al.}(1998)Schlegel, Finkbeiner, \&
  Davis}]{1998ApJ...500..525S}
Schlegel, D.~J., Finkbeiner, D.~P., \& Davis, M. 1998, \apj, 500, 525

\bibitem[{Scoville {et~al.}(2007)Scoville, Aussel, Brusa, Capak, Carollo,
  Elvis, Giavalisco, Guzzo, Hasinger, Impey, Kneib, Lefevre, Lilly, Mobasher,
  Renzini, Rich, Sanders, Schinnerer, Schminovich, Shopbell, Taniguchi, \&
  Tyson}]{Scoville:2007p12720}
Scoville, N., Aussel, H., Brusa, M., {et~al.} 2007, \apjs, 172, 1

\bibitem[{{Scoville} {et~al.}(2013){Scoville}, {Arnouts}, {Aussel}, {Benson},
  {Bongiorno}, {Bundy}, {Calvo}, {Capak}, {Carollo}, {Civano}, {Dunlop},
  {Elvis}, {Faisst}, {Finoguenov}, {Fu}, {Giavalisco}, {Guo}, {Ilbert},
  {Iovino}, {Kajisawa}, {Kartaltepe}, {Leauthaud}, {Le F{\`e}vre}, {LeFloch},
  {Lilly}, {Liu}, {Manohar}, {Massey}, {Masters}, {McCracken}, {Mobasher},
  {Peng}, {Renzini}, {Rhodes}, {Salvato}, {Sanders}, {Sarvestani}, {Scarlata},
  {Schinnerer}, {Sheth}, {Shopbell}, {Smol{\v c}i{\'c}}, {Taniguchi}, {Taylor},
  {White}, \& {Yan}}]{2013ApJS..206....3S}
{Scoville}, N., {Arnouts}, S., {Aussel}, H., {et~al.} 2013, \apjs, 206, 3

\bibitem[{{Silverman} {et~al.}(2015){Silverman}, {Kashino}, {Sanders},
  {Kartaltepe}, {Arimoto}, {Renzini}, {Rodighiero}, {Daddi}, {Zahid}, {Nagao},
  {Kewley}, {Lilly}, {Sugiyama}, {Baronchelli}, {Capak}, {Carollo}, {Chu},
  {Hasinger}, {Ilbert}, {Juneau}, {Kajisawa}, {Koekemoer}, {Kovac}, {Le
  F{\`e}vre}, {Masters}, {McCracken}, {Onodera}, {Schulze}, {Scoville},
  {Strazzullo}, \& {Taniguchi}}]{2015Silverman}
{Silverman}, J.~D., {Kashino}, D., {Sanders}, D., {et~al.} 2015, \apjs, 220, 12

\bibitem[{{Steinhardt} {et~al.}(2014){Steinhardt}, {Speagle}, {Capak},
  {Silverman}, {Carollo}, {Dunlop}, {Hashimoto}, {Hsieh}, {Ilbert}, {Le Fevre},
  {Le Floc'h}, {Lee}, {Lin}, {Lin}, {Masters}, {McCracken}, {Nagao}, {Petric},
  {Salvato}, {Sanders}, {Scoville}, {Sheth}, {Strauss}, \&
  {Taniguchi}}]{2014ApJ...791L..25S}
{Steinhardt}, C.~L., {Speagle}, J.~S., {Capak}, P., {et~al.} 2014, \apjl, 791,
  L25

\bibitem[{{Sutherland} \& {Saunders}(1992)}]{1992Sutherland}
{Sutherland}, W., \& {Saunders}, W. 1992, \mnras, 259, 413

\bibitem[{Szalay {et~al.}(1999)Szalay, Connolly, \& Szokoly}]{Szalay:1999p4804}
Szalay, A.~S., Connolly, A.~J., \& Szokoly, G.~P. 1999, \aj, 117, 68

\bibitem[{{Taniguchi} {et~al.}(2007){Taniguchi}, {Scoville}, {Murayama},
  {Sanders}, {Mobasher}, {Aussel}, {Capak}, {Ajiki}, {Miyazaki}, {Komiyama},
  {Shioya}, {Nagao}, {Sasaki}, {Koda}, {Carilli}, {Giavalisco}, {Guzzo},
  {Hasinger}, {Impey}, {LeFevre}, {Lilly}, {Renzini}, {Rich}, {Schinnerer},
  {Shopbell}, {Kaifu}, {Karoji}, {Arimoto}, {Okamura}, \&
  {Ohta}}]{2007ApJS..172....9T}
{Taniguchi}, Y., {Scoville}, N., {Murayama}, T., {et~al.} 2007, \apjs, 172, 9

\bibitem[{{Taniguchi} {et~al.}(2015){Taniguchi}, {Kajisawa}, {Kobayashi},
  {Shioya}, {Nagao}, {Capak}, {Aussel}, {Ichikawa}, {Murayama}, {Scoville},
  {Ilbert}, {Salvato}, {Sanders}, {Mobasher}, {Miyazaki}, {Komiyama}, {Le
  F{\`e}vre}, {Tasca}, {Lilly}, {Carollo}, {Renzini}, {Rich}, {Schinnerer},
  {Kaifu}, {Karoji}, {Arimoto}, {Okamura}, {Ohta}, {Shimasaku}, \&
  {Hayashino}}]{2015Taniguchi}
{Taniguchi}, Y., {Kajisawa}, M., {Kobayashi}, M.~A.~R., {et~al.} 2015, \pasj,
  67, 104

\bibitem[{{Tempel} \& {Libeskind}(2013)}]{2013ApJ...775L..42T}
{Tempel}, E., \& {Libeskind}, N.~I. 2013, \apjl, 775, L42

\bibitem[{{Vogelsberger} {et~al.}(2014){Vogelsberger}, {Genel}, {Springel},
  {Torrey}, {Sijacki}, {Xu}, {Snyder}, {Nelson}, \&
  {Hernquist}}]{2014MNRAS.444.1518V}
{Vogelsberger}, M., {Genel}, S., {Springel}, V., {et~al.} 2014, \mnras, 444,
  1518

\bibitem[{{Welker} {et~al.}(2014){Welker}, {Devriendt}, {Dubois}, {Pichon}, \&
  {Peirani}}]{2014MNRAS.445L..46W}
{Welker}, C., {Devriendt}, J., {Dubois}, Y., {Pichon}, C., \& {Peirani}, S.
  2014, \mnras, 445, L46

\bibitem[{{White} \& {Rees}(1978)}]{1978MNRAS.183..341W}
{White}, S.~D.~M., \& {Rees}, M.~J. 1978, \mnras, 183, 341

\bibitem[{Williams {et~al.}(2009)Williams, Quadri, Franx, van Dokkum, \&
  Labb{\'e}}]{Williams:2009p10339}
Williams, R., Quadri, R., Franx, M., van Dokkum, P., \& Labb{\'e}, I. 2009,
  \apj, 691, 1879

\bibitem[{Zabl(2015)}]{phdthesisZabl15}
Zabl, J. 2015, PhD thesis, The Niels Bohr Institute, Faculty of Science,
  University of Copenhagen

\bibitem[{{Zamojski} {et~al.}(2007){Zamojski}, {Schiminovich}, {Rich},
  {Mobasher}, {Koekemoer}, {Capak}, {Taniguchi}, {Sasaki}, {McCracken},
  {Mellier}, {Bertin}, {Aussel}, {Sanders}, {Le F{\`e}vre}, {Ilbert},
  {Salvato}, {Thompson}, {Kartaltepe}, {Scoville}, {Barlow}, {Forster},
  {Friedman}, {Martin}, {Morrissey}, {Neff}, {Seibert}, {Small}, {Wyder},
  {Bianchi}, {Donas}, {Heckman}, {Lee}, {Madore}, {Milliard}, {Szalay},
  {Welsh}, \& {Yi}}]{2007ApJS..172..468Z}
{Zamojski}, M.~A., {Schiminovich}, D., {Rich}, R.~M., {et~al.} 2007, \apjs,
  172, 468

\end{thebibliography}

\end{document}